\documentclass[a4paper,fleqn]{cas-sc}
\pdfoutput=1 
\usepackage[square,sort,comma,numbers]{natbib}

\def\tsc#1{\csdef{#1}{\textsc{\lowercase{#1}}\xspace}}
\tsc{WGM}
\tsc{QE}
\tsc{EP}
\tsc{PMS}
\tsc{BEC}
\tsc{DE}
\usepackage{graphicx}
\graphicspath{ {./images/} }
\usepackage{natbib}
\usepackage{hyperref}
\usepackage{picins}

\begin{document}
\let\WriteBookmarks\relax
\def\floatpagepagefraction{1}
\def\textpagefraction{.001}

\shorttitle{AI-based Fog and Edge Computing: A Systematic Review, Taxonomy and Future Directions}

\shortauthors{Sundas Iftikhar et~al.}

\title [mode = title]{AI-based Fog and Edge Computing: A Systematic Review, Taxonomy and Future Directions}

\author[1, 2]{Sundas Iftikhar} 

\ead{s.iftikhar@qmul.ac.uk}

\affiliation[1]{organization={School of Electronic Engineering and Computer Science, Queen Mary University of London, London, UK}}

\affiliation[2]{organization={University of Kotli Azad Jammu \& Kashmir, Kotli,  Azad Kashmir, Pakistan}}

\author[1]{Sukhpal Singh Gill*} [orcid=0000-0002-3913-0369] 

\ead{s.s.gill@qmul.ac.uk}

\author[3]{Chenghao Song} 

\ead{ch.song@siat.ac.cn}

\affiliation[3]{organization={Shenzhen Institute of Advanced Technology, Chinese Academy of Sciences, Shenzhen, China}}

\author[3]{Minxian Xu} 

\ead{mx.xu@siat.ac.cn}

\author[4,5]{Mohammad Sadegh Aslanpour} 

\ead{mohammad.aslanpour@monash.edu}

\affiliation[4]{organization={Department of Software Systems and Cybersecurity, Faculty of Information Technology, Monash University, Australia}}

\affiliation[5]{organization={CSIRO DATA61, Australia}}

\author[4]{Adel N. Toosi} 

\ead{adel.n.toosi@monash.edu}

\author[6]{Junhui Du} 

\ead{dujunhui_0325@tju.edu.cn} 

\affiliation[6]{organization={Center for Applied Mathematics, Tianjin University, Tianjin, China}}

\author[6]{Huaming Wu} 

\ead{whming@tju.edu.cn}

\author[7]{Shreya Ghosh} 

\ead{spg5897@psu.edu}

\affiliation[7]{organization={The Pennsylvania State University, Pennsylvania, USA}}

\author[8]{Deepraj Chowdhury} 

\ead{deepraj19101@iiitnr.edu.in}

\affiliation[8]{organization={Department of Electronics \& Communication Engineering, International Institute of Information Technology (IIIT), Naya Raipur, India}}

\author[1,9]{Muhammed Golec} 

\ead{m.golec@qmul.ac.uk}

\affiliation[9]{organization={Abdullah Gül University, Kayseri, Turkey}}

\author[10]{Mohit Kumar}

\ead{kumarmohit@nitj.ac.in}

\affiliation[9]{organization={Department of Information Technology, National Institute of Technology, Jalandhar, India}}

\author[1]{Ahmed M. Abdelmoniem} 

\ead{ahmed.sayed@qmul.ac.uk}

\author[12]{Felix Cuadrado} 

\ead{felix.cuadrado@upm.es}

\affiliation[12]{organization={Technical University of Madrid (UPM), Spain}}

\author[13]{Blesson Varghese} 

\ead{blesson@st-andrews.ac.uk}

\affiliation[13]{organization={School of Computer Science, University of St Andrews, Scotland, UK }}

\author[14]{Omer Rana} 

\ead{ranaof@cardiff.ac.uk}

\affiliation[14]{organization={School of Computer Science and Informatics, Cardiff University, Cardiff, UK}}

\author[15]{Schahram Dustdar} 

\ead{dustdar@dsg.tuwien.ac.at}

\affiliation[15]{organization={Distributed Systems Group, Vienna University of Technology, Vienna, Austria}}

\author[1]{Steve Uhlig} 

\ead{steve.uhlig@qmul.ac.uk}

\fntext[fn2]{Sukhpal Singh Gill co-led this work with first author.}

\cortext[cor1]{Corresponding author at: School of Electronic Engineering and Computer Science, Queen Mary University of London, London, E1 4NS, UK.}

\begin{abstract}
Resource management in computing is a very challenging problem that involves making sequential decisions. Resource limitations, resource heterogeneity, dynamic and diverse nature of workload, and the unpredictability of fog/edge computing environments have made resource management even more challenging to be considered in the fog landscape. Recently Artificial Intelligence (AI) and Machine Learning (ML) based solutions are adopted to solve this problem. AI/ML methods with the capability to make sequential decisions like reinforcement learning seem most promising for these type of problems. But these algorithms come with their own challenges such as high variance, explainability, and online training. The continuously changing fog/edge environment dynamics require solutions that learn online, adopting changing computing environment. In this paper, we used standard review methodology to conduct this Systematic Literature Review (SLR) to analyze the role of AI/ML algorithms and the challenges in the applicability of these algorithms for resource management in fog/edge computing environments. Further, various machine learning, deep learning and reinforcement learning techniques for edge AI management have been discussed. Furthermore, we have presented the background and current status of AI/ML-based Fog/Edge Computing. Moreover, a taxonomy of AI/ML-based resource management techniques for fog/edge computing has been proposed and compared the existing techniques based on the proposed taxonomy. Finally, open challenges and promising future research directions have been identified and discussed in the area of AI/ML-based fog/edge computing.

\end{abstract}

 \begin{NoHyper}
\maketitle
\end{NoHyper}

\section{Introduction}
Most modern web applications now follow the standard practice of tapping into the remote computing resources provided by cloud data centers \cite{zhong2022machine}. Mobile phones, wearables, and other user devices, as well as sensors in a smart city or factory, all create data that is often sent to remote clouds for processing and storage \cite{dai2022task}. Due to the likelihood of a rise in communication latencies when billions of devices are linked to the Internet, this computing architecture is impractical for the long term \cite{gill2019transformative}. The increased communication latencies will negatively affect applications and lower the Quality of Service (QoS) \cite{hazra2022cooperative}. Bringing computing resources nearer to end devices and sensors and employing them for data processing is an alternate computing strategy that can help with the aforementioned issue (even if only partially) \cite{chakrabortyjourney}. This might lessen the load placed on the cloud and speed up communications. The recent fashion in computing research is to implement this concept by moving part of the processing power currently housed in huge data centers to the network's periphery, where it will be closer to end-users and sensors \cite{singh2021fog}. Internet of Things (IoT) devices such as routers, gateways, and switches may be equipped with computer resources, or specialised ``micro'' data centers may be built within public/private infrastructure for the ease of access and security \cite{sri2021deedsp}. \textit{``Edge computing'' refers to a computing model that takes advantage of network edge resources. \textcolor{black}{``Fog computing'' refers to a paradigm that employs both on-premises hardware and cloud services \cite{pujol2021fog}. }}Edge resources differ from cloud resources in several ways \cite{iftikhar2022fogdlearner, karagiannis2021context, murturi2022utilizing}: (a) they are resource constrained, meaning they have fewer computational capabilities due to edge devices' smaller processors and lower power budgets; (b) they are heterogeneous, meaning that different processors use different configurations; and (c) their workloads adjustment and applications fight for them. Hence, one of the major difficulties in fog and edge computing is controlling resources. 

\subsection{Resource Management Issues in Fog/Edge computing}
In recent years, IoT applications (e.g. smart homes, self-driving cars, smart agriculture, smart healthcare)  have improved people’s quality of life \cite{ghobaei2020resource}. The increase in IoT applications has also increased a number of IoT devices such as sensors, smart CCTV cameras, smart gadgets, and other smart devices. These IoT applications generate a massive amount of data \cite{dehury2022securing}. According to a report by International Data Corporation, in 2025 data generation from IoT devices will reach 79.4 zettabytes \cite{yousefpour2019all}. Traditional cloud infrastructure is not designed to handle such a huge amount of data \cite{kansal2020introduction}.  The large amount of data generated from the actuators, mobile devices and sensors, has incorporated latency, network bandwidth and security challenges to cloud infrastructure for time-sensitive applications \cite{ding2022roadmap}. To overcome these challenges, the emerging distributed computing paradigm ``fog computing'' and ``edge computing'' as an extension of Cloud computing has drawn the attention of the industrial and research community \cite{deng2018workload}. Fog/edge computing provides computing, network and storage services and control close to the data origin by combining distanced resources between cloud and end devices \cite{lan2022task}. Though the resources in fog/edge are more limited in capability than cloud resources, they can play an important role in processing data for time-sensitive or real-time applications \cite{motlagh2022edge}. It enables location awareness, user mobility support, real-time interactions, low latency, high scalability, and interoperability that cloud-based systems could not support \cite{gill2022manifesto}. But the increase in IoT applications and limited resources in fog/edge computing environments has made efficient resource management very crucial.

\subsection{Need of AI/ML for Fog/Edge computing}
With the increase in the use of IoT and Machine Learning (ML), cloud and fog/edge workloads are becoming increasingly diverse and dynamic. The confluence of fog and AI for improvement in human quality of life necessitates the use of smart management of fog resources. In traditional cloud computing platforms, resource management is done using traditional heuristic approaches without considering diverse and dynamic workloads \cite{gill2022ai}. Most of these methods (e.g., Threshold-based method) are static heuristics configured offline to certain workload scenarios. They are not able to scale applications in and out at run time based on the pattern and behavior of workload \cite{tuli2021start}. The performance of heuristic methods can also be drastically downgraded when the system is scaled up. Resource contention is also a major problem in fog environments where co-located applications compete for shared resources in such policies and cause performance deterioration and Service Level Agreement (SLA) violation \cite{teoh2021iot}. The shift of application structure from monolithic applications to miro-services and serverless has also increased the complexity \cite{xu2022coscal}.  Dependency in micro-services may cause Service Level Objective (SLO) violations due to communication costs and higher resource demands in fog computing.

The fast-rising diversity of workloads, the complexity of applications and the near optimum requirement of QoS parameters of some IoT applications in Fog/Cloud environments, motivate the utilization of AI/ML techniques to optimize their resource management policies \cite{bianchini2020toward}. AI and ML models could be used to model and predict application and infrastructure level metrics that could also assist in task/resource orchestration by improving the quality of resource provisioning decisions \cite{shao2022iot}. Also, ML method can be directly used for resource management decisions for high accuracy and lower time overhead in large-scale systems \cite{tang2018migration}. ML algorithms e.g., Support Vector Machine (SVM), and polynomial regression, can be used to explore relations between performance metrics \cite{zhong2022machine}, the K-means algorithm can be used for the detection of abnormal system behaviors, Reinforcement learning models can be adopted for decision-making for resource provisioning, advanced Recurrent neural network can be used for the analysis of resource utilization or regression of application performance metrics and SVM can also be utilized for dependency analysis of application components \cite{yau2017survey}.

\subsubsection{Motivation}

In the fog/edge computing context, AI/ML-based solutions have been employed for a variety of goals, including resource efficiency, load balancing, energy-efficiency, SLA assurance, etc. Therefore in this article, we aim to investigate ``AI for fog/edge computing'' and its components for the realization of fog/edge-enabled AI. 
\begin{itemize}
  \item AI/ML-based resource management techniques have demonstrated potential for managing resources and deploying applications in cloud computing. Hence, aim to outline the evolution and principles of AI/ML-based resource management in fog/edge computing in recent studies.
  \item Existing survey papers do provide light on AI/ML-based resource management for fog/edge computing, but this area of study is rapidly growing as new AI/ML models are integrated. In order to uncover new research problems, trends, and potential future directions, a new Systematic Literature Review (SLR) of AI/ML-based resource management systems for fog/edge computing is required.
 
\end{itemize}

\subsection{Related Surveys and Contributions}

Many reviews/surveys have been conducted that discuss the role of AI/ML in fog/edge computing. Ghobaei-Arani \textit{et al.} \cite{ghobaei2020resource} reviewed solutions for resource management approaches in fog computing. They presented a taxonomy of resource management methods considering six dimensions, resource planning, load balancing, task offloading, resource allocation, resource provisioning, and application placement. They presented a thorough analysis of several case studies and their methodologies but focused on general approaches and partially discussed AI approaches. They did not provide any classification of AI-based solutions. Zhong \textit{et al.} \cite{zhong2022machine} presented a review of machine learning approaches for container orchestration issues from resource management. They proposed a taxonomy to classify current research by its common features. Thang Le \cite{duc2019machine} investigated machine learning-based resource provisioning in joint edge-fog-cloud environments, and surveys technologies, mechanisms, and ML-based methods that can be used to improve the reliability of distributed applications in diverse and heterogeneous network environments. Emiliano \textit{et al.}~\cite{casalicchio2019container} explored the problem of autonomic container orchestration and presented a taxonomy of container technology, container tools, and architecture, but they only provided a generalized discussion on container technology not specific to edge or fog. Another review ~\cite{deng2020edge} addressed the confluence of edge computing and AI. This work has two-dimensional agenda: the use of edge computing for AI and the use of AI for Edge. They only focused on computational offloading and mobility management with AI methods and only discussed a few AI-based works for these issues.Kansal \textit{et al.}~\cite{kansal2022classification} presented a review of data-driven approaches for fog management issues. They are classified based on the technology used, QoS factors, and data-driven strategies. However, they generically reviewed all the data-driven techniques and do not present any classification or taxonomy of AI techniques. 

Although existing survey articles provide new insights into AI/ML-based resource management for fog/edge computing, the research field is constantly expanding with the integration of new AI/ML models. Therefore, new reviews of AI/ML-based resource management approaches are needed to identify emerging research challenges and possible future directions. Further, none of existing surveys have used Systematic Literature Review (SLR) approach to conduct the survey. In this work, we followed a systematic review methodology as per the "Centre for Reviews and Dissemination (CRD) guidelines" given by Kitchenham \cite{kitchenham2004procedures} to conduct this review on AI/ML-based resource management in Fog/Edge computing. Table \ref{table:comparison} compares the related surveys with our SLR based on important key parameters. 

\subsubsection{Our Contributions}
The main contributions of this Systematic Literature Review (SLR) are summarized as follows:
\begin{itemize}
    \item  Review AI/ML approaches used for the realization of AI/ML for fog/edge computing.
    \item Offer  a comprehensive literature review to discuss the background and current status of AI/ML-based resource management approaches in fog/edge computing environments. 
    \item Propose a taxonomy of the most common AI algorithms used for resource management in fog/edge computing environments.  
    \item Compare existing studies using various parameters related to identified categories through the proposed taxonomy. 
    \item Identify open issues and future directions for the confluence of edge and AI as Edge AI. 
\end{itemize}

\begin{table*}[ht!]
\caption{Comparison of related surveys with our Systematic Literature Review (SLR)}
\small \resizebox{1\textwidth}{!}{
\begin{tabular}{|cl|l|l|l|l|l|l|l|l|l|l|l|l|l|}
\hline
\multicolumn{2}{|l|}{Work} & \cite{ghobaei2020resource} &\cite{casalicchio2019container} & \cite{deng2020edge}
 & \cite{zhong2022machine}  & \cite{yang2020task} & \cite{abdulkareem2019review}& \cite{atlam2018fog}& \cite{duc2019machine}& \cite{nayeri2021application} &\cite{tran2022reinforcement} & \cite{askar2021deep} & \cite{kansal2022classification} & Our SLR\\ \hline
\multicolumn{2}{|l|}{Year} & 2020 & 2019 & 2020 & 2022 & 2020 & 2019 & 2018 & 2019 & 2021 & 2022 &2021 &2022 &2022 \\ \hline
\multicolumn{1}{|c|}{\multirow{3}{*}{Environment}} & IoT &   & \checkmark &   &  &   & \checkmark & \checkmark & \checkmark & & &  & & \checkmark  \\ \cline{2-15}  
\multicolumn{1}{|c|}{} & Edge &  & & \checkmark &  &  & &  & \checkmark &  & &  & &   \checkmark\\ \cline{2-15}
\multicolumn{1}{|c|}{} & Fog & \checkmark &  & \checkmark &  & \checkmark & \checkmark & \checkmark &  & \checkmark & \checkmark & \checkmark & \checkmark & \checkmark \\ \cline{2-15}
\multicolumn{1}{|c|}{} & Cloud &  & \checkmark & & \checkmark &   &  &  & \checkmark & & & & & \checkmark\\ \hline
\multicolumn{1}{|c|}{\multirow{3}{*}{AI Method}} & Machine learning &  &  &  &  \checkmark&  & &  & \checkmark & \checkmark  &  &  &  & \checkmark  \\ \cline{2-15} 
\multicolumn{1}{|c|}{} & Deep learning &  &  &  &  &  &  &  & & \checkmark &  & \checkmark & & \checkmark\\ \cline{2-15} 
\multicolumn{1}{|c|}{} & \begin{tabular}[c]{@{}l@{}}Reinforcement \\learning \end{tabular} &  &  &  &  &  &  &  & & \checkmark & \checkmark & & \checkmark &\checkmark\\ \hline
\multicolumn{1}{|c|}{\multirow{8}{*}{\begin{tabular}[c]{@{}c@{}} AI for Fog/Edge\end{tabular}}} & \begin{tabular}[c]{@{}l@{}}Resource \\ Discovery \end{tabular}&  &  &  &  &  & &  & & &  &  & \checkmark  & \checkmark \\ \cline{2-15} 
\multicolumn{1}{|c|}{} & \begin{tabular}[c]{@{}l@{}}Resource \\ Estimation\end{tabular} & \checkmark &  & & &  &  &  &  & &  &  & \checkmark & \checkmark\\ \cline{2-15} 
\multicolumn{1}{|c|}{} & \begin{tabular}[c]{@{}l@{}}Application \\ Placement\end{tabular} & \checkmark & &  & & \checkmark & &  &  & \checkmark &  &  & \checkmark & \checkmark \\ \cline{2-15} 
\multicolumn{1}{|c|}{} & \begin{tabular}[c]{@{}l@{}}Resource \\ Orchestration\end{tabular} &  & \checkmark &  & \checkmark & &  &  &  & &  &  &\checkmark & \checkmark \\ \cline{2-15} 
\multicolumn{1}{|c|}{} & \begin{tabular}[c]{@{}l@{}}Resource \\ Scheduling\end{tabular}  & \checkmark  &  &  &  & & &  &  & &   &  & \checkmark& \checkmark \\ \cline{2-15} 
\multicolumn{1}{|c|}{} &  \begin{tabular}[c]{@{}l@{}}Resource \\ Provisioning\end{tabular}  & \checkmark & &  & \checkmark &  & \checkmark &  & \checkmark &  & &  &\checkmark & \checkmark \\ \cline{2-15} 
\multicolumn{1}{|c|}{} & Task offloading & \checkmark &  & \checkmark& & &  & & && & &\checkmark & \checkmark \\ \cline{2-15} 
\multicolumn{1}{|c|}{} & Load balancing & \checkmark &  & &  & && & &  & &  &\checkmark & \checkmark  \\ \hline
\multicolumn{2}{|l|}{Taxonomy} & \checkmark & \checkmark & \checkmark & \checkmark &  &  &  & \checkmark &  & \checkmark &  & & \checkmark\\ \hline
\multicolumn{2}{|l|}{Classification} & & & & \checkmark& & &  \checkmark& \checkmark& \checkmark& & \checkmark &  & \checkmark\\ \hline
 \multicolumn{2}{|l|}{Systematic Literature Review (SLR)} &  &  &  &  &  &  &  &  &  &  & & &\checkmark \\ \hline
\end{tabular}
}
\label{table:comparison} 
\end{table*}

\subsection{Article Organization}
\begin{figure*}
	\centering
	\includegraphics[width=1\textwidth]{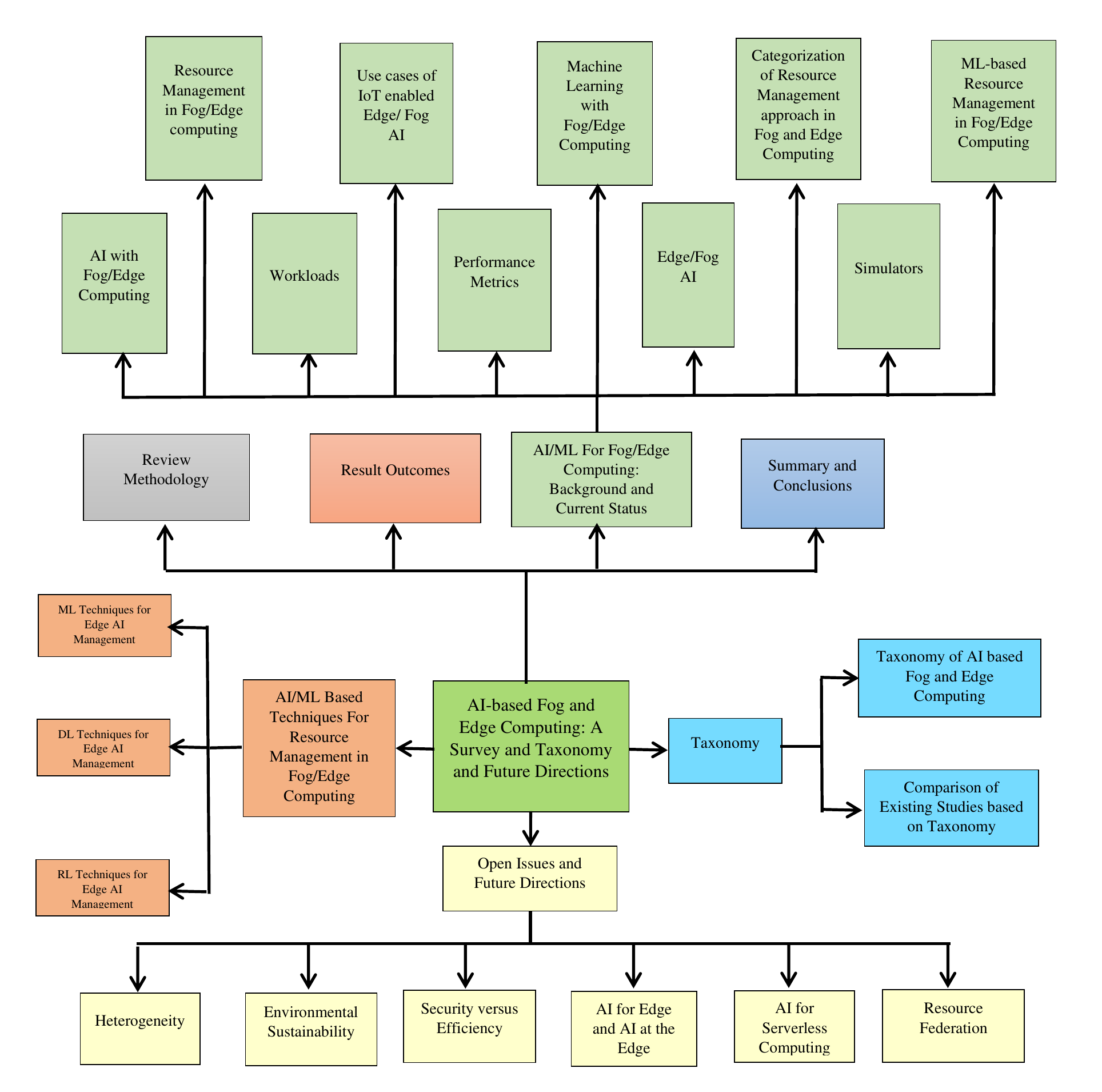}
	\caption [Caption for LOF] {The Organization of this Survey}
	\label{organization}
\end{figure*}

The rest of this article is structured as illustrated in Fig.~\ref{organization}. In Section~\ref{Methodology}, the review methodology is described. Section~\ref{background} presents the background and current status of resource management and AI approaches, Section~\ref{AI/ML Based Techniques For Resource Management: Current Statu} gives a detailed review of AI/ML-based techniques used for resource management issues and their current status. Section~\ref{Taxonomy} presents a taxonomy of frameworks and comparison analysis in AI-based edge and fog computing. Section~\ref{Result Outcomes} discusses result outcomes and Section ~\ref{Open Issues and Future Directions} provides open issues and future directions. Finally, Section~\ref{Summary and Conclusions} concludes the paper. 

\section{Review Methodology}
\label{Methodology} 
This work is a Systematic Literature Review (SLR) of AI/ML-based resource management in Fog/Edge computing. We followed a systematic review methodology as per the "Centre for Reviews and Dissemination (CRD) guidelines" given by Kitchenham \cite{kitchenham2004procedures} to collect the most relevant studies on this issue. The following steps are included in the process of reviewing this article: i) establishing the evaluation process; ii) describing the evaluation criteria; iii) creating the taxonomy; iv) performing the analysis; v) contrasting the different previous studies; vi) analyzing the finding and outcomes; and vii) emphasising promising research directions.

\subsection{Planning the review}

\begin{figure*}
	\centering
	\includegraphics[width=1\textwidth]{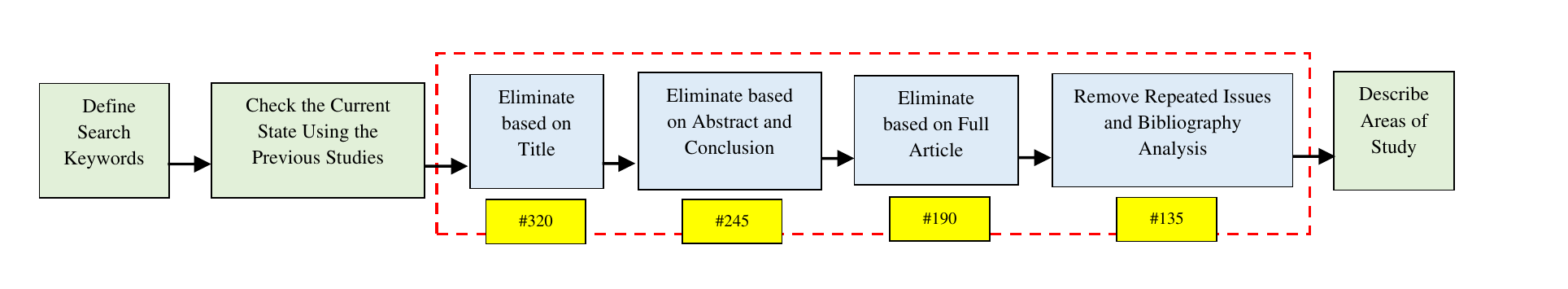}
	\caption [Caption for LOF] {Process of Review Methodology}
	\label{review}
\end{figure*}

Creating research questions is the first step in designing the rules of evaluation. We used these carefully constructed queries to do additional searches across a variety of data sources. The review method identifies and accumulates relevant data for the intended investigation. Articles are either taken into consideration or discarded heavily due to the evaluation procedure. The selection of this task by a single researcher might potentially instill bias in the study. This Systematic Literature Review was thus conducted by splitting among all of the contributors of this paper. Each author has written a document that outlines their thoughts on the review process and distributed it to other team members. Over a set period of time, this cycle has replicated itself. After much debate over several versions, the review guidelines have been completed. Several online databases have been combed thoroughly. Figure \ref{review} represents the evaluation process.

\subsection{Research questions}

In order to better understand AI/ML for fog/edge computing, we plan to do a comprehensive overview of the field. Researchers may use the results of this study to have a better grasp of the state of AI/ML-based fog/edge computing and to pinpoint fruitful avenues for further investigation. Planning the review procedure requires the Research Questions (RQ). Table \ref{table:tabRQ} displays the research questions, the rationale behind them, and the relationships between various parts and subsections of our literature review to demonstrate how we are addressing these RQs using Systematic Literature Review (SLR).

\begin{table*}[ht]
\caption{Research questions, Motivation, Category and Mapping Sections} 
\centering \resizebox{1\textwidth}{!}{
\begin{tabular} {| p{1cm} | p{5cm} | p {5.5cm}| p{3cm} | p{1.5cm} |}   \hline 
\textbf{Sr. No.} & \textbf{Research question} & \textbf{Motivation} &\textbf{Category} & \textbf{Mapping Section}\\  \hline
RQ1 & What is the current status of AI/ML-based fog/edge computing?  & The research question investigates the many different subareas within AI/ML-based fog/edge computing.  &  Current Status \& Background and Result Outcomes & Section 3 and 6
\\ \hline

RQ2 & In fog/edge computing, what resource management methods are available that are based on AI and ML? & The purpose of this question is to delve into the numerous methods employed in either the simulation or real-time application of AI/ML-based fog/edge computing. & AI For Resource Management in Fog/Edge and Taxonomy & Section 4 and 5  \\ \hline

RQ3 & What are the most important sub-fields of AI/ML-powered fog/edge computing? & This question is useful for determining the nature of research that has been conducted utilising AI/ML-based fog/edge computing. & Current Status \& Background and AI For Resource Management in Fog/Edge & Section 3 and 4
\\ \hline

RQ4 & Where are AI/ML-based fog/edge computing frameworks stand right now?  & This inquiry probes the Multiple models for AI/ML-driven fog/edge computing that have been established by the scholars for use in certain IoT use cases. & AI For Resource Management in Fog/Edge and Taxonomy & Section 4 and 5 \\ \hline

RQ5 & How can the efficiency of AI/ML-based fog/edge computing be measured, and what metrics are used for this purpose? & The effectiveness of AI/ML-based Resource Management Techniques for fog/edge computing is measured in terms of delay, cost, and power usage, among others, by the researchers.& Performance Metrics  &  Section 3.8  \\ \hline

RQ6 & What kinds of workloads are utilised to evaluate the efficacy of AI/ML-based fog/edge computing frameworks? & The survey identifies and mentions the workloads utilised by the fog/edge computing system. & Workloads  & Section 3.10 \\ \hline

RQ7 & Which simulators are utilized for fog/edge computing that is based on AI/ML? & The paper identifies and discusses the simulators utilised in the fog/edge computing architecture for AI/ML-based Resource Management techniques. &  Simulators  & Section \ref{simulators}  \\ \hline

RQ8 & What are the most common applications of IoT-enabled Edge/Fog AI? & Use cases for IoT-enabled Edge/Fog AI are discovered and discussed in the paper. &    Edge/Fog AI and Usecases of IoT enabled Edge/Fog AI  & Section 3.1 and 3.2 \\ \hline

RQ9 & What are ML/DL/RL, Online and Offline Learning Techniques for Edge AI Management? & Various techniques for Edge AI management are discovered and discussed in the paper. & ML with Fog/Edge and ML-based Resource Management & Section 3.4 and 3.7 \\ \hline

RQ10 & What are the techniques for Edge AI Management? & Various Deep Learning/Reinforcement Learning techniques for Edge AI management are discovered and discussed in the paper. &  AI For Resource Management in Fog/Edge  & Section 4 \\ \hline

RQ11 & How will AI and machine learning impact fog and edge computing in the future? & Finding out where fog/edge computing research is headed and what problems remain unanswered is crucial.  &   Open Challenges and Research Directions & Section 7 \\
\hline 
\end{tabular}
}
\label{table:tabRQ} 
\end{table*}

\subsection{Sources of information}

A comprehensive search of electronic sources is essential for a thorough literature evaluation. In an effort to improve the probability of locating relevant research publications, we have selected the following collection of data sources:

\begin{itemize}
\item Wiley Interscience (www3.interscience.wiley.com)
\item Springer (www.springerlink.com)
\item ACM Digital Library (www.acm.org/dl)
\item IEEE eXplore (ieeexplore.ieee.org)
\item ScienceDirect (www.sciencedirect.com)
\item Semantic Scholar (www.semanticscholar.org)\\
\end{itemize}

\textbf{Additional Sources:} In order to broaden our search for relevant research, we also consulted the following supplementary sources: 
\begin{itemize}
\item Investigated the original sources included in the reference list.
\item Technical Reports
\item Edited Books and Text Books 
\end{itemize}

\subsection{Search criteria}

The determined search method from various online sources is presented in Table \ref{table:tabSS}. The research articles featured here were gathered using the most widely-used Internet resources in the field of AI/ML for fog/edge computing. ScienceDirect, IEEE Xplore, Springer, Taylor \& Francis (T\&F), ACM, Sage, Wiley, InderScience, and Google Scholar are only a few of the digital libraries from which the papers were retrieved. Finding relevant studies in the literature relies heavily on ``Search string construction'' and ``Search keywords choice''. Search terms like ``fog computing'' and ``edge computing'' and related terms like ``Artificial Intelligence'' and ``Machine Learning'' and ``Deep Learning'' revealed relevant items. Combining the keywords with the boolean operators AND and OR produced the final string for search. The following sequence has the following specified format:

\begin{verbatim}
[(Application placement) OR (Service placement) OR 
(Task scheduling) OR (Container placement) ) OR 
(VM placement) OR (Resource management) AND 
((( Artificial Intelligence) OR (Machine Learning)) AND
(Challenges) OR (Metrics) OR (simulators) OR (Workload) OR
(Algorithms) OR (Methods))) AND (Edge computing) OR 
(Fog computing) Or (Cloud computing)]
\end{verbatim}

Firstly, we constructed a search query based on the formulated research questions in Table \ref{table:tabRQ}. Table \ref{table:tabSS} details the evaluation process's search strings.

\begin{table*}[ht]
\caption{Search strings for e-resources} 
\centering \resizebox{1\textwidth}{!}{
\begin{tabular} {| p{1cm} | p {3.5cm}| p {3.5cm}| p {2cm}| p {2cm}|p {3cm}|}   \hline 
\textbf{Sr. No.} & \textbf{E-resource}& \textbf{Search String}& \textbf{Dates}& \textbf{Source Type}& \textbf{Subjects} \\  \hline  
1 & ieeexplore.ieee.org & Abstract: Artificial Intelligence or Machine Learning for Fog Computing or Edge Computing & 2016 - 2022 & Conferences, Journals, Magazines and Transactions & Fog Computing, Edge Computing, Machine Learning, Deep Learning, Artificial Intelligence \\
2 & www.springerlink.com & Abstract: Artificial Intelligence or Machine Learning for Fog Computing, Edge Computing & 2016 - 2022 & Conferences, Journals and Magazines & Fog Computing, AI, ML, DRL, RL, Edge Computing  \\
3 & www.sciencedirect.com & Abstract: Artificial Intelligence or Machine Learning for Fog Computing, Edge Computing & 2016 - 2022 & All sources & Fog Computing, Deep Learning, Artificial Intelligence, Edge Computing  \\
4 & www.onlinelibrary.wiley.com/ & Abstract: Artificial Intelligence or Machine Learning for Fog Computing & 2016 - 2022 & Conferences, Journals, Magazines and Transactions & Fog Computing, Edge Computing, Machine Learning, Deep Learning, Artificial Intelligence \\
5 & www.acm.org/dl & Abstract: Article Title: Fog, Full Text/Abstract: Artificial Intelligence or Machine Learning for fog or edge & 2016 - 2022 & Conferences, Journals, Magazines and Transactions & AI/ML, Fog, Cloud, Edge  \\
6 & www.taylorandfrancis.com/ & Abstract: Artificial Intelligence or Machine Learning for Fog/Edge Computing & 2016 - 2022 & Conferences and Journals & Edge, Fog, AI, ML, DRL  \\
7 & www.inderscience.com/ & Abstract: Artificial Intelligence or Machine Learning for Fog Computing & 2016 - 2022 & Journals & All Subjects \\
8 & www.semanticscholar.org & Abstract: AI/ML for Fog/Edge Computing & 2016 - 2022 & arXiv Preprints & Fog Computing, Edge Computing, Machine Learning, Deep Learning, Artificial Intelligence  \\
9 & Other Publishers & Article Title: Fog, Full Text/Abstract: AI/Ml for fog or edge & 2016 - 2022 & All sources & Edge, Fog, AI, ML, DRL\\

\hline 
\end{tabular}
}
\label{table:tabSS} 
\end{table*}

\subsection{Inclusion and exclusion criteria}

AI/ML-based Fog/edge computing is a relatively new area of study, and only a small number of papers have addressed the key questions surrounding them prior to 2015. As a result, the number of articles covering the topic before 2015 was quite low. Figure \ref{review} displays the selection procedure of research papers from the Internet and digital library databases. The aforementioned search terms and string combinations were utilized to narrow the available databases down to the most pertinent articles. Starting with publications that were not peer-reviewed or indexed by ISI, 320+ papers were chosen for the first phase. To find quality publications, a research screening method has been done to exclude brief publications, non-peer-reviewed papers, book chapters, and low-quality studies that weren't capable of delivering any technical knowledge and scientific argument. By the end of the process, 135 articles from prestigious journals and conferences had been hand-picked for this evaluation. In Section~\ref{Taxonomy}, the suggested taxonomy is explained alongside an analysis of each work that fits into it.

The elimination of research was performed using the following criteria to pick the rigorous quality publications: 

 \begin{itemize}
 	\item Neither the journal nor the conference are indexed.  
 	\item The articles present any survey and analysis work.
 	\item These are the documents that were not written using the English language. 
 	\item Works that do not undergo a rigorous peer review procedure.
 \end{itemize}

\subsection{Quality assessment}

There are several research publications on AI/ML for fog/edge computing in a wide variety of journals and proceedings from conferences. After applying the exclusion and inclusion criteria, we conducted a quality evaluation of the selected papers to choose the most relevant ones for further consideration. To evaluate the studies' overall quality, we checked them against critical factors such as objectivity, internal consistency, and bias using the CRD recommendations \cite{kitchenham2004procedures}. We have established quality evaluation forms as presented in \textbf{Appendix A} to evaluate high-quality research papers for this systematic review of the literature. We've asked both broad, exploratory questions and in-depth, exploratory ones. Preliminary Examining questions are a helpful tool for locating high-level research publications that are associated with AI/ML in Fog/Edge Computing. In addition, specific questions are used to choose the research papers that are the most pertinent to the primary context of AI/ML in Fog/Edge Computing.

 \subsection{Data extraction}
The methodology for extracting data from the 135 research papers included in this analysis was detailed in \textbf{Appendix B}. Initiating the data-gathering process inspired us to create this data extraction form in order to answer the research questions. Our carefully stated selection criteria allowed us to identify the best works on AI/ML for fog/edge computing from a wide range of prestigious journals and conferences as listed in \textbf{Appendix C}. In addition, we have reached out to a number of authors in order to collect the necessary information regarding scholarly works. 
In this SLR, we used this procedure to retrieve the data:

\begin{itemize}
 	\item A set of authors read through all 135 publications to collect the necessary information.

	\item Other authors used random samples to verify the accuracy of data collection.

	\item Any issue that arose throughout the cross-checking procedure was discussed and settled in a number of meetings.
	
\end{itemize}

\subsection{Acronyms}
Abbreviations utilized in the systematic literature evaluation are given in \textbf{Appendix D}.

\section{AI/ML For Fog/Edge Computing: Background and Current Status}
\label{background} 
\begin{figure}[ht!]
    \centering
    \includegraphics[width=1\linewidth]{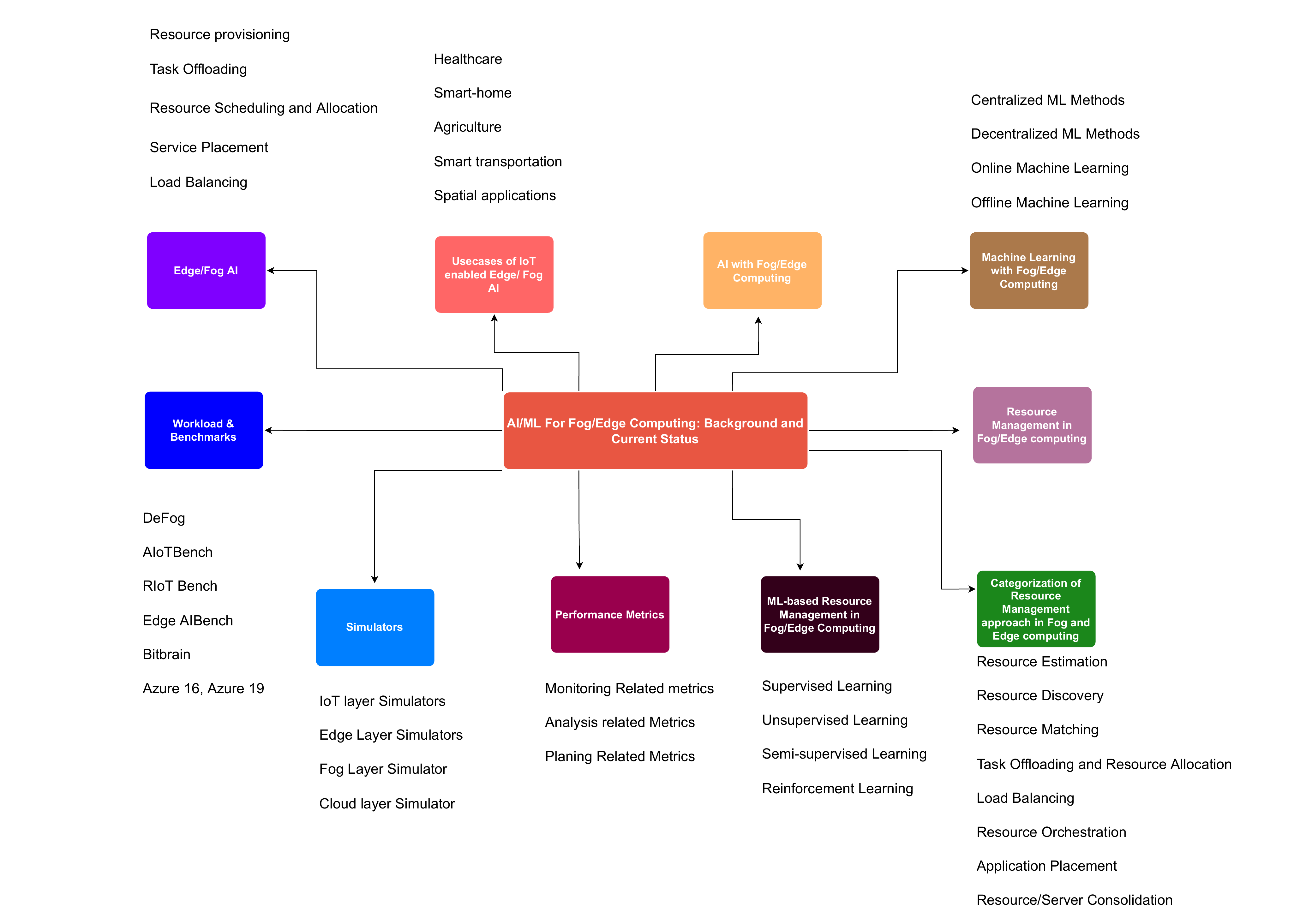}
    \caption{Taxonomy of AI/ML for Fog/Edge Computing}
    \label{fig:summary}
\end{figure}

In this section, we discuss the background concepts, including AI and ML with Fog/Edge Computing. Further, this section presents other concepts such as Resource Management in Fog/Edge computing, Categorization of Resource Management approach in Fog and Edge computing, IoT Applications, Performance Metrics, workloads and simulators. Figure \ref{fig:summary} represents a broad taxonomy of AI/ML for Fog/Edge Computing.

\subsection{Edge/Fog AI}
The emerging computing model named fog and edge computing can alleviate the problem of bringing the computational resources closer to the end user. These computing models offered the services to several latency-sensitive IoT applications such as vehicular networks, agriculture, healthcare, smart home, and transportation system \cite{kumari2022task}, where cloud models fall behind in handling the services with minimum response time \cite{gill2018fog}. The fog/edge paradigm supports low latency, high mobility, and interoperability with resource constraints for IoT applications \cite{n2022greenfog}. The contemporary research trend resides in the decentralization of resources towards the edge of the network. In contrast to cloud resources, edge resources need distinctive managerial techniques because of underlying heterogeneous resources, dynamic workload, scalable data centers, and last but not least, unpredictability, fluctuating interactions and multi-tenancy across end users \cite{jennings2015resource}. The dynamic workloads make the process even more complex when real-time applications are competing for limited resources \cite{tuli2020healthfog}. Failure recovery, data redundancy, high cost, power consumption, and privacy are still issues with the emerging computing paradigm, it necessitates the management of fog/edge resources is considered one of the significant challenges and needs to be addressed by the intelligent solution to improve the performance metrics and resolve the mentioned issues \cite{kok2022fogai}. 
Resource provisioning, task offloading, resource scheduling and allocation, service placement, and load balancing are the components of resource management \cite{shakarami2022resource}. Each component of resource management and its related issues are discussed briefly. 

\subsubsection{Resource provisioning}
\textcolor{black}{Resource provisioning is defined as selection, deployment, and run time management of software and hardware resources for the efficient performance of applications.} There are fluctuations in IoT devices' workload that leads to the issue of over and under provisioning. In the case of overprovisioning, a greater number of resources are allocated as compared with the required IoT workload, and IoT users must pay more for the services used \cite{tesco}. In case of under provisioning, a smaller number of resources are allocated for IoT services, as per the requirement of IoT workload and it increases the possibility of SLA violations \cite{lindsay2021evolution}. Hence, an efficient mechanism is needed to overcome the mentioned challenges and provide the resources based on the service demands.

\subsubsection{Task Offloading}
It is problematic to take the offloading decision at runtime due to the complex architecture of fog and edge networks with resource constraints and allocate the best possible resource (cloud or fog) for computation-intensive tasks. The most common applications it supports are virtual reality, vehicular networks, and multimedia delivery \cite{ghobaei2020resource}. We required an intelligent agent to decide, where the IoT devices-based workload will be processed and return the results within the deadline. The offloading decision are depending upon several factors like types of workloads, deadlines, priority, communication link, the capacity of fog nodes, and IoT devices. The main aim is to, utilize the link, and improve the latency and power consumption.  

\subsubsection{Resource Scheduling and Allocation}
The number of fog/edge nodes are available to process the IoT requests, but required an efficient scheduling technique that will search the optimal resources for the upcoming workload and execute it within the deadline \cite{gill2019router}. The scheduling of IoT service requests with objective function over the heterogeneous fog/edge nodes belongs to the class of NP-Complete and it is difficult to find the exact solution to the problem \cite{hunter}. Resource scheduling and allocation is a different problem in the edge/fog paradigm with additional entities as compared with the cloud paradigm, and becomes a double mapping problem under-provisioning and IoT services demands \cite{nabavi2022tractor}.   

\subsubsection{Service Placement}
The objective of service placement is to look at the optimum resources for IoT services, and deploy over the virtualized edge/fog nodes to enhance the QoS metrics, while maintaining the SLA \cite{nayeri2021application}. An IoT application can be mapped with more than one fog/edge node or multiple services can be placed over a single fog node, depending upon the requirement of services and computation capacity of resources. 

\subsubsection{Load Balancing}
It is one of the vital issues to distribute the workload over the virtualized fog nodes in balancing mode and avoid the possibility of over or underutilization of resources \cite{souri2022artificial}. The goal of load balancing in edge/fog computing is to reduce the response time for latency-sensitive applications and address the challenges like network delay, high waiting time, and scalability to improve the system performance with potential solutions \cite{hunterplus}.

There are many resource management solutions are existed for the cloud paradigm, but cannot apply the same for fog/edge computing due to different network conditions, and characteristics, more distributed infrastructure, and processing capabilities of nodes \cite{hunterplus, hunter}. Hence, it is more challenging to address the issues of resource management in the fog/edge platform, as compared to the cloud platform \cite{souri2022computational}. Several researchers are working in this direction to manage the resources of a heterogeneous network, accurate offloading decisions, optimal provisioning, intelligent scheduling techniques, best-effort service placement and efficient workload sharing for load balancing, but no one has explored all the mentioned challenges entirely despite its importance for real-time applications, hence it opened the door for the new researchers to propose a novel solution for the existing issues. 

\subsection{Usecases of IoT enabled Edge/Fog AI}
IoT has attracted significant research interests from both industry and academia and facilitates varied novel applications, including smart home, surveillance, smart healthcare and so on. Here, we present a brief summarization of different types of IoT applications.  

\subsubsection{Healthcare} IoT solutions have been considered and deployed for health management systems by efficiently tracing agents (patients, medical practitioners, medical resources), automatic data sensing and authentication and it is defined as \textit{Internet of Health Things (IoHT) \cite{tuli2020next}}. IoHT technology has redefined the healthcare system by health monitoring of patients anytime and anywhere for post-discharge care, elderly health management and several other emergency situations like pandemic~\cite{habibzadeh2019survey}. Wearable sensor is one of the major components in IoHT where health-related parameters are collected in different time intervals and processed for smart e-healthcare applications~\cite{wu2020rigid,esmaeili2020priority}. A framework with IoT-based wearable sensors coupled with machine learning methods has been proposed for monitoring sport's person health conditions by collecting health parameters and exercise traces~\cite{huifeng2020continuous}. Carlos \textit{et al.}~\cite{dourado2020open} presented an IoHT-based deep learning framework for medical image (cerebral vascular accident image, lung nodule and skin images) classifications~\cite{dourado2020open}. Another work by Ray \textit{et al.}~\cite{ray2019novel} designed a prototype of a cost-effective and low-power sensor system that is conducive to monitoring real-time intravenous (IV) fluid bag levels in e-healthcare applications. A collaborative edge-IoT framework, named RESCUE is proposed in~\cite{das2022rescue} for provisioning healthcare services specifically in exigency time by collecting patient's location, and health condition and predicting the route of nearby healthcare centers. The framework also devises latency-aware and power-aware frameworks using IoT devices. Several research works have been carried out to mitigate COVID-19 by leveraging IoT-based solutions using AI/ML \cite{kumar2021drone, 2020predicting}. Khan \textit{et al.}~\cite{khan2022dca} present \textit{DCA-IoMT}, a location-aware knowledge-graph-based recommendation framework for an alert generation against COVID-19. FairHealth, an Internet of Medical Things (IoMT) framework is proposed in~\cite{lin2022fairhealth}, where the fairness-aware resource scheduling method is deployed in 5G edge healthcare. Another imperative issue in the healthcare domain is the privacy aspect since such collected health data is sensitive in nature \cite{desai2022healthcloud}. To mitigate such issues, a secure Internet of Medical Things framework is presented leveraging blockchain~\cite{dewangan2022patient}. In particular, when IoMT devices send data using a patient’s Personal Digital Assistant (PDA), the data is transacted on the blockchain by the cloud server. Similarly, in the context of COVID-19, a blockchain-based privacy-preserving algorithm is proposed by Lv \textit{et al.}~\cite{lv2020towards} for contact tracing. The authors investigate several practical challenges including protecting data security and location privacy, dynamically and effectively deploying short-range communication IoT) for activity-tracking and location-based services in large areas. The utility of IoT is explored for vaccine supply chain distribution in India~\cite{kumar2022impact}. 

\subsubsection{Smart-home} IoT solutions can provision smart home services including automatic control of domestic appliances, alarm generation, security controls and developing an Internet-connected system. Gavrila \textit{et al.}~\cite{gavrila2020suitability} present an IoT-based framework for seamless integration with a Hybrid broadcast broadband TV-enabled television set in a smart home environment for a better user experience. A multi-objective and smart residential load management framework is presented for energy management in smart-homes~\cite{chatterjee2022multi}. Specifically, an IoT based controller manages the home loads and generates alerts if any malfunction in the household loads is detected. In the smart-home context, cyberattacks cause potential harm to the occupants and compromise their safety. In this regard, Yamauchi \textit{et al.}~\cite{yamauchi2020anomaly} devised a novel method to detect such attacks by learning occupants' behavior as sequences of events such as the operation of home IoT devices and activities along with environmental variables (temperature, humidity, time of the day). The method compares learned sequences and current sequences when an operation command is activated, and an anomaly is detected. Kratos+, a multi-user and multi-device-aware access control mechanism is proposed in a similar context for allowing smart home users to specify the access control demands~\cite{sikder2022s}. Li \textit{et al.}~\cite{li2021gpfs} present a human pose forecasting system for smart homes leveraging graph convolutional neural network on the IoT edge for online learning. IoT Meta-Control Firewall (IMCF+) is proposed to mitigate energy consumption and CO2 emission issues while also maintaining user comfort~\cite{constantinou2022green}. 

\subsubsection{Agriculture} IoT has brought dramatic improvements in agricultural production by enhancing the quality of agricultural products, reducing labor costs, and effective farm management~\cite{liu2020industry}. Alahi \textit{et al.}~\cite{alahi2017temperature} design a smart nitrate sensor that monitors nitrate concentrations in ground and surface water. The system is supported by WiFi-based IoT that can send data directly to an IoT-based web server and serves as a distributed monitoring system \cite{sengupta2021mobile}. A cyber-physical system for crop evapotranspiration estimation is proposed~\cite{wang2022hybrid}. A gradient-boosting decision tree along with a fuzzy granulation method is used on IoT data from Xi’an Fruit Technology Promotion Center in Shaanxi Province, China for cherry tree evapotranspiration estimation and the proposed system achieved promising accuracy \cite{singh2020agri}. The continuous monitoring of crop growth is one of the most important aspects of precision agriculture. Bauer \textit{et al.}~\cite{bauer2020towards} design a complementary framework for low-cost crop sensing leveraging temporal variations of the signal strength of low-power IoT radio communication \cite{gill2017iot}. MDFC–ResNet framework~\cite{hu2020mdfc} identifies fine-grained crop disease using IoT technology and deep learning method. 

\subsubsection{Smart transportation} IoT demonstrates a promising future in Intelligent Transportation Systems (ITS) by collecting, analyzing traffic/mobility-related data and developing a smart, safe, reliable and sustainable ITS~\cite{zhu2019parallel}. A smart parking surveillance system (detecting parking occupancy) is proposed by using edge computing and real-time video feed~\cite{ke2020smart}. Bansal \textit{et al.}~\cite{bansal2020deepbus} propose \textit{DeepBus} for identifying surface irregularities (e.g., potholes) on roads using IoT sensor and machine learning methods. The system centrally hosts a map and alerts users and authorities regarding pothole locations. Philip \textit{et al.}~\cite{philip2018distributed} designed an IoT-based smart traffic control system where a group of self-driving cars interact with road-side units and independently decide their lane velocities. IoT-based energy efficient ITS framework is presented that can reduce energy consumption, noise pollution, waiting time and greenhouse gas emissions in smart city environment~\cite{chavhan2022edge}. Wan \textit{et al.}~\cite{wan2021machine} proposed a framework consisting IoTs of vehicles for vehicle number estimation which in turn helps in vehicle localization. A predictive framework is designed for forecasting the parking space occupancy leveraging deep learning-based ensemble technique~\cite{piccialli2021predictive} in IoT environment. The system specifically reduces the search time for parking and the optimization of the flow of cars helps in better traffic management in congested areas of a city.

\subsubsection{Spatial applications} 
Internet of Spatial Things (IoST) integrates spatial or location information in the core IoT architecture to facilitate location-aware services~\cite{eldrandaly2019internet,sarwat2020spatial}. Ghosh \textit{et al.}~\cite{ghosh2019mobi} presented a mobility-aware IoST framework for time-critical applications (e.g., ambulance service, disaster relief) for predicting optimal paths with less delay. Koh \textit{et al.}~\cite{koh2016geo} proposed a new location spoofing detection algorithm that can be used for spatial tagging and location-based services in an IoT environment. A spatial-data driven IoT framework, \textit{STOPPAGE} is developed for predicting COVID-19 hotspot zones and efficient medical resource management in varied regions~\cite{ghosh2021stoppage}.

\subsection{AI with Fog/Edge Computing}  
  
IoT is a communication network created by objects that can connect to the Internet and communicate with each other \cite{golec2020biosec}. It has started to be used everywhere, from healthcare applications to military applications, and it is estimated that the number of IoT devices will reach approximately 30 billion by 2030 \cite{vailshery_2022}. Along with the vertical increase in the number of IoTs, the amount of data that needs to be processed and produced by sensors has reached gigantic proportions. Processing this data in the cloud seems like a logical solution at first because of its advantages, such as high processing power and storage capacity \cite{golec2022aiblock}. However, problems such as \cite{golec2021ifaasbus} latency may occur in time-sensitive IoT applications such as instant patient follow-up.

Fog computing can be seen as an inspiring development to solve problems such as latency, power consumption, and network traffic in Cloud-based IoT systems \cite{miah2018enhanced}. Unlike cloud data centers, Fog nodes are located close to the IoT layer. Thus, execution time and bandwidth issues can be reduced \cite{iftikhar2022fog}. On the other hand, fog nodes do not consist of devices with powerful processing power and large storage ability such as cloud data centers \cite{ghobaei2020resource}. Therefore, one of the difficulties that need to be solved in fog computing is resource management, which consists of subheadings such as resource scheduling, task offloading and resource provisioning \cite{iftikhar2022fogdlearner}.

Resource management issue for Edge/Fog AI is addressed using diverse techniques. One of these methods is AI-based techniques that have been gaining popularity recently. AI-based techniques used to solve resource management problems in Fog/Edge computing can be summarized as Deep Learning (DL), Machine Learning (ML), Reinforcement Learning (RL), and Deep Reinforcement Learning (DRL). AI-based techniques are very effective in dynamic resource scheduling \cite{kansal2022classification}. In particular, DRL has been shown to be very successful in dynamic complex problems and dynamic task offloading \cite{kansal2022classification}. In addition, AI-based techniques such as neural networking and reinforcement learning were found to be more popular in resource estimation than mathematical models \cite{guo2020demand}.

\subsection{Machine Learning with Fog/Edge Computing} 
AI and ML became an integral part of everyday application decision-making. It is used by recommender systems for tech giants such as Google, Amazon, Netflix, and Facebook and in more complicated use cases such as self-driving cars~\cite{Li2018-driving}, earthquake prediction~\cite{Wang2020-earthquake}, and smart healthcare~\cite{Amin2019-health}. Due to the abundance of data sources at the edge, Fog/Edge computing received increasing attention as an enabler of Machine Learning methods. In this section, we examine Centralized vs Decentralized ML Methods and Online vs Offline ML for fog/edge computing.
\subsubsection{Centralized ML Methods}
AI and ML models feed on a tremendous amount of data generated by thousands to millions of mobile and IoT devices. Typically, these devices continuously stream the generated data into the cloud applications to be stored for later processing and analysis. These data are analyzed to extract certain features to help train AI/ML models. These models are trained on high-performance servers residing in the data centers of the cloud. Google Cloud, Microsoft Azure, Amazon AWS are the most common providers for ML-as-a-service where models can be trained on large amounts of data at scale. The interactions between the various services in the Fog/Edge are another source of training data that can be leveraged to enable more intelligent ML-based applications to be deployed and enhance the service for the users. However, the major concern with this setting is the security and privacy of the collected data used for training which may contain private and sensitive information. Other major problems for centralized Fog/Edge-based ML methods are latency and communication transfer costs~\cite{Xu2021-grace,DC2-INFOCOM21,SIDCo-MLSys21}. 

There are many centralized learning methods for the purposes of workload prediction to aid with the resource allocation problem in literature~\cite{ai_cloud_edge_book,distribued_edge,cloud_edge,market_resourcealloc}. Wang \textit{et al.}~\cite{distribued_edge} proposed a feasible solution for edge cloud resource allocation over time based on an online algorithm to solve sub-problems with logarithmic objectives. The algorithm is shown to achieve a parameterized competitive ratio, without requiring any a priori knowledge of the resource price or the user mobility. The results with real-world and synthetic data confirm the effectiveness of the proposed algorithm. Rosendo \textit{et al.}~\cite{cloud_edge} provided an overview of the main state-of-the-art libraries and frameworks for ML and data analytics on the Edge-to-Cloud Continuum. This work also covers the main simulation, emulation, deployment systems, and testbeds. In addition, a holistic understanding of the performance optimization of applications and efficient deployment of AI/ML workflows is given. Nguyen \textit{et al.}~\cite{market_resourcealloc} proposed a market-based resource allocation framework in which the services act as buyers and fog resources act as divisible goods in the market. The aim is to compute a Market Equilibrium (ME) solution at which every service obtains its favorite resource bundle under the budget constraint, while the system achieves high resource utilization. The work discusses both centralized and privacy-preserving distributed solutions.

\subsubsection{Decentralized ML Methods}

Centralized learning (CL) for learning ML models is becoming obsolete because it requires the collection of decentralized user data imposing security and privacy risks and expensive data transfer \cite{Bonawitz19,cartel-Daga2019,kairouz2019advances,FLanalysis-22,REFL-2022}. Hence, decentralized paradigms are being explored as alternatives. Several techniques leverage decentralized learning methods for the purposes of workload prediction to aid with the resource allocation problem~\cite{fed_ddqn,fed_loadpredict,fed_sensors}.

\begin{itemize}
  \item \textbf{Federated Learning (FL): }
In FL architecture, the learners are end-user devices such as smartphones, sensors, or IoT devices; training data is owned and stored at these devices; the learners train a global model collaboratively with the assistance of a centralized FL (or aggregation) server~\cite{mcmahan2017,Bonawitz19,AQFL-2021,FLanalysis-22, REFL-2022}. As described in~\cite{mcmahan2017,Bonawitz19,REFL-2022}, the training of the global model occurs over a series of rounds until the model converges to a satisfactory accuracy. In each round, a few clients are sampled to update the model and a new model is produced. But, due to the server's central role, FL faces challenges of synchronization, reliability, and expensive communication~\cite{kairouz2019advances}. At the start of each round, the server waits for available devices to check-in. The server selects a subset of these devices which meet certain conditions, such as being idle and connected to WiFi and a power source. Then, the server sends the global model along with the necessary configurations (i.e., hyper-parameter settings) to the selected clients. The learners perform the same number of local optimization steps as set by the server. Then, the learners send their updated models (or the delta) to the server. Finally, the server aggregates, with the global model, the model updates sent by the clients, and then checkpoints the new global model to the local storage~\cite{mcmahan2017}.  One of the main challenges in FL use cases is the heterogeneity of the environment~\cite{Bonawitz19,kairouz2019advances} which is studied and addressed by several works~\cite{FLanalysis-22,AQFL-2021,REFL-2022}. 

\item \textbf{Decentralized Learning (DL):} It is an alternative approach for training common models on decentralized data, typically in environments consisting of edge devices~\cite{defog-jonathan2019,cartel-Daga2019}. In DL, the learners, without centralized coordination, engage in the learning process to train a model tailored to their common tasks and coordinate among themselves via peer-to-peer communication. Thus, device groups can train a common model while each device preserves its data. However, due to a lack of central coordination, the devices need to be available at the same time to iterate over the training process in a lock-step fashion, causing training to be as fast as the slowest device. This hinders the scalability and efficiency as devices can not train at their own pace without being held back by slow learners~\cite{cartel-Daga2019,kairouz2019advances}.

\end{itemize}

There has been recent interest in techniques that leverage the non-conventional decentralized learning methods for the purposes of workload prediction to aid with the resource allocation problem~\cite{fed_ddqn,fed_loadpredict,fed_sensors}. Zarandi \textit{et al.}~\cite{fed_ddqn} provides an optimization of the offloading decisions, computation resource allocation, and transmit power allocation for Edge IoT networks. The problem is presented as a multi-agent Distributed Deep Reinforcement Learning (DDRL) problem which is addressed via double deep Q-network (DDQN), where the actions are offloading decisions. Then, federated deep learning (FDL) is used to enhance the learning speed of IoT devices (agents) by creating a context for cooperation between agents with minimal impact on their privacy. Fantacci \textit{et al.}~\cite{fed_loadpredict} applies FL to train models for demand prediction. The proposed method achieves high accuracy levels on the predicted application demand via aggregating the feedback received from the user models. Chen \textit{et al.}~\cite{fed_sensors} propose a two-timescale federated deep reinforcement learning based on Deep Deterministic Policy Gradient (DDPG) to solve the joint optimization problem of task offloading and resource allocation to minimize the energy consumption of all IoT devices subject to delay threshold and limited resources. The simulation results show that the proposed algorithm can greatly reduce the energy consumption of all IoT devices.

\subsubsection{Online Machine Learning }

\begin{figure}[ht!]
    \centering
    \includegraphics[width=.8\linewidth]{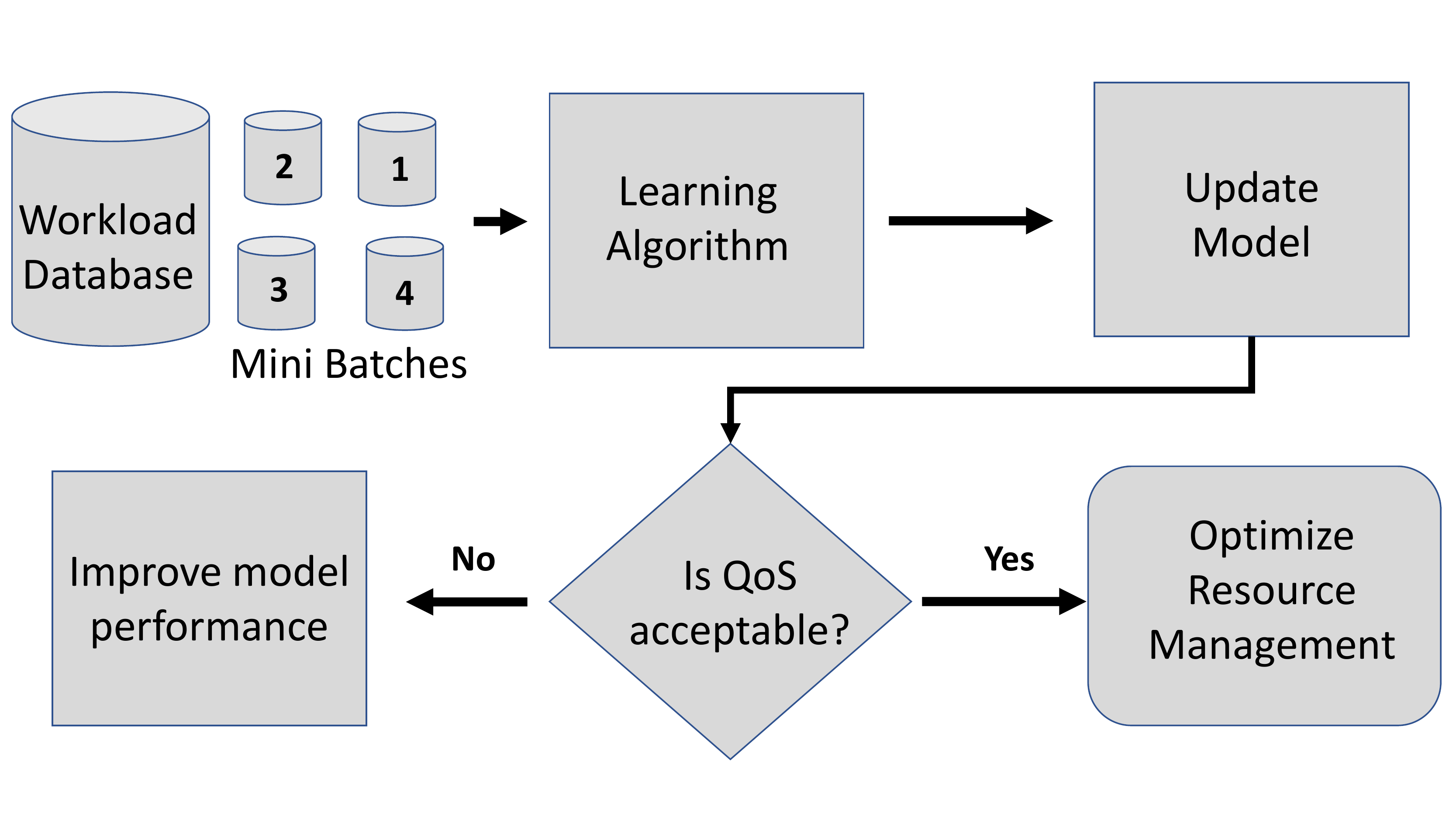}
    \caption{Online Machine Learning General Scheme for Fog/Edge Computing}
    \label{fig:onlineml}
\end{figure}

One of the design options used when modeling ML method is Online ML. In this model, the learning algorithm is constantly updated using new data \cite{fontenla2013online}. Therefore, real-time data must be used in scenarios where Online ML is used for fog/edge computing. An example is models that predict the stock market \cite{bisong2019batch}. Figure \ref{fig:onlineml} shows the working scheme of Online ML \cite{bisong2019batch}. ML parameters are updated by being trained by a new set of data each time. The learning step continues as new data comes in, and this process is quite fast and inexpensive. Online ML can be a suitable design option for scenarios where data flow is intense and constantly changing.

\subsubsection{Offline Machine Learning}

\begin{figure}[ht!]
    \centering
    \includegraphics[width=.8\linewidth]{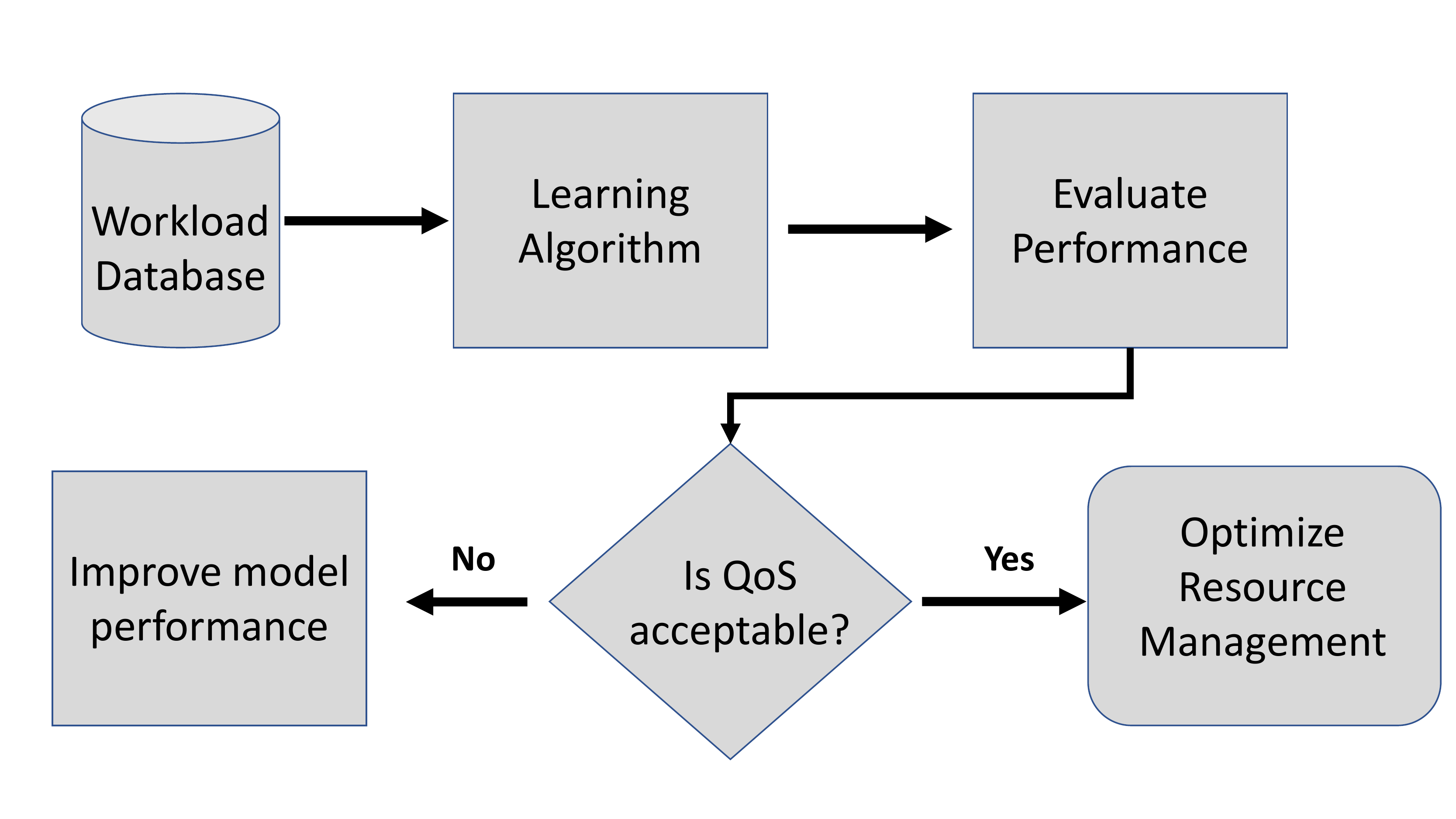}
    \caption{Offline Machine Learning General Scheme for Fog/Edge Computing}
    \label{fig:offlineml}
\end{figure}

Unlike Online ML, there is no continuous data flow in Offline ML or Batch Learning. The ML model is trained using a certain number of data. After the model is trained, the test set performance is checked. If the test set performance is good enough, the learning phase ends. In case the model needs to be trained using new data, old and new data are used together. Figure \ref{fig:offlineml} gives the working diagram of Offline ML \cite{bisong2019batch}. Compared to online ML, the amount of data used to train the model is larger. Therefore, it is obvious that more CPU and RAM will be needed to train a model in Offline ML. In addition, with a large amount of data, it will take longer to train the models. In Fog/Edge Computing, offline ML methods are often used to solve offloading problems.

\subsection{Resource Management in Fog/Edge computing}
This computer architecture is not sustainable in the long run because of the expected increase in communication latencies when billions of devices are connected to the Internet. Application performance will suffer and QoS may drop as a result of longer connection delays \cite{moqurrab2022deep}. An alternative computing method that can aid with this problem is to bring computer resources closer to end devices and sensors and use them for data processing. Communications might be sped up and cloud resources used could be reduced \cite{singh2022machine}. In recent years, there has been a trend in Computer Science to put this theory into practice by relocating some of the computing capacity now located in massive data centers to the network's perimeter, making it more accessible to end-users and sensors~\cite{dhillon2022iotpulse}. The Internet's routers, gateways, and switches may have access to computing power, or ``micro'' data centers may be set up in existing public and private networks for convenience and safety. Computing models that take advantage of network edge resources are known as ``edge computing''. Fog computing is the practice of combining local hardware with remote cloud resources. Edge resources are distinct from cloud resources in that they are resource limited. This means that they have less computing capability than cloud resources because of the smaller processing units and reduced energy constraints of edge devices. They also employ various configurations for different CPUs, making them heterogeneous~\cite{tuli2020ithermofog}.

\subsection{Categorization of Resource Management Approach in Fog and Edge computing}
\begin{figure}[!ht]
\centering
\includegraphics[width=6in]{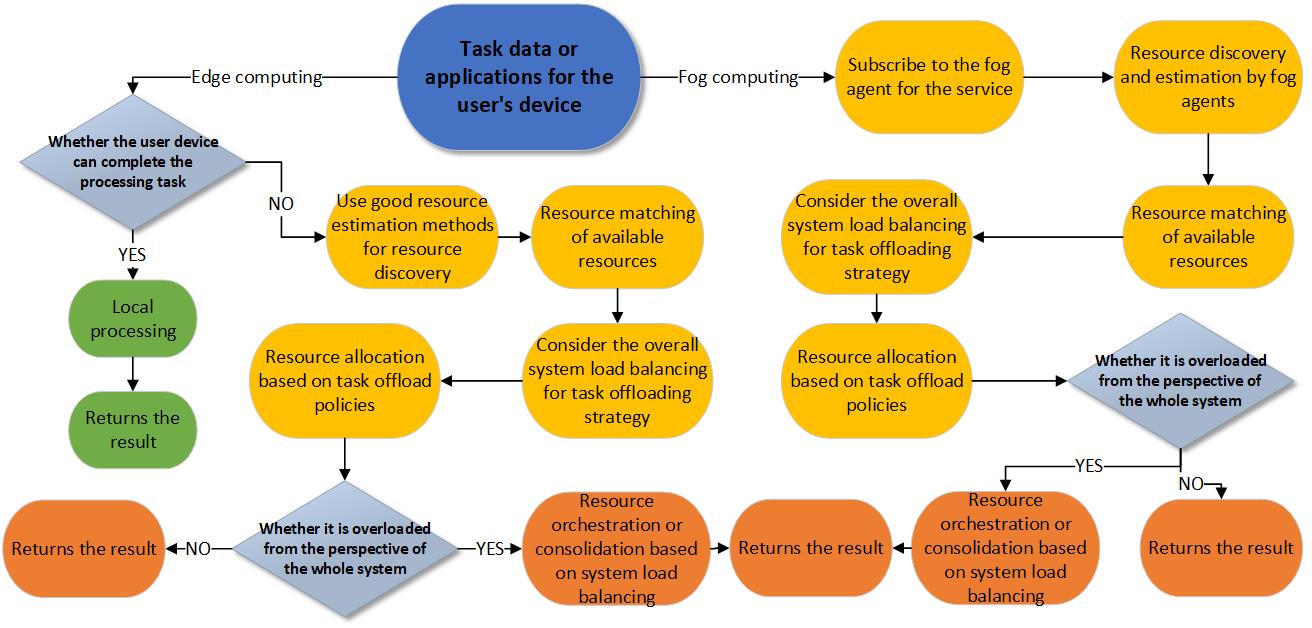}
\caption{The flow chart of resource management in fog and edge computing.}
\label{flow-chart-rm}
\end{figure}

For the edge computing paradigm and the fog computing paradigm~\cite{2018All}, the common denominator of the two is to sink the computing resources in the cloud to the user side, and provide better services for those user devices that do not have enough resources at a lower latency and energy consumption. To do this, we need to offload task data or place applications on another device or multiple devices, these devices usually have more computing resources or fewer energy constraints than the user device~\cite{ghafouri2022mobile}. Generally, resource management is closely related to task offloading, in order to make better offloading decisions, we need to understand in detail the different resources in fog computing and edge computing scenarios, and these are all provided by resource management technologies~\cite{sriraghavendra2022dosp}. For example, the estimation, discovery, and matching for resources can be used to make offload decisions, while resource allocation techniques can be used to perform offload decisions, and load balancing and resource orchestration or consolidation are designed to improve resource utilization and speed up response across the system after offloading tasks~\cite{gill2021quantum}. All in all, a better approach to resource management is to better offload task data or application placement to better serve users. Figure \ref{flow-chart-rm} shows the flow of resource management approaches in the edge and fog continuum for realizing Edge AI.

\subsubsection{Resource Estimation}
Our estimates of resources under the two computational paradigms focus on the following five aspects~\cite{2018A}: computational resources~\cite{2015Dynamic} (e.g., CPU computational frequency, the number of CPU cycles required for computing one bit data, etc.), communication resources~\cite{2015Fog} (e.g., spectrum resources under Frequency division multiple access (FDMA)~\cite{2006Single}, length of time allocated per user under Time-division multiple access (TDMA)~\cite{2009Rhee}, etc.), storage resources (e.g., memory for devices, flash memory, etc.), data resources (e.g., some popular content), energy resources (e.g., battery power, virtual energy queues, etc.).

\subsubsection{Resource Discovery}
\textcolor{black}{For resource discovery, it's mostly about discovering which resources are available, where are they located, and how long can they be used (especially devices with batteries). Regarding the implementation of resource discovery, there are two main ways: centralized and distributed}. Centralized \cite{2016Resource, Hamid2015A} refers to selecting a device as a Cluster Header (CH) to record resources on other devices in a cluster of many devices, or setting up a central resource agent as the CH to record resource information for other devices. Once the user has a need, the user sends a message requesting the service to a nearby node, and then the requested device node will check whether it meets the user's needs, if it does not meet, the node will send the user's request packet to the CH of the cluster where the node is located, and then the CH will retrieve the resource record table on it to find a node that meets the user's needs for the user; Distributed \cite{2014Adaptive,2018Agent} refers to the fact that there is no CH to record the resources of other devices, when there is user demand, send a request message to the surrounding agent nodes, and then the requested node checks whether it meets the user needs, if not, the agent node (or the mobile device itself) sends resource request packets directly to all surrounding nodes by broadcasting to "discover" the required resources.

\subsubsection{Resource Matching}
With the continuous development of the era of big data, different types of sensors, mobile devices, edge servers, and fog nodes will be connected to the core network, and the number of devices connected together will be in the hundreds of millions. In the face of so many resource-rich devices and edge nodes to choose from, we should not directly take all the resource nodes found as input before making an offload decision, which will increase the complexity of the offload optimization algorithm and make it difficult to converge. As with a complex neural network model \cite{J2015Deep}, it is better to preprocess the collected raw data first, rather than directly using the collected data as input to the neural network. Resource matching plays the role of ``data preprocessing''.

For two different computing paradigms, the first thing we have to do is to identify malicious nodes~\cite{2019Trust}, exclude malicious nodes by judging the data integrity of the nodes found, and then because the user needs of the two computing paradigms are not only computing, but also storage, acceleration networks, etc., so in addition to the initial matching and screening of resources such as computing, communication, energy and other resources on nodes~\cite{2018Method}, it is also necessary to filter out devices with insufficient processing power or insufficient energy, thereby reducing the dimensionality of input data for offloading decisions. Then, we also need to match and screen these nodes for reliability, security, social ratings, etc.

\subsubsection{Task Offloading and Resource Allocation}
\textcolor{black}{Task offloading is the transfer of resource-intensive computational tasks to an external, resource-rich platform. Partial or full task offloading is usually done to accelerate resource-intensive and latency-sensitive applications \cite{saeik2021task}}. Resource allocation is usually directly associated with task offloading, for the edge computing paradigm, usually we not only have to give offload decisions (Binary offloading~\cite{2016Energy,2017Computation} or partial offloading~\cite{2021Shi,2017Energy}), but also under the response time, energy constraints, or other constraints, give the resource allocation scheme of all devices \cite{2022Combining,2021Computation,2020Task}, in order to meet the needs of users with different preferences.

\subsubsection{Load Balancing}
For task offloading, we generally formulate the offload strategy from the user's point of view, in order to respond to the user's needs faster and reduce the energy consumption of the user's device \cite{lu2022analytical}. However, the reality is that there may be many users who choose the same edge server or fog node for task offload in a period of time. Due to the resource heterogeneity of each device node, in the case of many task requests, there may be some resource-rich nodes with a too-heavy load, and some nodes will have a too-light load, then there will be a waste of resources for the entire system, and may lead to many user processes waiting for too long a time \cite{2020Dynamic}. Then, in order to improve resource utilization more effectively and speed up the response, we must fully consider the load of the system when making the offload decision, transfer the task data from each user device to all edge servers or fog nodes equally, or optimize the processing sequence of the task data of each user \cite{2015Credit}, which can not only alleviate the waste of resources, but also shorten the waiting time of many processes and achieve load balancing of the entire system \cite{2012Load,2020Dynamic}. \textcolor{black}{The load we generally consider can be CPU load, amount of memory used, latency, or network load. Load balancing is defined as a technique that divides workloads into multiple devices (such as edge servers or fog nodes), so load balancing not only considers the needs of users, but also improves the resource utilization of the entire system from the perspective of the system}.

\subsubsection{Resource Orchestration}
A lot of research work is to take load balancing into account before making offload decisions, but the reality may be that after the offload decision, some nodes are still selected by many user devices at the same time, resulting in high latency and low bandwidth of the entire system, for example, when some edge servers or fog nodes have a good signal-to-noise ratio, or contain a lot of popular cache content and high processing power \cite{malik2021effort}. These servers or nodes are often used by a large number of user devices. Therefore, we need to perform resource orchestration of task offloading between nodes~\cite{2020DynamicVu} to improve their service capabilities and the load balance of the entire system. \textcolor{black}{Resource orchestration refers to the coordination of resource allocation of the entire system by migrating offloaded user task data, etc., to each node}.

\subsubsection{Application Placement}
In addition to the user's data needing to be offloaded, sometimes we also need to place the application or model on the user's device on the edge server or fog node, such as some latency-sensitive IoT applications: interactive online games, face recognition, etc. \textcolor{black}{ Application placement~\cite{2017Learning} means that all or some of the compute-intensive components of an IoT application (e.g., services, modules, applications, or models) can be placed (i.e., offloaded) executed and stored on edge servers or fog nodes to reduce the execution time of IoT applications and the energy consumption of IoT devices}.

\subsubsection{Resource/Server Consolidation}
In order to ensure that the placement of applications and the offloading of computing tasks can improve the performance of the entire system on the basis of completing user needs, we can not only re-orchestrate the user's computing tasks or applications, but also consolidate the resources of the entire system, such as server consolidation with the help of Virtual Machine (VM) migration technology~\cite{2019Machine}; Save more energy for the entire system at the cost of increasing the latency of single or multiple users~\cite{2017Computation}; Or from the perspective of resource utilization~\cite{2016Machine}, when the resource utilization is reduced to a certain threshold, resource migration is carried out to achieve the purpose of consolidating resources.

\subsection{ML-based Resource Management in Fog/Edge Computing}
In general, ML algorithms can be broadly classified into (i) supervised learning and (ii) unsupervised learning. Supervised learning aims to develop a model from a collection of training instances ($(X_1,Y_1), (X_2, Y_2), \dots  (X_i, Y_i)$) where $X_i$ and $Y_i$ represents the predictor and label respectively. In unsupervised learning, the algorithms discover hidden patterns and learn the structure of the training data. In the context of offline learning, the models learn over all the observations in a dataset at a go. First, we discuss the problems related to resource management followed by different AI/ML-based offline learning techniques.

\subsubsection{Supervised Learning}

It works by predicting Y outputs using the X inputs given to the algorithm to learn from \cite{cunningham2008supervised}. In short, it consists of ML methods that generate functions with training data. It is generally classified in two ways classification and regression algorithms \cite{nasteski2017overview}. It gives better results in complex problems than unsupervised ML algorithms. On the other hand, since the prediction results depend on the training data, the prediction success rate will decrease when there is bad training data. Intelligent Offloading problems in Edge computing can be solved using Supervised Learning \cite{cao2019intelligent}.

\subsubsection{Unsupervised Learning}
Contrary to supervised learning, is the learning of correlations in data without input-output tags between data \cite{barlow1989unsupervised}. In short, they are ML methods that produce functions according to the densities of the data and their neighborhood relations. It consists of two main approaches: Dimensionality reduction and Cluster analysis \cite{han2020effect}. Dimensionality reduction is also converted to a low dimensional space \cite{dash1997dimensionality}, as high data will require more processing load. Cluster analysis involves grouping clusters of objects with higher correlations to each other \cite{kassambara2017practical}. Resource management in Edge and Fog computing using Unsupervised Learning is still an open research area.

\begin{itemize}
  \item \textbf{K-means Clustering:}  
K-means clustering refers to the method of vector quantization to partition $m$ observations into $k$ clusters where each observation belongs to the cluster with the nearest mean (cluster centers or cluster centroid). K-means clustering is one of the popular methods in resource allocation that can be used for clustering different types of devices based on the available resources in fog/edge computing environments. Such resources can be allocated according to the QoS requirements of each cluster.
\end{itemize}
\subsubsection{Semi-supervised Learning}
\textcolor{black}{It is a machine learning method used to combine lowly labeled data with high rates of unlabeled data. It is generally used where Natural Language Process (NLP) is used. It is frequently used for computation offloading problems in Edge and Fog computing \cite{carvalho2020computation}.}
\begin{itemize}
  \item \textbf{Graph Neural Network (GNN): }
GNN analyzes data represented as graphs for extracting inferences on node-level and edge-level. Graph theory can be adopted where the network can be represented as graph topology. Chen \textit{et al.}~\cite{chen2021gnn} propose a GNN-based framework for resource allocation in wireless IoT networks. The framework specifically deals with the computational and time complexity for conventional resource allocation and outperforms two tasks, namely, link scheduling in Device-to-Device (D2D) networks and joint channel and power allocation. Wang \textit{et al.}~\cite{wang2022learning} present aggregation graph neural network for resource allocation in decentralized wireless networks. 

\end{itemize}

\subsubsection{Reinforcement Learning}
Standard RL is based on an agent being in connection with the environment by way of perception and action. The agent performs an action based on the environment. The RL model is trained in an iterative manner. The agent upon receiving an input (I) along with an indication of the current state of the environment (S), the agent then chooses apt action, which is triggered as output \cite{yang2018applied}. The agent works with the objective to maximize reward points. The state can be defined as the snapshot of the environment at that instant particular in time. In the past decade, RL has expeditiously drawn interest amongst the machine learning and artificial intelligence communities. It is one of the dominant and potential techniques, which is immensely being utilized in several domains, including industry and manufacturing. Q-Learning and State-Action-Reward-State-Action (SARSA) are prominent algorithms of this category that have been widely preferred by researchers in the arena of Fog/Edge computing \cite{chen2018optimized}. Fog nodes often face challenges in context to mobility amongst VMs/Containers and location awareness. Concurrently, it becomes expensive to move the VMs to a new location for which RL provides an efficient solution \cite{vemireddy2021fuzzy}.

\textbf{Resource Allocation Strategies:} RL components such as action space, state space, reward, and Markov Decision Process (MDP) emphasize decisions in different computing paradigms. The algorithms for predicting and deciding which resource to be allocated and when, i.e. optimized resource allocation can be done by following algorithms:

\begin{itemize}
  \item \textit{Deep Neural Network (DNN)}: 
A DNN is a category of an Artificial Neural Network (ANN) with many hidden layers placed between the Output and Input layers \cite{tuli2021gosh}. It can perform real-time allocation of resources as it requires a simple operation. The decisions are made based on experiences and learning, it is different from neural networks in terms of creativity and complications and, hence, gives a global solution with minimal input data \cite{tuli2021mcds}.
\item \textit{Deep Q-Learning (DQN)}: 
Deep Q-Learning is a simple form of RL that utilizes action or Q-values that enhance the behavior of a learning agent iteratively. In Deep Q-learning, the initial state is input to the neural network which in return output all possible Q-values. It was developed by DeepMind in 2015 giving the benefits of both reinforcement learning and deep neural networks \cite{chen2020minimizing}.
\item \textit{Double Deep Q-Network (DDQN)}: 
Double DQN  uses Double Q-Learning to minimize overestimation by breaking down the max operation in the target to action selection and evaluation. The difference between DDQN and DQN is that DDQN uses the main value network for selecting an action \cite{gazori2020saving}.
\item \textit{Deep Reinforcement Learning (DRL)}:
It is a sub-field of ML that combines the benefits of both Deep Learning and Reinforcement Learning. It is able to input large data sets and predicts what action to perform for optimizing an action. The two sub-algorithms are used in this paradigm, namely, model-based and model-free reinforcement learning algorithms \cite{shi2020priority}.

\end{itemize} 
Table \ref{tab:taqx} shows the summary of RL-based resource allocation strategies for Fog/Edge Computing.

\begin{table}[ht!]
\caption{RL-based Resource Allocation Strategies for Fog Computing (FC) and Edge Computing (EC)}
\label{tab:taqx}
\small \resizebox{1\textwidth}{!}{
\begin{tabular}{p{0.04\linewidth} | p{0.2\linewidth}| p{0.075\linewidth} | p{0.2\linewidth} |p{0.2\linewidth} | p{0.2\linewidth}}
\hline
\textbf{Work} & \textbf{Research Focus/Application Area}                                               & \textbf{Paradigm} & \textbf{Method/ Algorithm}                                                                                   & \textbf{Parameter}                        & \textbf{Result}                                                                                                                             \\ \hline
\cite{vemireddy2021fuzzy}      & Task offloading energy efficiently in Vehicular Fog Computing (VFC) for smart villages & FC               & Fuzzy Reinforcement Learning, Integrated on-policy reinforcement learning technique (SARSA) and Greedy heuristic                                       & Total task service time, energy consumption, and average response time  & Outperforms over other algorithms up to 15.38\% and 46.73\% in terms of query response time and energy consumption respectively.\\ \hline
\cite{chen2018optimized}    & Computation offloading in Virtual Edge computing systems (Sliced Radio Access Networks)   & EC             & Integrated Double Deep Q-Network with Q-Function decomposition technique (online Deep-state-action-reward-state-action-based RL algorithm (Deep-SARL)) & Maximizing Long term utility performance & Outperforms over three baseline schemes, namely, mobile execution, server execution, and greedy execution \\ \hline

\cite{ning2019deep}      & Intelligent offloading system for vehicular networks                                     & EC             & Mobility-Aware Double DQN (MADD), Dynamic V2I Matching Algorithm                                                                                         & Task scheduling and resource allocation (Quality of Experience)                                                                                    & Proposed MADD algorithm performance is 20\% and 12\% higher than greedy and DQN method, respectively\\ \hline
\cite{conti2017battery}     & Green Fog Computing (Battery management)                                                 & FC              & Markov-Based analytical model integrated with reinforcement learning process                                                                             & Job Loss Probability                                                                                                                               & Effect of Battery Energy Storage System (BESS) varies on the system according to the number of servers                         \\ \hline
\cite{yang2018deep}     & Resource allocation edge computing network for multiple user                       & EC              & Deep Q-Learning                                                                                                                                          & Data packet size, Channel quality, and waiting time                                                                                                & Deep Q-learning outperforms the random and equal scheduling                                                                      \\ \hline
\cite{chen2019iraf}    & Intelligent   Resource Allocation Framework (iRAF) for Edge paradigm                       & EC            & Deep   Neural Network for prediction and Monte Carlo Tree Search (MCTS) approach for generating training data                                            & Network states and task characteristics like utilization of edge network resources, the channel quality, latency requirement of services, etc & iRAF achieves 51.71\% and 59.27\% performance over deep learning and greedy search methods respectively                        \\ \hline
\cite{shi2020priority} & Task offloading scheme on priority basis for vehicular Fog Computing & FC              & Soft Actor-Critic (SAC) based Deep reinforcement learning algorithm                                                                                      & Entropy of policy and Expected utility to be maximized                                                                                             & High priority task completed preferentially while having better performance of task completion and ratio offloading delay       \\ \hline
\cite{liu2019resource}      & Resource allocation in Internet-of-Things network  & EC            & $\epsilon$-greedy Q-learning   & Long term weighted sum cost (task execution latency and power consumption )     & Achieved a better trade-off between edge and local computing modes\\ \hline
\end{tabular}
}
\end{table}

\subsection{Performance Metrics}
Optimisation and comparison of any AI-based fog and edge computing architectures are done on the basis of one or more performance metrics, hence these metrics play an important role in the analysis of this architecture and also help to define the merits and demerits of an architecture \cite{aslanpour2020performance}. Performance metrics are mostly dependent on the type of layers/ computing model where the architecture is performed. In general, any Fog/Edge computing architecture can be separated on the basis of the following 4 layers.

\begin{itemize}
\item \textbf{IoT Layer}: This layer is regarded as the first layer of any architecture. This layer is defined as where the IoT devices like Raspberry pi or Arduino can do computation and can coordinate with other sensor nodes and forms a mesh topology-based network. In this layer, the devices are responsible for sensing the data from the sensors and doing some minor operations. This layer can be implemented without any interaction with edge, fog or cloud layers.

\item \textbf{Edge Computing Layer}: This layer comes next to the IoT layer. This layer consists of switches and routers which are generally termed gateways. This layer acts as an entry point to the fog and cloud layers. It is responsible for workload distribution and traffic monitoring. It is also responsible for a few less expensive operations resulting in minimizing the response time and optimizing the latency. 

\item \textbf{Fog Computing Layer}: Fog can be defined as the combination of edge and cloud. This layer is located near to edge and IoT layer, and has the capability to perform the expensive operation in comparison to the edge layer. This layer helps to respond to the request faster by computing the work rather than sending the request to the cloud.
\item \textbf{Cloud Computing Layer}: This layer is the ultimate layer and most powerful layer. The operations which are highly expensive and cannot be performed by any of the previous layers are performed in this layer.
\end{itemize}
The performance of the above-mentioned layers is measured in multiple terms.
\subsubsection{Monitoring Related metrics}
These metrics are responsible for monitoring the performance of the entire architecture. Few such metrics are

\begin{enumerate}
\item \textbf{\textit{Resource Utilisation}}: This metric is defined as the amount or the percentage of the resource used or occupied by a specific incoming workload.
\item  \textbf{\textit{Throughput}}: It can be considered as a ratio of the number of tasks arrived at to the number of tasks processed for a certain interval of time.
\item  \textbf{\textit{Resource Load}}: It is defined as the measure of the number of tasks waiting in the queue to be executed along with the number of tasks running.
\item \textit{\textbf{Latency}}: It is the amount of time gap between actual response time and desired response time.
\item \textbf{\textit{Maximum Running Resource}}: It is the highest number of resources used.
\item \textbf{\textit{Virtual Machine Runtime}}: It is the time for which the VM is borrowed.
\item \textbf{\textit{SLA Violation}}: It is defined as the number of tasks that have been delayed more than the time conceded.
\item \textbf{\textit{Energy Consumption}}: It is described as the measurement  of the  energy required by a source to finish the execution of a certain workload.
\item \textbf{\textit{Fault tolerance}}: It can be defined as a ratio of the number of faults detected to the number of faults that exist.
\end{enumerate}

\subsubsection{Analysis related Metrics}

Analysis-related metrics are used for the analysis of the performance using monitoring-related metrics. Statistical methods like machine learning or deep learning can be used for this purpose.
\begin{enumerate}
\item \textbf{\textit{Statistical Analysis}}: This is the process where a huge amount of time series data is statistically analyzed and some meaningful information is extracted.
\end{enumerate}
\subsubsection{Planing Related Metrics}
Planning is the phase in which decision regarding optimization is taken such as VM migration and VM placement.
\begin{enumerate}
\item \textbf{\textit{Decision Number}}: It is referred to the total number of decisions taken.
\item \textbf{\textit{Contradictory Decision}}: It is the number of times an already made decision is reversed, due to an incorrect decision.
\item  \textbf{\textit{Completion Ratio}}: It is referred to as the ratio in which sources compete for resources.
\item \textbf{\textit{Cache Hit Ratio}}: It is referred to as the success of a service caching system in reducing the data transmission across the network.
\end{enumerate}

\subsection{Simulators}
\label{simulators}
Simulators are the first experimental setup in which architecture is tested before deployment. As any architecture has multiple layers, simulators differ from layer to layer.
\subsubsection{IoT layer Simulators}
IoT layers initially experiment in this environment. Two Popularly known simulators are
\begin{enumerate}
\item  \textbf{\textit{SysML4IoT}}: Abstractions are provided by SysML4IoT to precisely specify various hardware and software services, data flows, and personnel \cite{costa2016modeling}. 
\item \textbf{\textit{IOTSim}}: IOTSim \cite{zeng2017iotsim} is a simulator that uses the MapReduce model, for IoT Big Data processing and simulations in the cloud computing environment. Using this simulator makes the work easier and more cost-effective instead of renting entire large-scale data centers.
\end{enumerate}
\subsubsection{Edge Layer Simulators}
The edge layer is the next layer to the IoT layer, this layer consist of the Gateways and switches.
A few well-known simulators used for the simulation of Edge layers are
\begin{enumerate}
\item \textbf{\textit{PureEdgeSim}}: PureEdgeSim \cite{mechalikh2019pureedgesim} is a large-scale simulation framework for studying the IoT as a distributed, dynamic, and highly heterogeneous infrastructure, as well as the applications that run on these things. It includes realistic infrastructure models, allowing for research on the edge-to-cloud continuum. It covers all aspects of edge computing modeling and simulation. It has a modular design, with each module addressing a different aspect of the simulation. For example, the Datacenters Manager module is concerned with the creation of data centers, servers, and end devices, as well as their heterogeneity. The Location Manager module, on the other hand, handles their geo-distribution and mobility. Similarly, the Network Module is in-charge of allocating bandwidth and data transfer.

\item \textbf{\textit{IoTsimEdge}}: IoTsimEdge \cite{jha2020iotsim} is the extension of IoTsim, which provides the testing of the Edge layer of architecture. This helps its user to test the heterogeneous edge and IoT layer in a configurable manner. It is very user-friendly and easy to use.
\item \textbf{SimEdgeIntel}: 
SimEdgeIntel \cite{wang2021simedgeintel} is an edge simulator that provides the facility to easily deploy mobile with edge intelligence. It provides researchers with detailed configuration options such as customized mobility models, caching algorithms and switching strategies to test their resource management techniques.

\end{enumerate}
\subsubsection{Fog Layer Simulator}
A few majorly known Fog layer simulators are:
\begin{enumerate}
\item \textbf{\textit{iFogSim}}:
iFogSim \cite{gupta2017ifogsim} allow researchers to develop, deploy and test their IoT applications in fog-cloud infrastructure to test custom-made resource management
strategies. It provides a hierarchical fog architecture simulation with the first layer made of sensors and actuators for the generation of data, and other layers simulate fog and cloud computing, network,  and storage resources. 
\item
\textbf{
iFogSim2}: iFogSim2 \cite{mahmud2022ifogsim2} is an advanced version of iFogSim \cite{gupta2017ifogsim}. It offers advanced features like migration, mobility support, dynamic distribution, and microservice orchestration with resource management.
\item \textbf{RelIoT}:
RelIoT \cite{ergun2020reliot} is NS-3 simulator-based reliability framework for IoT networks. It enables researchers to design customized network reliability management strategies by providing reliability-oriented analysis and predictions early in the design cycle.  
\item
\textbf{YAFS}:
YAFS (Yet Another Fog Simulator) is a discrete event-based simulator \cite{lera2019yafs} used to model complex IoT application scenarios in fog infrastructure. With the placement, scheduling, and routing strategy modeling facility, it also provides dynamic module allocation and user movement features.

\item\textbf{COSCO}: COSCO  \cite{tuli2021cosco} is a python-based simulator that provides the facility to develop and test scheduling policies for an edge, fog, and cloud-integrated environment. It provides seamless integration of scheduling policies with a simulated back-end for enhanced decision-making. It also supports real deployment in real-world applications.

\item \textbf{DeepFogSim}:
It is an extension of VirtFogSim which provides the execution of applications described by generally directed Application Graphs \cite{scarpiniti2021deepfogsim}. DeepFogSim simulates the  Conditional Neural Networks (CNNs) with early exit on customized fog topology and performance of dynamic joint optimization and tracking of the energy and delay performance of Mobile-Fog-Cloud systems.
\item \textbf{\textit{iThermoFog}}: This simulator is used to measure the heat or temperature of Cloud Data Centers (CDC) \cite{tuli2020ithermofog}. This simulator uses a Gaussian model to approximate the thermal characteristics of the fog layer server, and optimize the average temperature by scheduling the task.
\item \textbf{FogNetSim++}:
It is an extension of FogNetSim and provides the facility to simulate both networks with computing modeling aspects of fog computing \cite{qayyum2018fognetsim++}. It supports low-level network details such as switching and packet routing. 
\end{enumerate}
\subsubsection{Cloud layer Simulator} Cloud layer is the final layer of any architecture, which is responsible for storage and high-capacity computation. A few major Cloud layer simulators are
\begin{enumerate}
\item \textbf{\textit{CloudSim}}: CloudSim  \cite{calheiros2011cloudsim} is the most widely used simulator for the simulation of Cloud layer architecture. Many modifications and advances of CloudSim are brought like Dynamic Cloudsim, Container CloudSim, Network CloudSim, and CloudSim Plus. The most recent version of CloudSim is CloudSim5, which combines various releases including containers, VM extensions with performance monitoring features and modeling of Web applications on multi-clouds. This version of CloudSim can also work with other Software-Defined Networking (SDN)/Service Function Chaining (SFC) simulation functions.
\item  \textit{\textbf{ThermoSim}}: ThermoSim \cite{gill2020thermosim} is similar to works similarly to iThermoFog, but the CDC whose thermal profile optimization is the cloud layer CDC. This simulator reduces the temperature of cloud CDC by proper scheduling of tasks.
\item \textbf{IoTSim}: 
IoTSim is built on top of cloudSim simulator, it simulates the processing of IoT big data using the MapReduce model in cloud \cite{zeng2017iotsim}.
\end{enumerate}

\subsection{Workloads}
Different researchers have used multiple execution traces and benchmarks for the simulation of workloads for AI applications.
Some of the well-known workload traces and benchmarks  are : 
\begin{enumerate}

\item \textbf{\textit{DeFog}}:
DeFog consists of five computation-intensive AI applications. These applications cover a diversity of workloads such as deep learning-based object classification applications (YOLO), speech-to-text conversion applications (PocketSphinx), geo-location based online mobile game applications (ipoke-Mon), IoT edge gateway applications (FogLamp), real-time face detection from video streams application (RealFD)  and Text audio synchronization or forced alignment (Aeneas). This benchmark captures application-specific system performance metrics for
different application domains \cite{defog-jonathan2019}. 

\item \textbf{\textit{AIOT BENCH}}:
This benchmark is designed for evaluating IoT device intelligence. It covers different application domains such as image recognition, speech recognition, and natural language processing \cite{luo2018aiot}. 
\item \textbf{\textit{RIoTBench}}:
It is a real-time suit that captures both system level such as CPU, memory, network, and storage performance metrics, and application-specific system performance metrics for different application domains. It evaluates distributed stream processing systems for streaming IoT applications. It contains 27 IoT tasks classified across multiple categories \cite{shukla2017riotbench}.

\item \textbf{\textit{Edge AIBench}}:
This benchmark model four application scenarios namely: Smart home, autonomous vehicle, ICU patient monitoring, and  Surveillance Camera for workload collaboration between three layers. Given models can be executed using a federated learning framework available publically \cite{hao2018edge}. 
\item \textbf{\textit{Bitbrain}}:
Bitbrain traces consist of performance metrics of more than one thousand hosts of the heterogeneous cloud data center \cite{shen2015statistical}. These traces are categorized into two categories: RND and fast storage. They consist of models of CPU, RAM, Disk and Bandwidth utilization characteristics. These traces are related to  business-critical workloads.

\item \textbf{\textit{Azure2016}}:
This dataset contains VM workload traces captured from November 16, 2016 to February 16,
2017. The information captured in this dataset includes VM id, its subscription, VM role name, cores, memory, and disk allocations, and minimum, average, and maximum VM resource utilization \cite{cortez2017resource}.

\end{enumerate}

\section{AI Based Techniques For Resource Management in Fog/Edge Computing}
\label{AI/ML Based Techniques For Resource Management: Current Statu} 
\begin{figure*}
	\centering
	\includegraphics[width=.7\textwidth]{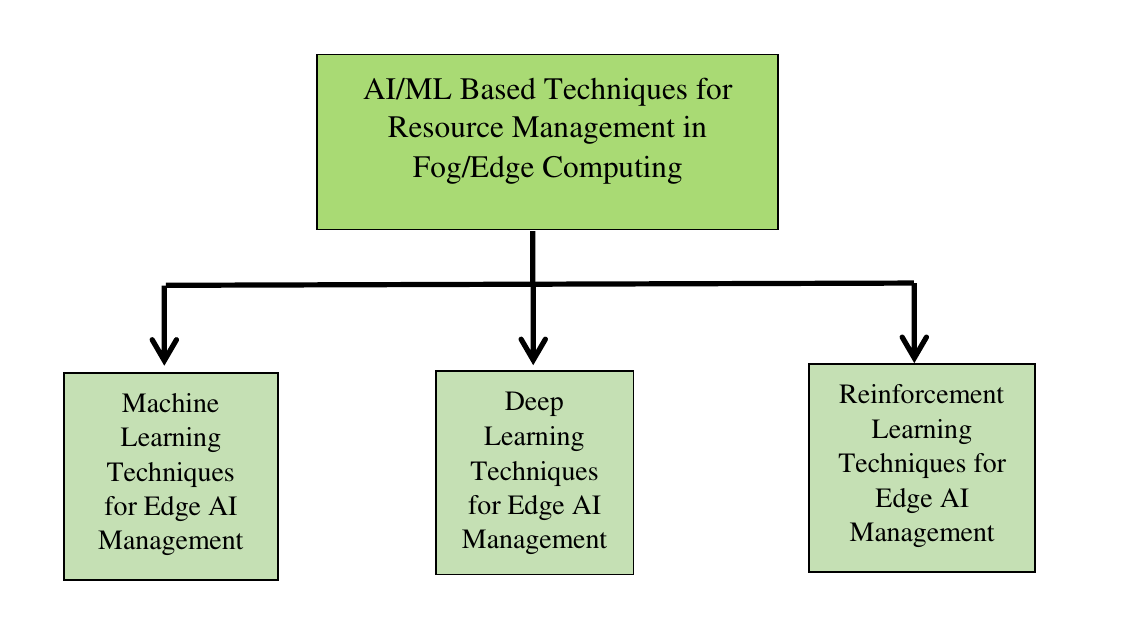}
	\caption [Caption for LOF] { \centering Taxonomy of AI-Based Techniques For Resource Management in Fog/Edge Computing}
	\label{TaxonomyRM}
\end{figure*}

In this section, we discuss resource management-focused AI/ML techniques for enabling Edge/Fog AI and present a summary of AI/ML-based resource management techniques for Fog/Edge Computing. Figure \ref{TaxonomyRM} shows the taxonomy of AI-Based Techniques For Resource Management in Fog/Edge Computing.

\subsection{Machine Learning Techniques for Edge AI Management}
\label{Machine Learning Techniques for Edge AI Management}
Machine learning is showing remarkable results in various fields. The promising results of machine learning in various domains are also attracting researchers' attention to the use of ML for modeling, classification, prediction, and forecasting related to resource management in fog computing for enabling Edge AI \cite{merenda2020edge}. 
When we discuss multi-tenant environments where infrastructure is used by many applications which have different requirements, it is very obvious that tunning one application may have impacts on other applications \cite{murshed2021machine}. Also adding complexity to the problem, these applications are deployed on multiple heterogeneous distributed resources.  When dealing with different workloads, and heterogeneous distributed resources, it is better to analyze the workload from application and infrastructure perspectives. Understanding workload behavior can lessen the complexity of the problem and improve the performance of an application in edge AI \cite{lee2018techology}. Different works in the literature considered workload analysis using ML algorithms. The IoT-enabled edge AI is also prone to temporal effects. There can be an increase in workloads on certain resources at certain times. The prediction of workload behavior or spatio-temporal effects in advance can ease the resource orchestration process.
To facilitate auto scaling of resources, Liu \textit{et al.} \cite{liu2017adaptive} proposed a workload pattern discrimination-based adaptive prediction approach for infrastructure as a service. Due to the speed of workload change, they classified workload into two groups: fast time scale data and slow time scale data. Fast time scale data had the feature of stochasticity and nonlinearity while slow time scale data had the feature of linearity.  For two datasets they used Support Vector Machine (SVM) and Linear regression (LR)  for the prediction of workload.
Another work proposed use of auto-correlation measurement and similarity clustering for CPU workload prediction on VMs \cite{bankole2016predicting}. A combination of random forest, SVM and neural network is used to predict future workloads to reduce training time and increase the accuracy of the model \cite{didona2015enhancing}. Work in reference \cite{marcus2016workload} addressed the issue of workload management using  decision trees.
As fog environment is distributed and heterogeneous and diverse IoT-based AI applications with different resource requirements make a selection of optimal nodes for application placement to satisfy QoS and Quality of Experience (QoE) constraints,  more challenging. Addressing the application placement problem in mobile fog, Rahbari \textit{et al. } proposed to use of classification and regression trees. In order to manage power consumption in edge/fog-based smart building services, work in \cite{ferrandez2018deployment} used KNN and decision tree algorithms.

\begin{figure*}
	\centering
	\includegraphics[width=1\textwidth]{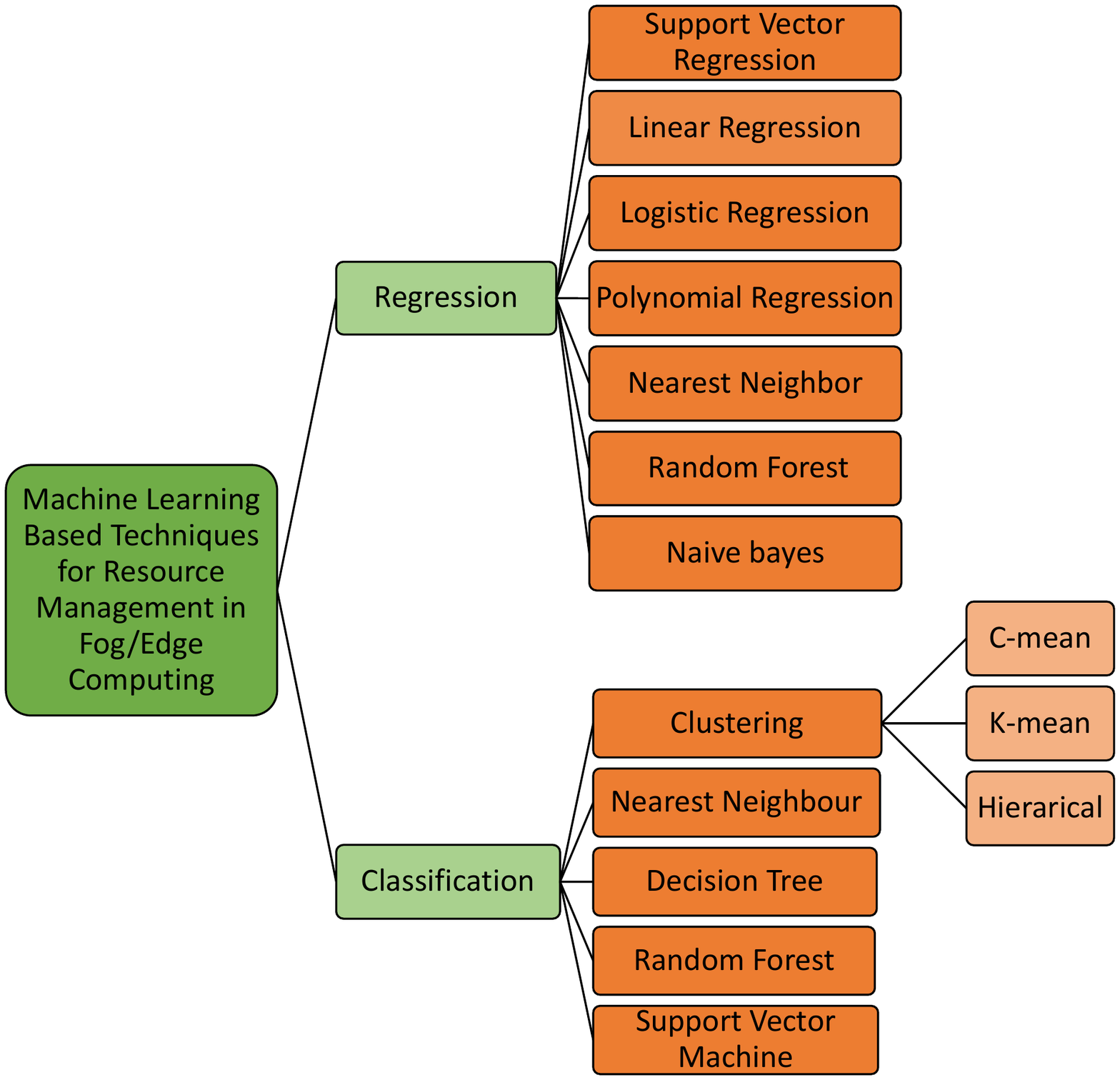}
	\caption [Caption for LOF] { \centering Summary of Machine Learning-based Resource Management for Fog/Edge Computing}
	\label{MLTaxonomyRM}
\end{figure*}

In another work, they addressed the application placement problem for smart city applications. They employed logistic regression and support vector machine for job completion time approximation \cite{he2017multitier}. Addressing the resource scheduling problem, Liu \textit{et al.} ~\cite{liu2019resource} combined fuzzy c-mean clustering with particle swarm optimization. Using optimized fuzzy c-mean clustering they tried to reduce the scale of resource search. They compared the proposed work with Fuzzy c-mean clustering and the objective function value of optimized fuzzy c-mean showed faster convergence speed than the Fuzzy c-mean algorithm. Yadav \textit{et al.} 
 \cite{2019Trust} also used fuzzy c-mean clustering for task allocation in distributed systems to minimize the cost of the system. To minimize delay for IoT-based applications in fog environment, Shooshtarian \textit{et al.}~\cite{shooshtarian2019clustering} used hierarchical clustering to find the nearest neighbour node to the IoT device to solve the resource allocation problem. 
Container orchestration is also investigated using ML methods by many researchers. Researchers in reference \cite{meng2016crupa} used a time series analysis model (ARIMA) combined with the docker container technique for resource utilization prediction in containerized applications. Another work explored performance analysis of containerized applications using polynomial regression and k-means clustering. They classified multi-layer container execution structures based on the application performance requirement in distributed resources \cite{venkateswaran2019fitness}. 
Authors in \cite{yang2018intelligent} used SVM, Boosting decision tree, Random forest and Naive Bayes for node performance prediction to improve resource scheduling decisions.
Podolskiy \textit{et al.} \cite{podolskiy2019maintaining} used Lesso for the self-adaptive problem of vertical elasticity for co-located containerized applications.
 Figure \ref{MLTaxonomyRM} presents a summary of machine learning algorithms that are used for the analysis and prediction of workload and resource usage to aid resource management in edge/fog computing. 

\subsection{Deep Learning Techniques for Edge AI Management}
\label{Deep Learning Techniques for Edge AI Management}
\begin{figure*}
	\centering
	\includegraphics[width=1\textwidth]{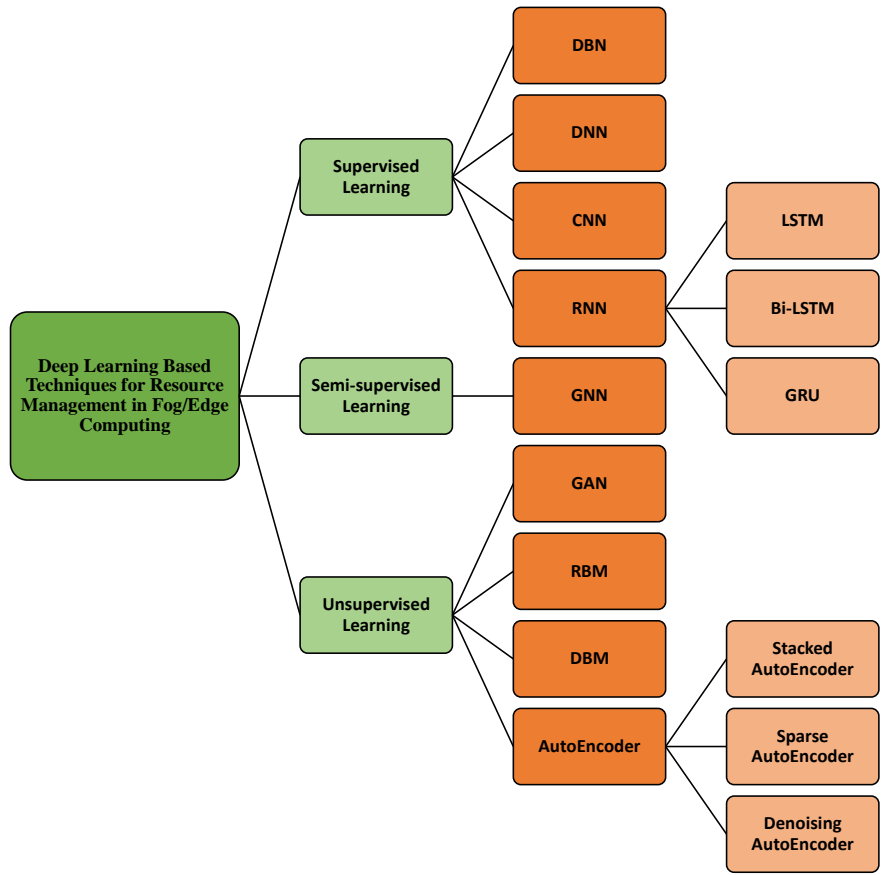}
	\caption [Caption for LOF] { \centering Summary of Deep Learning-based Resource Management for Fog/Edge Computing}
	\label{DLTaxonomyRM}
\end{figure*}

Currently, deep learning-based prediction models are the most promising architectures for computational intelligence. It shows good performance in various problems such as workload prediction, where traditional machine learning algorithms fail. CNNs can be used to model wide spatial dependencies by extracting local features by adopting layers with convolutional filters \cite{mozo2018forecasting}. Long Short-Term Memory (LSTM) can be utilized for the prediction of fluctuating and volatile workload time series due to its capability to capture long-term temporal dependencies \cite{kumar2018long}.
 For time series analysis, authors in \cite{zhang2018efficient} have presented  a deep learning model based on the canonical polyadic decomposition for workload prediction for industry informatics. Sima and Saeed \cite{jeddi2019water} used CNN for predicting future cloud workload in advance for optimized resource allocation.
For dynamic management of network resources, Bega \textit{et al.} \cite{bega2019deepcog} also used CNN in their work. Their proposed strategy returns a cost-aware capacity forecast, which can be directly used by network operators to take re-allocation decisions that maximize their revenues. Authors in reference \cite{levy2020predictive, chen2019iraf} addressed the resource provisioning issue with Fully Convolutional Networks (FCNs). Tuli \textit{et al.} 
 \cite{tuli2021start} focused on straggler detection for system's QoS and used an encoder LSTM network for the analysis of tasks. Their proposed model analyzes the tasks and predicts which tasks can be straggler. Work in \cite{feng2019content} also used Bi-LSTM to address the scheduling issue of fog-enabled Radio Access Networks (F-RAN). For optimal performance, they used Bi-LSTM for the prediction of content popularity. Some of the recent work also used DNN as a surrogate model for QoS prediction to make scheduling decisions \cite{tuli2021gosh, tuli2021mcds}.
Considering accurate resource requests prediction essential for achieving efficient task scheduling and load balancing, Zhang \textit{et al.} \cite{zhang2017resource} used  Deep Brief Network (DBN) for day and hour scale predictions of CPU and memory utilization. They evaluated their proposed technique with the Google dataset. 
Many works also improved prediction performance by ensembling multiple algorithms for workload prediction for fog computing \cite{yazdanian2021e2lg}. To provide real time responses to vehicular applications, such as traffic and accident warnings in the highly dynamic IoV environment. Lee \textit{et al.} \cite{lee2020resource} used LSTM-based Deep Neural Networks (DNNs) to predict mobility behaviour and movements of  vehicles. They combined RL with LSTM-based DNN for resource allocation in Vehicular Fog Computing (VFC). In their other work, they also used Recurrent Neural Network (RNN) for resource allocation problems. To extract the time and space-based pattern of resource availability, they integrated the RNN into the DNN of the proximal policy optimization algorithm \cite{lee2020resource}. 
Another work \cite{ouhame2021efficient} used a hybrid CNN-LSTM model for the prediction of multivariate workload in an attempt to extract complex features of the VM usage components, then model temporal information of irregular trends in the time series components. They evaluated the proposed model for resource provisioning using bit brains data and compared it with other predictive models. 
To minimize the complexity and non-linearity of the prediction model, Yazdanian \textit{et al.}~\cite{yazdanian2021e2lg} decomposed workload time series into its constituent components in different frequency bands and used ensemble  Generative Adversarial Networks (GAN)/LSTM for prediction of each sub-band workload time-series. The proposed model employs stacked LSTM blocks as its generator and 1D ConvNets as the discriminator.
Graph Neural Network (GNN) analyzes data represented as graphs for extracting inferences on node-level and edge-level. Graph theory can be adopted where the network can be represented as graph topology. Chen \textit{et al.}~\cite{chen2021gnn} propose a GNN-based framework for resource allocation in wireless IoT networks. The framework specifically deals with the computational and time complexity for conventional resource allocation and outperforms two tasks, namely, link scheduling in D2D networks and joint channel and power allocation. Wang \textit{et al.}~\cite{wang2022learning} present aggregation graph neural network for resource allocation in decentralized wireless networks. Figure \ref{DLTaxonomyRM} presents a detailed classification of deep learning based algorithms used for Resource Management in Edge/Fog computing.

\subsection{Reinforcement Learning Techniques for Edge AI Management}
\label{Reinforcement Learning Techniques for Edge AI Management}
As resources are heterogeneous and capacity-constrained in edge, smart resource allocation is considered one of the important factors in enabling Edge AI. Considering edge/fog a pool of different resources like CPU, GPU, storage etc. efficient resource allocation necessitates resource sharing. Reinforcement learning algorithms are the most experimented algorithms for resource sharing and allocation decision-making at the moment. Usually, MDP type problems are solved using policy gradient methods, tabular RL and deep Q-learning methods. Tabular methods such as Q and SARSA are not preferred by researchers due to their low scalability for modeling computing systems with thousands of devices.

In ~\cite{shi2020priority}, author presented a deep reinforcement learning (DRL)-based scheme for task offloading for VFC application. They proposed a soft-actor critic for the maximization of the policy of entropy and anticipated reward.  Fu et al. \cite{fu2020soft} utilized a maximum entropy framework-based soft actor-critic DRL algorithm in VFC-enabled Internet of vehicles (IoV) for providing low bitrate variance live streaming service for vehicles. 
To reduce the vehicle’s long-term mean cost with promising reliability and latency performance in VFC, a Deep Q Network (DQN) is presented for the switching problem. They designed the mobile network operator (MNO) preference and switching problem by simultaneously analyzing switching cost, cost variation by MNO and fog servers, and QoS variation within MNOs ~\cite{zhang2020deep}. 
In another work~\cite{van2019optimal}, a dual neural network of Deep Q-Learning method is implemented for resource slicing management. They formulated a semi-MDP for the simultaneous allocation of resources.  
Considering computational offloading an important factor for enabling edge AI, another work considered energy-efficient vehicle scheduling for task offloading in VFC. To resolve the high dimensionality issue caused by the increased number of vehicles in road-side units (RSU) coverage, an on-policy reinforcement learning-based scheduling algorithm combined with a fuzzy logic-based greedy heuristic, named Fuzzy Reinforcement Learning (FRL) is proposed. This greedy heuristic not only accelerates the learning process, but also improves long-term reward when compared to Q-learning algorithm~\cite{vemireddy2021fuzzy}
Chen \textit{et al.} ~\cite{chen2018optimized} addressed offloading issue of virtual edge computing. They formulated the offloading problem in a sliced radio access network as MDP. They resort to DNN based function approximator and drive a double deep q network for making offloading decisions. Cheng \textit{et al.} \cite{cheng2021multiagent} proposed a policy gradient learning-based scheduler for task scheduling in edge devices. The same approach Multi-agent Deep Deterministic Policy Gradient (DDPG)-based scheduling is adapted for joint task partitioning and power control in fog computing networks.  

Ning \textit{et al.}~\cite{ning2019deep} explored deep reinforcement learning for optimization of task scheduling and resource
allocation in vehicular networks. They divided the problem into two sub-optimization problems. First is deciding the priority of the vehicles for the quality of experience of users using a utility function. The second subproblem of resource allocation is formulated as the DRL problem. A deep Q network is improved by applying dropout regularization and double deep Q networks to deal with the defect of overestimation.
To address the resource provisioning issue in fog, work in reference \cite{sami2021ai,xu2020recarl} used Deep RL based on DQN. In addition, some authors experimented with Policy gradient learning for efficient resource provisioning resources \cite{chen2021deep}. One other work used A3C (Asynchronous Advantage Actor-Critic) and residual neural network for scheduling stochastic edge-cloud environment \cite{tuli2020dynamic}. some work also used the same RL model for workflow scheduling \cite{hu2019learning, ghosal2020deep}.

Liu \textit{et al.}~\cite{liu2019resource} also addressed the resource allocation problem for IoT-enabled edge AI and proposed a $\epsilon-$ greedy Q-learning-based optimum offloading algorithm. The problem is formulated as a weighted sum cost minimization problem with its objective function including the task execution latency and the power consumption of both the edge device and the end device.
\begin{figure*}
	\centering
	\includegraphics[width=1\textwidth]{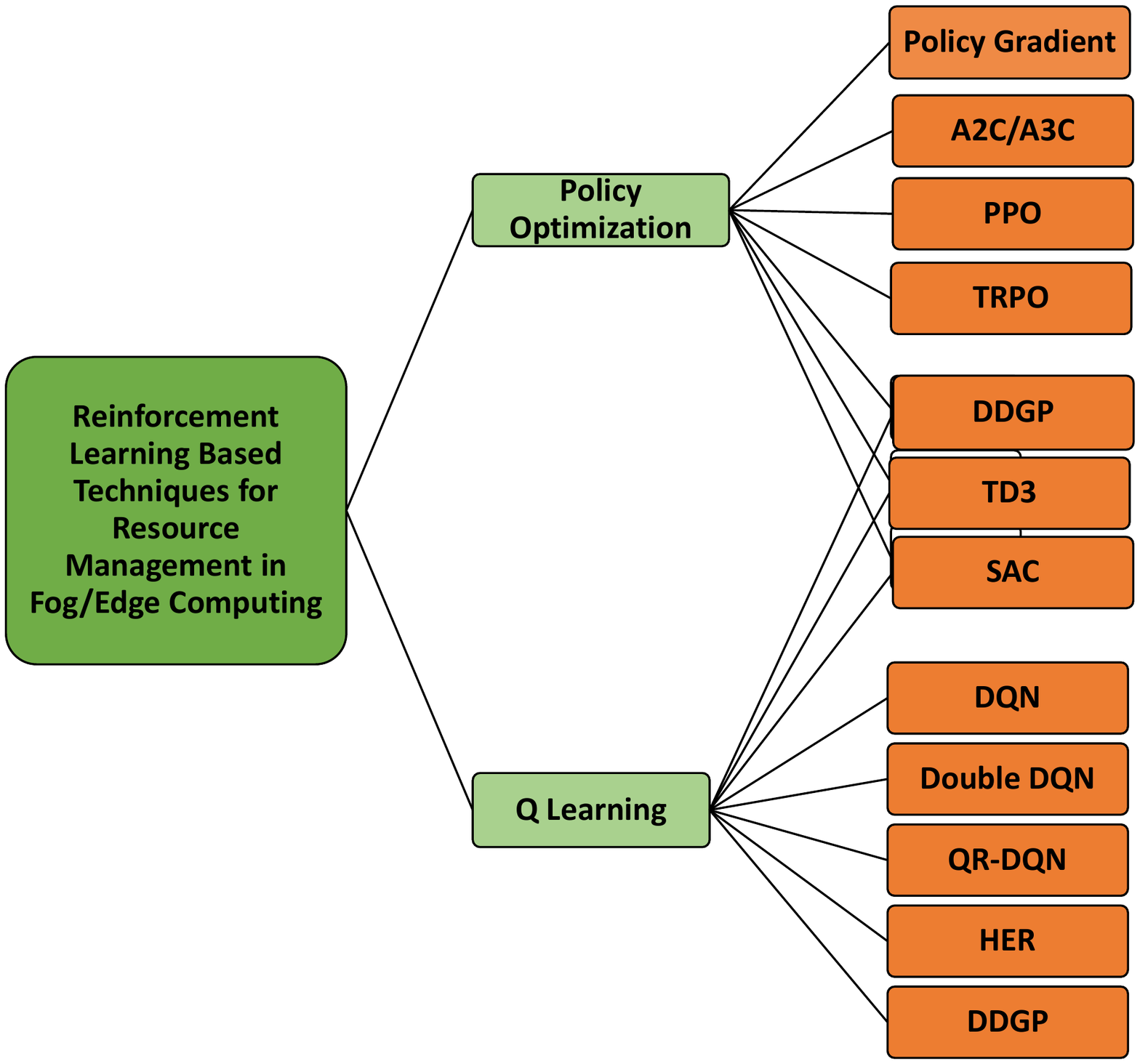}
	\caption [Caption for LOF] { \centering Summary of Reinforcement Learning-based Resource Management for Fog/Edge Computing}
	\label{RLTaxonomy-RM}
\end{figure*}

Since in the majority of IoT-based Edge AI systems, sensors produce a lot of data that needs to be processed within the deadline of applications, the inherent lack of information in tasks arrival of such systems necessitates adaptive task scheduling. Intelligent task scheduling not only minimizes task execution delays but also improves system key performance indicators (KPI) like reduced energy consumption, and load balancing. DRL methods have shown some promising results in decision-making problems. There are several works in the literature that explored RL for adaptive task scheduling. To minimize computation costs and long-term service delays, a Double deep Q learning (DDQL) is proposed in work~\cite{gazori2020saving}. In order to achieve optimal action selection, each agent used  two separate models for action selection and Q-value calculation. Each RL agent was embedded in the gateway device to schedule tasks and allocate resources to tasks. The RL agent tries to  maximize cumulative reward to achieve reduced end-to-end delay. To cease fluctuation in the results, they integrated the target network and experience replay mechanism in the DDQL-based scheduling policy. In order to maximize the long-term value of QoE, Sheng \textit{et al.} ~\cite{sheng2021deep} designed an intelligent task scheduling system using a model-free DRL algorithm. They formulated a task scheduling problem on heterogeneous virtual machines as an MDP problem and solved it with policy-based DRL. This work considered task satisfaction degree as reward and action is represented as a pair of tasks and VM. They decoupled real-time steps from scheduling steps in MDP formulation to make action space linear with a product of the number of virtual VMs and queue size and to schedule multiple tasks in a single time step. In order to achieve ultra-low latency and fairness in resource sharing, Bian \textit{et al.}~\cite{bian2019online} proposed FairTS that ensures fairness between tasks and with ultra-low average task latency. 
One other factor that can degrade the performance of Edge AI applications is an imbalance in workload distribution between resources in the system. The solution to this issue is offloading or redistribution of tasks. RL is investigated for offloading decision-making in ~\cite{van2018deep}. They formulated Offloading problem as MDP and proposed a DRL-based scheme to make users enable to make near-optimal decisions by considering  uncertainties in the user device and cloudlet movements and resource availabilities. Another work used Deep Q Network (DQN) for making optimal actions on how main tasks will be offloaded and how many processed locally ~\cite{mnih2015human}. Chen \textit{et al.}~\cite{fed_sensors} propose a two-timescale federated deep reinforcement learning based on Deep Deterministic Policy Gradient (DDPG) to solve the joint optimization problem of task offloading and resource allocation to minimize the energy consumption of all IoT devices subject to delay threshold and limited resources. The simulation results show that the proposed algorithm can greatly reduce the energy consumption of all IoT devices.
 Lee and Lee \cite{lee2020resource} utilized proximal policy optimization (PPO) RL for offloading problems in order to provide real-time responses for vehicular applications. PPO with the ability to continuously learn dynamic environments can easily adjust to make resource allocation decisions accordingly.
 Some works \cite{chen2021deep, hu2020mdfc} used Policy gradient learning for the deployment of DNN in Edge AI. For efficient real-time resource allocation and offloading in internet of vehicles, Hazarika \textit{et al.} \cite{hazarika2022drl} utilized DDPG
and twin delayed DDPG (TD3) algorithms. They compared the proposed technique Soft Actor Critic (SAC) and DDPG. Another work formulated resource allocation in Mobile Edge Computing (MEC) as an MDP problem in order to minimize system delay and solved it with hindsight experience replay (HER) improved DQN \cite{cen2022resource}. Figure \ref{RLTaxonomy-RM} presents a summary of reinforcement learning based algorithms that are used separately or in hybrid fashion with DL for resource management in Edge/Fog computing.

\section{Taxonomy}
\label{Taxonomy}
This section discusses the proposed taxonomy of frameworks and comparison analysis in AI-based edge and fog computing

\subsection{Taxonomy of AI-based Fog and Edge Computing}

\begin{figure}[!ht]
\centering
\includegraphics[width=6in]{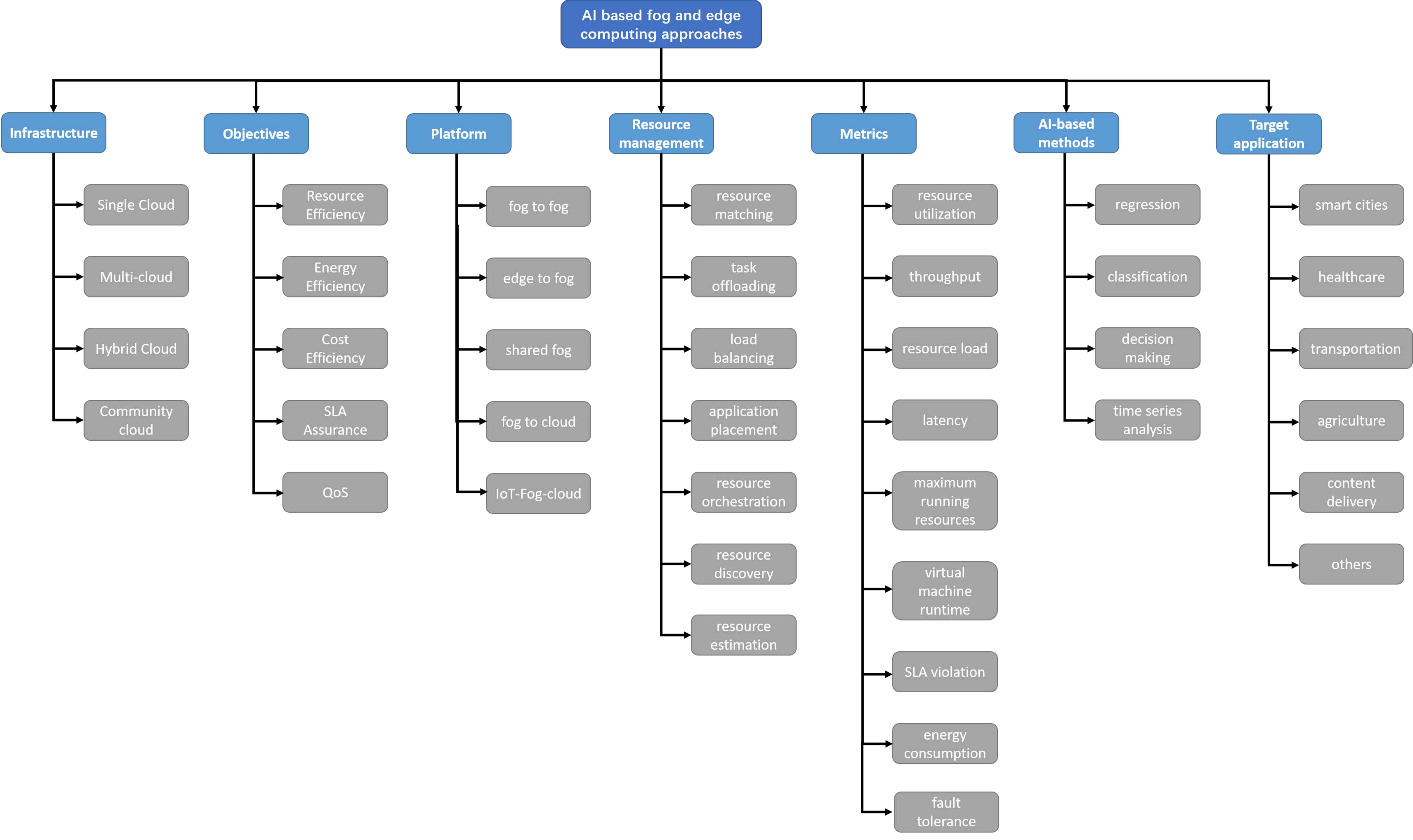}
\caption{Taxonomy of AI-based Fog and Edge Computing Frameworks}
\label{fig_tax}
\end{figure}

In this section, a comprehensive taxonomy of AI-based fog and edge computing approaches is proposed based on the existing studies following a systematic review. The taxonomy of the framework is shown in Fig.~\ref{fig_tax}, which includes infrastructure that supports the platform, objectives that the proposed approach aims to achieve, deployed platform, the mechanism for resource management, metrics for performance evaluations, the category of AI-based methods, and target application. Each taxonomy is further classified into the detailed study of AI-based fog and edge computing framework. 

\textit{Infrastructure:} the AI-based fog and edge computing approaches can be supported by different infrastructure models including single cloud, multi-cloud, hybrid cloud and community cloud with different focuses. For example, multiple clouds can work collaboratively to complete the partitioned deep learning tasks in fog to edge environment. 

\textit{Objectives:} based on our investigation, we notice that the existing major optimization objectives in AI-based fog and edge computing approaches are improving resource efficiency, reducing energy consumption, decreasing cost efficiency, assuring SLA and ensuring QoS.

\textit{Platform:} the platform indicates how the fog and edge computing approach is deployed. The current mainstream platforms include fog-to-fog, edge-to-fog, shared fog, fog-to-cloud and IoT-Fog-cloud. 

\textit{Resource Management:} one of the key challenges in fog and edge computing environments is managing the resources efficiently. The current research has been conducted for resource matching that maps the suitable amount of resources to tasks, task offloading that processes task collaboratively among fog and edge, load balancing that balances the workloads for different nodes, application placement that deploys the fog/edge applications to devices, resource orchestration that automates the resource allocation, resource discovery that provides the naming services of available services and resource estimation that predicts the amount the required resources. 

\textit{Metrics:} multiple metrics have been utilized to evaluate the performance of proposed approaches. The dominant metrics include resource utilization, throughput, resource load, latency, maximum running resources, virtual machine runtime, SLA violations, energy consumption and fault tolerance. 

\textit{AI-based methods:} the fog and edge environment has adopted AI-based methods to assist their management. Some existing categories of AI-based methods include regression, classification, decision making and time series analysis, which can be applied to workloads prediction, application feature analysis and making resource scheduling policies. 

\textit{Target applications:} the investigated approaches have been applied to support IoT applications in different areas including smart cities, healthcare, transportation, agriculture, content delivery and etc.

\subsection{Comparison of Existing Studies based on Taxonomy}
Table \ref{tab:tax} summarizes and compares the selected studies of AI-based fog and edge computing frameworks discussed in previous sections in terms of infrastructure, objectives, platform, resource management, metrics, AI-based methods and target application. For example, Golec \textit{et al.} \cite{golec2021ifaasbus} applied their approach in a multi-cloud environment and aimed to improve resource efficiency under the IoT-Fog-Cloud paradigm. They also utilized an AI-based approach for classification in resource orchestration to improve resource utilization. Wu \textit{et al.} \cite{wu2020collaborate} exploited the task offloading technique to reduce energy consumption and the proposed approach-based AI can help to make optimized decisions on when and how to manage offloaded tasks. Vu \textit{et al.} \cite{2020Dynamic} considered their scenario for smart cities with a hybrid cloud model to improve resource utilization under fog to cloud environment. Based on our investigation and comparison, we can notice that AI-based approaches have been comprehensively applied in fog and edge environments and more applications can be further incorporated into this paradigm. 

To summarize, the existing research works have covered all the types of dominant infrastructures including single cloud, multi-cloud and hybrid cloud. In terms of optimization objectives, most of the works focus on improving resource efficiency and QoS. As for the deployed platforms, Edge-to-Fog, IoT-Fog-Cloud, and Fog-to-Fog are the mainstream ones to support applications. The techniques applied to optimize resource management are diverse, including load balancing, resource matching, resource orchestration, application placement and task offloading. There are several metrics have been widely utilized to measure the performance of the proposed approach from different perspectives, including resource utilization, latency, energy consumption, throughput and SLA violation. The AI-based methods have been exploited for two main objectives, including classification, prediction and decision-making.

\begin{table*}[]
\caption{Comparison of Existing Studies based on Taxonomy}
\label{tab:tax}
\resizebox{\textwidth}{!}{
\begin{tabular}{|c|c|c|c|c|c|c|c|}
\hline

\textbf{Approach} & \textbf{Infrastructure} & \textbf{Objectives} & \textbf{Platform} & \textbf{Resource management} & \textbf{Metrics} & \textbf{AI-based methods} & \textbf{Target application} \\ \hline
Rafique \textit{et al.} \cite{rafique2019novel} & Multi-cloud & Resource Efficiency & edge to fog & load balancing & resource utilization & None & others \\ \hline
Golec \textit{et al.} \cite{golec2020biosec} & Single Cloud & Others (safety) & IoT-Fog-cloud & resource matching & latency & None & others \\ \hline
Golec \textit{et al.} \cite{golec2021ifaasbus} & Hybrid Cloud & Resource Efficiency & IoT-Fog-cloud & resource orchestration & running resources & classification & healthcare \\ \hline
Iftikhar \textit{et al.} \cite{iftikhar2022fog} & Multi-cloud & \begin{tabular}[c]{@{}c@{}}Resource Efficiency \end{tabular} & edge to fog & resource orchestration & resource utilization & decision making & others \\ \hline
McChesney \textit{et al.} \cite{defog-jonathan2019} & Hybrid Cloud & Resource Efficiency & edge to fog & application placement & latency & None & others \\ \hline
Aazam \textit{et al.} \cite{2015Dynamic} & Multi-cloud & Cost Efficiency & edge to fog & resource estimation & energy consumption & decision making & others \\ \hline
Aazam \textit{et al.} \cite{2015Fog} & Hybrid Cloud & QoS; Cost Efficiency & IoT-Dog-cloud & resource estimation & resource utilization & decision making & others \\ \hline
Ahmed \textit{et al.} \cite{abdelmoniem2021dc2} & Multi-Cloud & Resource Efficiency & fog to cloud & resource orchestration & throughput & classification & others \\ \hline
Bi \textit{et al.} \cite{2017Computation} & Single Cloud & Resource Efficiency & IoT-Dog-cloud & task offloading & resource load & regression & others \\ \hline
Sim \textit{et al.} \cite{2018Agent} & Multi-cloud & Resource Efficiency & fog to fog & resource orchestration & None & None & others \\ \hline
Vu \textit{et al.} \cite{2020Dynamic} & Hybrid Cloud & Resource Efficiency & fog to cloud & resource orchestration & resource utilization & None & smart cities \\ \hline
Yadav \textit{et al.} \cite{2019Trust} & Multi-cloud & QoS & fog to fog & application placement & running resources & decision making & others \\ \hline
Yao \textit{et al.} \cite{2020Task} & Multi-cloud & QoS & edge to fog & task offloading & latency & decision making & others \\ \hline
Wu \textit{et al.} \cite{wu2020eedto} & Hybrid Cloud & QoS & edge to fog & resource orchestration & SLA violation & decision making & others \\ \hline
Xue \textit{et al.} \cite{xue2021eosdnn} & Hybrid Cloud & Energy Efficiency & edge to fog & task offloading & energy consumption & decision making & others \\ \hline
Wu \textit{et al.} \cite{wu2020collaborate} & Hybrid Cloud & Energy Efficiency & edge to fog & task offloading & energy consumption & decision making & others \\ \hline
Liu \textit{et al.} \cite{liu2017adaptive} & Single Cloud & Resource Efficiency & edge to Cloud & Resource Provisioning & Cost, Resource Utilization & decision making & others \\ \hline

\end{tabular}
}
\end{table*}

\section{Result Outcomes}
\label{Result Outcomes}
\begin{figure*}[ht!]
    \centering
    \includegraphics[width=0.8\linewidth]{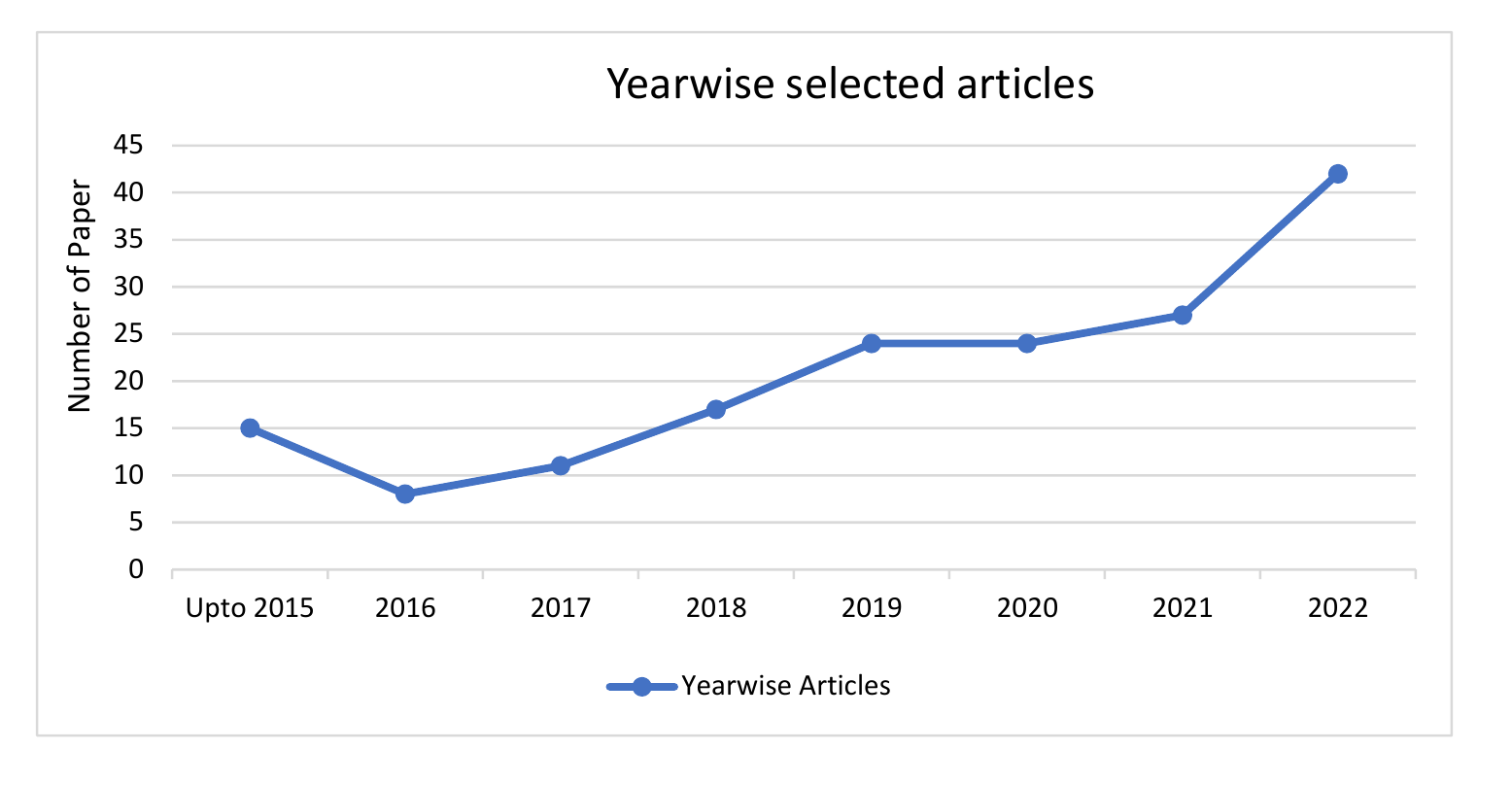}
    \caption{Year-wise Publications of AI/ML based Fog/Edge computing papers}
    \label{fig:Year-wise Publicationl}
\end{figure*}

Our study highlights the systematic review of various articles in order to understand the prevailing status of fog/edge computing. The extensive study comprises various driving forces responsible for making an impact on emerging paradigms in the form of open issues and future work. In total, we collected 320 articles, out of which 135 were shortlisted after the iterative selection process. The articles emphasize the state-of-art work done in the domain of research management in fog/edge and how the implication of intelligent paradigms like Artificial Intelligence, Machine Learning is inciting researchers. The taxonomy of our study has been designed with references to articles from the year 2015 to 2022. As depicted in Figure \ref{fig:Year-wise Publicationl}, the majority of our referred papers are from the year 2022. This accentuates the fact that our survey includes the latest work done by the research community. 

\begin{figure*}[ht!]
    \centering
    \includegraphics[width=.8\linewidth]{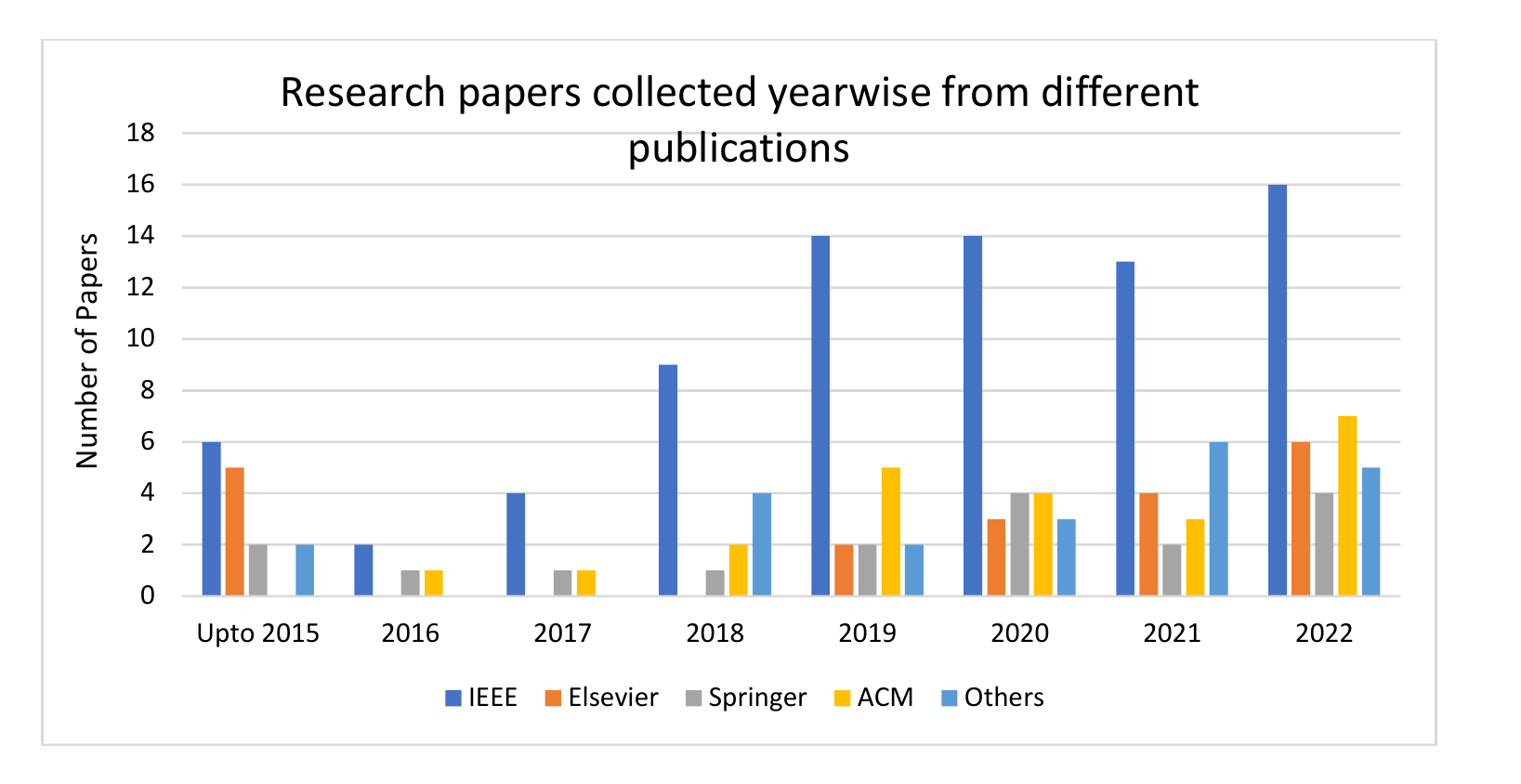}
    \caption{Bifurcation of research papers on the basis of publishers}
    \label{fig:collected yearwise}
\end{figure*}

\begin{figure*}[ht!]
    \centering
    \includegraphics[width=.8\linewidth]{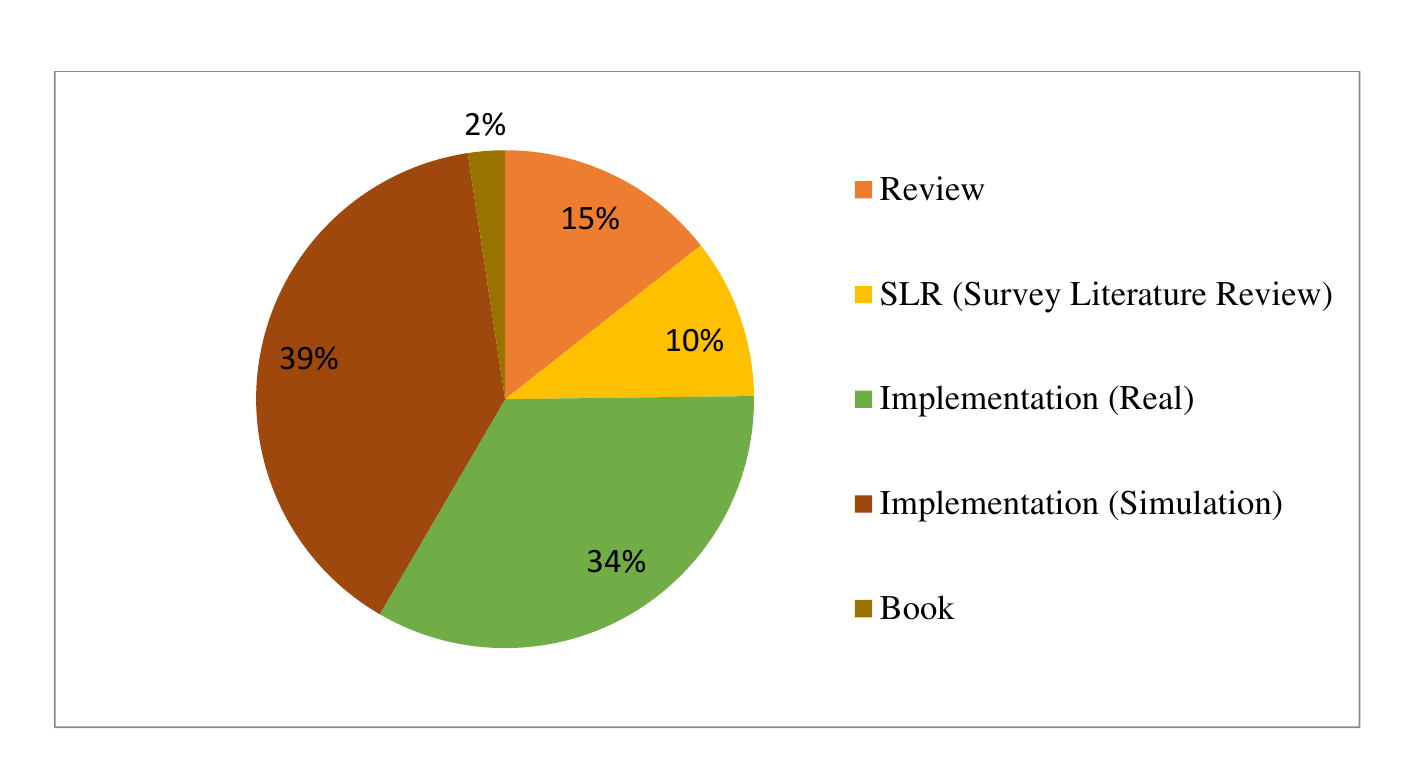}
    \caption{Study type of research paper}
    \label{fig:Study type}
\end{figure*}

The structure and methodology of the survey are inspired by the Systematic literature review (SLR) procedure by Kitchenham \cite{kitchenham2004procedures}. Furthermore, the identification of research questions channelizes the process flow of reviewing methodology. In a research review, the search process comprises the research topic which plays a significant role. The content in this paper has been accumulated from various sources including ACM Digital Library, IEEE Xplore, Springer Link, and other resources like Scopus, National Digital Library and electronic scientific research databases. Figure \ref{fig:collected yearwise} describes the yearly bifurcation of various sources in the form of paper count from different publications that represent most of the articles are from IEEE journals, transactions and conferences as compared with other publishers. Further, we have rigorously reviewed every article and divided it into five sections review, survey literature review, Implementation in real and simulation environments and book chapters as shown in Figure \ref{fig:Study type}. 

\begin{figure*}[ht!]
    \centering
    \includegraphics[width=0.8\linewidth]{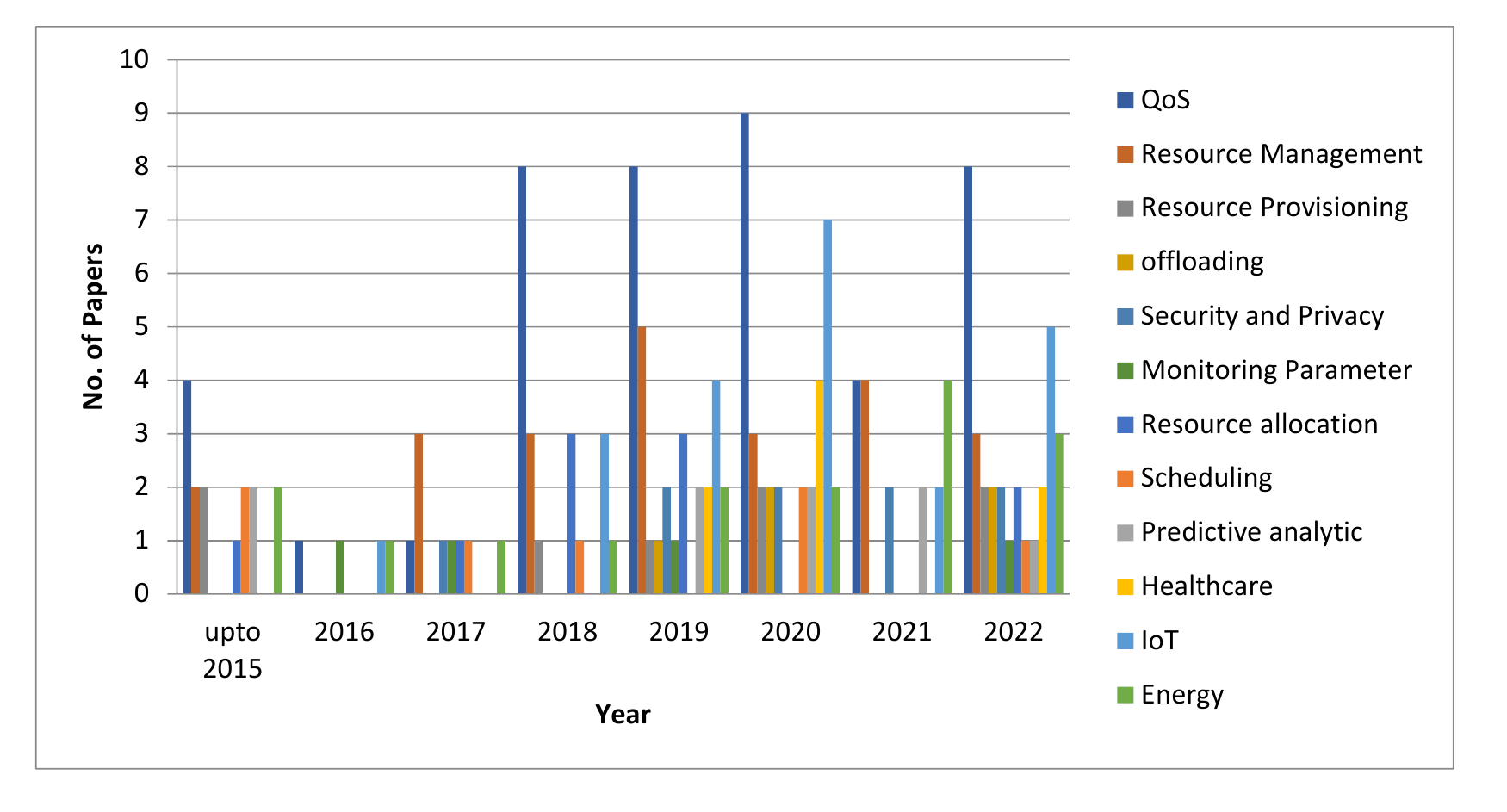}
    \caption{Categorization of papers based on factors relating to fog/edge computing }
    \label{fig:Categorization of papers}
\end{figure*}

This study considers various aspects of fog/edge computing which have been categorized into resource relating aspects of resource management further categorized as resource provisioning, task offloading and resource allocation), QoS parameters, and concerning other factors relating to real-world challenges like IoT, healthcare, security and privacy as shown in Figure \ref{fig:Categorization of papers}. A major chunk of our survey is inspired by the resource aspect comprising 61 papers and QoS parameters including 49 papers. The studies have been demonstrated in chronological order similar to other studies for the identification of state-of-art work in an effective manner. 

\section{Open Issues and Future Directions}
\label{Open Issues and Future Directions}
Despite the fact that a significant amount of progress has been made in AI and ML thanks to fog and edge computing. Despite this, there are a great number of problems and obstacles in this area that need to be solved. We have compiled a list of outstanding challenges in this field based on the existing literature.
\subsection{Heterogeneity}
Fog and edge computing are meant to support IoT applications that will emerge in different programming languages (e.g., C, Python, and Java), hardware architecture (e.g., ARM and AMD), processing units (e.g., CPU, GPU, and TPU), etc.
Such heterogeneity appears a highly challenging concern for the problem of resource management. That is, a resource manager, e.g., a resource balancer, requires understanding such differences in decision-making. Otherwise, traditional homogeneous-based solutions would result in considerable resource wastage and inappropriate decisions made by the resource manager. For instance, if some edge devices are ARM-based and others are AMD-based architecture, the resource manager must adhere to this difference since IoT applications may be incompatible with counterpart architectures. Another example is when fog devices provide unequal computation capacities. In this case, the resource manager requires an understanding of the devices' capacity to treat them proportionally. Otherwise, resources will be wasted and the QoS will degrade. When AI-based solutions are approached, the heterogeneity becomes even further challenging since AI models are agnostic to the heterogeneity while they perform completely differently on different devices.  For instance, if an edge device to run AI models is enabled with accelerators such as GPU which can provide higher precision and shorter latency, the resource manager requires an understanding of such differences to other edge devices that lack accelerators. Otherwise, costly accelerators will be undermined. Hence, heterogeneous resources at the edge will bring several opportunities for AI-based resource management, only if the traditional solutions are redesigned to support them.  

\subsection{Environmental Sustainability}
Sustainability appears to be a first-class requirement of IoT applications, since many of them rely on renewable energy sources such as solar irradiation. Sustainability is further important since edge devices are presumably constrained in computation resources. With such characteristics, edge devices strive for resource efficiency, in terms of energy or computations. However, this appears to conflict with the resource-hungry nature of AI models that demand a considerable amount of resources to be able to perform as expected. 

With the sustainability requirements of edge and resource-hungry features of AI models, it is very challenging to welcome AI at the edge. The hardware sector at the edge side and the software developing sector on the AI have to progress towards this ambition. However, as far as AI-based resource management is concerned, certain considerations can be assessed by the community. That is, a trade off between how much benefit the AI-based solution, as compared to a non-AI-based solution, can achieve and how much resource is consumed is a key question.

Another direction is, instead of asking about using AI-based solutions or not, what sort of AI-based solutions in terms of precision should be used? For instance, models can train to perform on lightweight frameworks to remain edge-friendly, but provide weaker precision. For instance, a resource manager can perform on the TensorFlow framework to make precise decisions, but consume a considerable amount of resources; or can perform on TensorFlow Lite to consume fewer resources upon inferences, but provide weaker precision. Hence, the open question is when and how to utilize AI-based solutions to satisfy the requirements of IoT applications, while achieving desirable precision.


\subsection{Security versus Efficiency}
AI-based resource management solutions for IoT applications require to adhere to security concerns. While edge computing is to not necessarily rely on the cloud, it brings its own limitations such as security. IoT applications in many domains such as Smart Home or Smart Healthcare, require the edge platform to adhere to such. The resource manager is a key component of an edge platform, hence, requires re-architect to satisfy that. This appears challenging due to the distributed nature of edge platforms. However, when AI-based solutions are enabled, this becomes even further challenging since AI-based models continuously require observations of the IoT applications to obtain sufficient data for the training or inference. Hence, the open issue here is how to satisfy AI requirements without degrading security. From another perspective, the security requirement itself can be driven by AI itself where the resource manager can utilize AI to learn patterns, outliers and features that can affect the security of an AI-based resource manager. An open question here would be how to leave such responsibilities to the resource manager through AI. 

\subsection{AI for Edge and AI at the Edge}
AI-based resource management represents a perfect example of AI for an edge while AI-based applications, such as object detection, classification, etc, represent AI-based applications serving at the edge. Throughout this paper, AI for edge was discussed, but the cohabitation of AI applications and AI managers would raise several new opportunities and challenges. 
A key open issue is how to shift the resource manager's focus from tuning the resources, e.g., by offloading, scheduling, etc, to tuning the AI application to achieve the adaptation. 
This is important because AI applications can be tuned to consume different amounts of resources from CPU to accelerators. Even on CPUs, they can perform differently such that they affect the energy and computation resources variably. Hence, with AI both for and at the edge, there are many opportunities that require investigations.

\subsection{AI for Serverless Computing}
While new AI applications are being used at the edge for certain use cases such as object detection for a smart traffic light, new application deployment and development models are also entering the edge. Serverless computing with its Function-as-a-Service (FaaS) is one of the major technologies in this context. FaaS is an application development model that turns bulky applications into single-purpose execution units, called functions, that are deployed upon event-based invocations and terminated after executions to save cost and resources. However, if long-running AI-based applications and resource managers stem into this area, several questions would raise that require investigation. For instance, would FaaS, whose billing model highly relies on the execution duration, still be cost-efficient? would this ephemerality (executing and terminating) be a bonus for the AI-based resource manager to remain resource efficient? Would the cold start of a function, the time duration from invoking to launching, deteriorate for AI-based applicants that require loading presumable heavy run-times?

\subsection{Resource Federation}
Distributed edge and fog devices require shared data and computation to function properly. Using AI-based managers for inference requires such data to be collected regularly and may be of a bigger size than typical raw data. For instance, if a traditional manager collects CPU usage of devices, an AI-based solution may collect other forms of data such as objects. Using AI-based managers for training also requires much more data. Add to this demand the scatteredness and scale of a network in a distributed edge that can span up to thousands of devices. Given a such scale, research areas around decentralized and federated AI-based resource management appear highly important. The decentralization means each resource manager is in charge of a portion of the cluster. The federation means while each portion works in isolation, they can collaborate with peers to use resources or to achieve a collective goal. Such areas have already commenced in edge computing and AI, but little effort has been made, particularly in the area of resource management which requires consideration.

\section{Summary and Conclusions}
\label{Summary and Conclusions}
In this work, we have conducted a comprehensive literature review on how machine learning and artificial learning-based solution are utilized for the resource management problem in fog and edge computing environments. Recent research works have witnessed a quickly growing trend of adopting AI-based methods to address the limitations of traditional heuristic approaches without sufficient consideration of diverse and dynamic factors in the environment. Compared with most traditional heuristic methods, AI-based approaches can be used to make accurate resource management decisions with lower time overhead, model and predict application and infrastructure metrics to improve the quality of services. Our work also advances the relevant surveys by considering fog computing and edge computing together with extensive comparisons 

To summarize, we have noticed that AI-based methods have been applied in a wide range of scenarios, including resource estimation, resource discovery, resource matching, task offloading, load balancing, resource orchestration, application placement, and resource consolidation. We also observe that the applications deployed on fog and edge computing environment ranges from healthcare, smart home, agriculture, smart transportation, and spatial. Significant efforts have been made to utilize advanced AI-based approaches, e.g. DNN, Q-learning, DQN, and reinforcement learning-based algorithms, to optimize resource utilization, throughput, SLA violations, energy consumption, and fault tolerance. 

In conclusion, although the relevant research progresses fast, there is no systematic literature review that combines fog and edge computing with an AI-based optimization framework in charge of the whole resource management process. Employing microservice and serverless can be a promising approach to further optimize the application and system performance with fine-grained resource control. This taxonomy work will assist the researcher to find the important research directions in edge and fog computing and will also help to choose the most suitable AI-based methods for efficient resource management under the hybrid paradigm under a dynamic environment.

\section*{Acknowledgements}
Sundas Iftikhar would like to express her thanks to the Higher Education Commission (HEC) Pakistan, for their support and funding. This work is partially funded by Chinese Academy of Sciences President’s International Fellowship Initiative (Grant No. 2023VTC0006), and Shenzhen Science and Technology Program (Grant No. RCBS20210609104609044).

\section*{Conflict of Interest}
On behalf of all authors, the corresponding author states that there is no conflict of interest.

\section*{Appendix A: A quality assessment forms}  
A.1. Preliminary Examining Questions: Table \ref{A11}  represents the list of questions used during the preliminary examination. 

\begin{table*}[ht]
\centering
\caption{Preliminary Examining questions} \label{A11}
\begin{tabular} {| p{14cm} | p {1cm}| | p {1cm}|}   \hline 
\textbf{Question} & \textbf{Yes} & \textbf{No}\\  \hline
\textbf{Q1. Does the article discuss the use of AI/ML in Fog/Edge Computing?}  \newline
This report compiles findings from studies conducted on AI/ML in Fog/Edge Computing. This survey takes into consideration all of the research publications, including case studies, experimental studies, and so on. &  & \\ \hline
\textbf{Q2. Is the primary emphasis of this paper the AI/ML-based management of resources in Fog/Edge Computing?}  \newline
Does this paper provide a method, approach, system, or framework for resource management that could be used for AI/ML in Fog/Edge Computing? \newline
Is the validity of this investigation ensured by utilising a simulated testbed for fog/edge computing? \newline
Is the validity of this investigation ensured by utilising a real testbed for fog/edge computing?&  & \\ \hline

\end{tabular}
\end{table*}

A.2. Specific questions: Table \ref{A12} represents the list of questions used during evaluations. \\
\begin{table*}[ht]
\centering
\caption{Specific questions} \label{A12}
\begin{tabular} {| p{14cm} | p {1cm}| | p {1cm}|}   \hline 
\textbf{Question} & \textbf{Yes} & \textbf{No}\\  \hline
\textbf{Q1: }what resource management methods are available that are based on artificial intelligence and machine learning?  &  & \\  \hline
\textbf{Q2: }Where are AI/ML-based fog/edge computing frameworks stand right now??  &  & \\ \hline
\textbf{Q3: }How can the efficiency of AI/ML-based fog/edge computing be measured, and what metrics are used for this purpose?  &  & \\ \hline 
\textbf{Q4: }Which simulators are utilized for fog/edge computing that is based on AI/ML?  &  & \\ \hline
\textbf{Q5: }What are the most common applications of IoT-enabled Edge/Fog AI?  &  & \\ \hline
\textbf{Q6: }What kinds of workloads are utilised to evaluate the efficacy of AI/ML-based fog/edge computing frameworks?  &  & \\ 
\hline 
\end{tabular}
\end{table*}

\section*{Appendix B:  Data items extracted from all articles} 
Table \ref{AB} shows the data items extracted from all articles.\\ 
\begin{table*}[ht]
\centering
\caption{Data items extracted from all articles} \label{AB}
\begin{tabular} {| p{5cm} | p {10cm}|}   \hline 
\textbf{Data Item} & \textbf{Description} \\  \hline
Paper identifier & Digital Object Identifier (DOI) \\
Online Publication Date & Publication Year \\
Bibliographic Information & Author(s) Name(s), Publication Date, Article Title, and Journal Name\\
 Type of Article & Conference, Workshop and Journal  \\
Motivation  & What exactly are the primary aims of this work?  \\
Innovation  & Mechanism and context/application \\
 What is the Problem Statement    & The problem, as well as a description of it, is addressed and resolved in the research. \\
 What is the method for managing resources? & AI/ML-based Resource Management Technique for Fog/Edge Computing \\
  Implementation Environment    & The technique is carried out utilising either a simulated or actual setting. \\
 Performance Evaluation    & Which constraints were taken into account when the technique was analysed?  \\
 Workload Type & How do you create a dataset for use in experiments? \\
 Performance Metrics  & How are the results of a research evaluated using what kind of QoS metrics? \\
Drawbacks  & Where do you see the field of research going in the future? \\
                    
\hline 

\hline 
\end{tabular}
\end{table*}

\section*{Appendix C: Journals and Conferences for publishing articles about AI/ML in Fog/Edge Computing.} 
Table \ref{AC}   lists the top journals and conferences for publishing articles about AI/ML in fog/edge computing. \textbf{Notations: }J – Journal (including IEEE/ACM Transactions), C – Conference, W – Workshop, N – The total number of papers that reported AI/ML-based Resource Management Technique for Fog/Edge Computing as their primary research focus, \# – The total number of publications examined.\\
\begin{table*}[ht]
\centering
\caption{Appendix C: Journals, Workshops and Conferences} \label{AC}
\resizebox{1\textwidth}{!}{
\begin{tabular} {| p{12cm} | p {2cm}| | p {1cm}| | p {1cm}|}   \hline 
\textbf{Publication Venue} & \textbf{J/C/W} & \textbf{\#} & \textbf{N}\\  \hline
IEEE Transactions on  Parallel and Distributed Systems   & J   & 4   & 4  \\
 IEEE Transactions on Cloud Computing  &  J & 6   & 11  \\
 IEEE Transactions on Services Computing & J & 4 & 6 \\
IEEE Internet of Things Journal & J & 17 & 35 \\
IEEE Transactions on Industrial Informatics   &  J & 7  & 13  \\
IEEE Transactions on Vehicular Technology   & J   & 2   & 6  \\
IEEE Transactions on Network and Service Management  & J & 2  & 3  \\
IEEE Transactions on Sustainable Computing   & J  & 1  & 3  \\
IEEE/ACM Transactions on Networking & J & 1  & 2  \\
IEEE Transactions on Mobile Computing & J & 3  & 6  \\
IEEE Transactions on Wireless Communications & J & 2  & 5  \\
IEEE Transactions on Green Communications and Networking & J & 2  & 4  \\
IEEE Transactions on Computational Social Systems  & J & 1  & 1  \\
IEEE Transactions on Network Science and Engineering   & J  & 2  & 2  \\
IEEE Transactions on Consumer Electronics    & J & 1  & 1  \\
IEEE Transactions on Industry Applications     & J & 1  & 1  \\
IEEE Transactions on Broadcasting & J & 1  & 1  \\
IEEE Transactions on Intelligent Transportation Systems & J & 2  & 2  \\
ACM Transactions on Internet Technology  & J  & 4  & 6  \\
ACM Transactions on Internet of Things & J & 4  & 6  \\
ACM Transactions on Sensor Networks & J & 1  & 1  \\
IEEE Access & J & 8 & 41  \\
Future Generation Computer Systems   & J  & 4  & 9  \\
Journal of Parallel and Distributed Computing   & J  &  2 & 4  \\
Journal of Systems and Software   & J  & 4  & 6  \\
Software: Practice and Experience & J & 9 & 25 \\
Journal of Network and Computer Applications   &  J & 3  & 5  \\
Transactions on Emerging Telecommunications Technologies  & J    & 1  & 2   \\
Internet of Things (Elsevier)  & J  &  7 &  12 \\
IEEE International Symposium on Cluster, Cloud and Internet Computing (CCGrid)   & C  & 1  & 2  \\
 Euromicro Conference on Software Engineering and Advanced Applications  &  C & 1  &  2 \\

IEEE International Conference on Distributed Computing Systems (ICDCS)   &  C  & 2  &  4 \\
 IEEE International Conference on Communications  & C  & 3  &  5 \\
Australasian Computer Science Week Multiconference   & C  & 2  &  4 \\

IEEE/ACM International Conference on Utility and Cloud Computing (UCC) & C  & 1  &  1 \\
International Conference on Service-Oriented Computing & C  & 1  &  3 \\
IEEE International Conference on Pervasive Computing and Communication (PerCom) Workshop & W  & 1  &  2 \\
IEEE International Conference on Advanced Networks and Telecommunications Systems (ANTS) & C  & 1  &  2 \\
International Conference on Internet of Things & C  & 1  &  4 \\
IEEE International Conference on Networking, Architecture and Storage (NAS) & C  & 1  &  2 \\
\hline 

\hline 
\end{tabular}
}
\end{table*}

\section*{Appendix D. List of Acronyms} 
Table \ref{AD} shows the list of acronyms. 
\begin{table}[ht]
\caption{List of Acronyms} \label{AD}
\begin{tabular} {| p{2cm} | p {6cm}|}   \hline 
\textbf{Abbreviation} & \textbf{Description} \\  \hline
IoT & Internet of Things \\
QoS & Quality of Service \\
SLA &   Service-Level Agreement   \\
VM & Virtual Machines    \\
ML  &  Machine Learning  \\
AI   &  Artificial Intelligence \\
SLO  &   Service Level Objectives \\
 RQ &   Research Questions \\
 FC   & Fog Computing  \\
   EC   & Edge Computing  \\ 
   IoHT     &  Internet of Health Things \\
PDA          &   Personal Digital Assistant\\
     IMCF+       &   IoT Meta-Control Firewall \\
  ITS   & Intelligent Transportation Systems   \\
  DL   &  Deep Learning \\ 
    RL    &  Reinforcement Learning \\
 DRL & Deep Reinforcement Learning  \\
CL  &  Centralized learning \\
 FDMA & Frequency Division Multiple Access  \\
 TDMA & Time Division Multiple Access   \\
CDC  &  Cloud Data Centers \\
 VFC & Vehicular Fog Computing  \\
 DQN &  Deep Q Network \\
MNO  &  Mobile Network Operator \\
 FRL &  Fuzzy Reinforcement Learning \\
 DDQL &  Double Deep Q Learning \\
 MDP & Markov Decision Process  \\
 ANN & Artificial Neural Network  \\
GNN  &  Graph neural network \\
 GPU  & Graphical Processing Unit  \\
 TPU &  Tensor Processing Unit \\
  
\hline 

\hline 
\end{tabular}
\end{table}

\bibliographystyle{IEEEtran}
\bibliography{ms}
\section*{Authors Biography}

\parpic{\includegraphics[width=0.8in,clip,keepaspectratio]{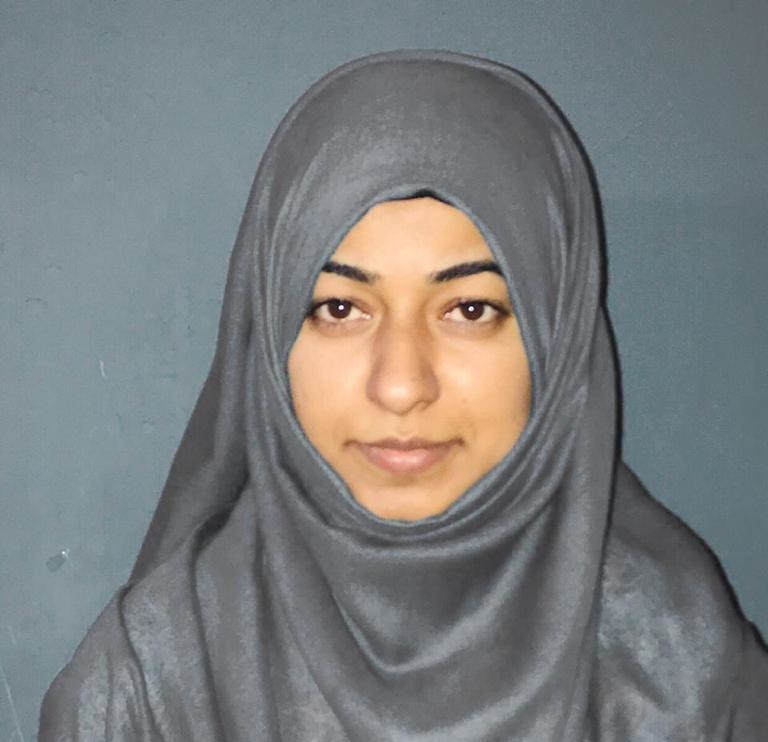}}
\noindent {\footnotesize \sffamily{\bf  Sundas Iftikhar}
is a Ph.D. Scholar at the School of Electronic Engineering and Computer Science, Queen Mary University of London. Prior to this, she held positions as Research associate and Lecturer at University of Kotli Azad Jammu and Kashmir, Azad Kashmir, Pakistan. She did her Masters in computer software engineering from National University of Science and Technology, Pakistan. Her research interest include Cloud computing, Fog computing, and resource Management in Fog.} \\

\parpic{\includegraphics[width=0.8in,clip,keepaspectratio]{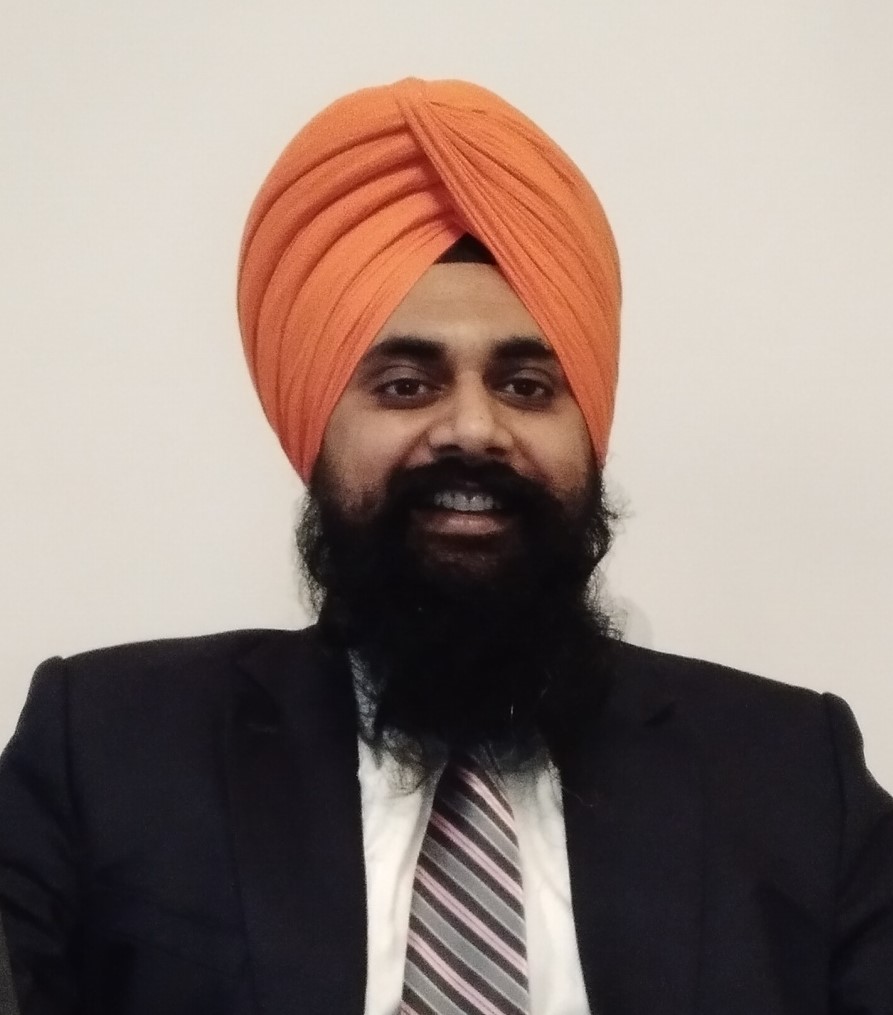}}
\noindent {\footnotesize \sffamily{\bf Sukhpal Singh Gill} is a Lecturer (Assistant Professor) in Cloud Computing at the School of Electronic Engineering and Computer Science, Queen Mary University of London, UK. Prior to his present stint, Dr. Gill has held positions as a Research Associate at the School of Computing and Communications, Lancaster University, UK and also as a Postdoctoral Research Fellow at CLOUDS Laboratory, The University of Melbourne, Australia. Dr. Gill is serving as an Associate Editor in Wiley ETT and IET Networks Journal. He has co-authored 70+ peer-reviewed papers (with H-index 30+) and has published in prominent international journals and conferences such as IEEE TCC, IEEE TSC, IEEE TII, IEEE TNSM, IEEE IoT Journal, Elsevier JSS, ACM UCC and IEEE CCGRID. He has received several awards, including the Distinguished Reviewer Award from SPE (Wiley), 2018, Best Paper Award AusPDC at ACSW 2021 and has also served as the PC member for venues such as PerCom, UCC, CCGRID, CLOUDS, ICFEC, AusPDC. His research interests include Cloud Computing, Fog Computing, Software Engineering, Internet of Things and Energy Efficiency. For further information, please visit \url{http://www.ssgill.me}}. \\

\parpic{\includegraphics[width=0.8in,clip,keepaspectratio]{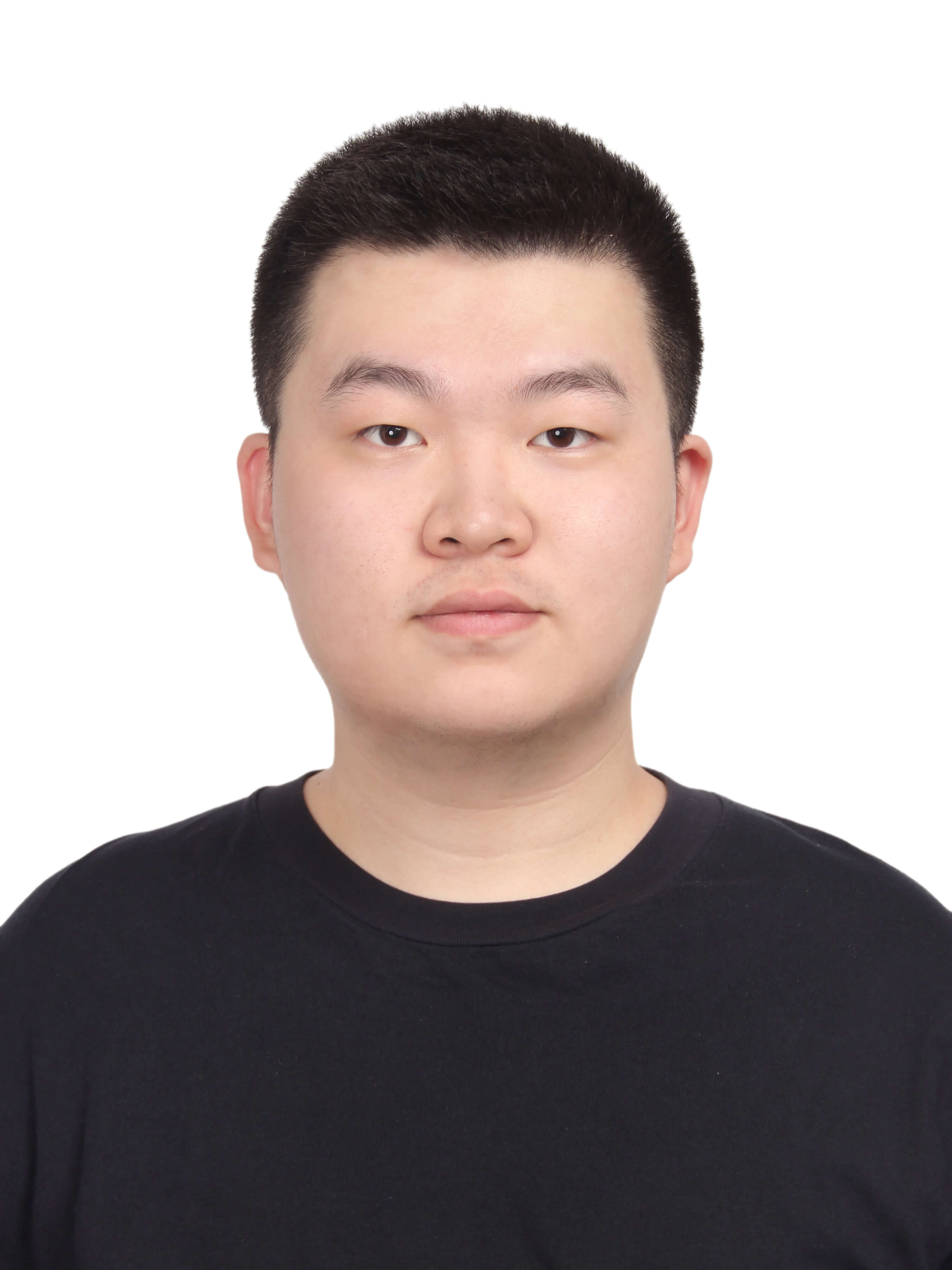}}
\noindent {\footnotesize \sffamily{\bf Chenghao Song}
received his BSc degree from University of Electronic Science and Technology of China. Now he is a master student at University of Melbourne, he is also a visiting student at Shenzhen Institutes of Advanced Technology, Chinese Academy of Science. His research interest includes deep learning for cloud resource optimization.
}\\ \\ \\

\parpic{\includegraphics[width=0.8in,clip,keepaspectratio]{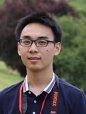}}
\noindent {\footnotesize \sffamily{\bf Minxian Xu} is currently an associate professor at Shenzhen Institute of Advanced Technology, Chinese Academy of Sciences. He received the B.Sc. degree and the M.Sc. degree, both in software engineering from University of Electronic Science and Technology of China. He obtained his Ph.D. degree from the University of Melbourne in 2019. His research interests include resource scheduling and optimization in cloud computing. He has co-authored 30+ peer-reviewed papers published in prominent international journals and conferences, such as ACM Computing Surveys, IEEE Transactions on Sustainable Computing, IEEE Transactions on Cloud Computing, Journal of Parallel and Distributed Computing, Software: Practice and Experience, International Conference on Service-Oriented Computing. His Ph.D. Thesis was awarded the 2019 IEEE TCSC Outstanding Ph.D. Dissertation Award. He is member of CCF and IEEE. More information can be found at: \url{http://www.minxianxu.info}}. \\

\parpic{\includegraphics[width=0.8in,clip,keepaspectratio]{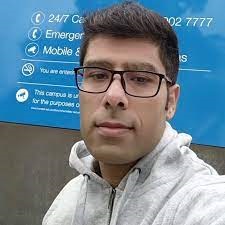}}
\noindent {\footnotesize \sffamily{\bf Mohammad
Sadegh Aslanpour} is a PhD student in Monash University and CSIRO’s DATA61, Australia. He obtained his MSc degree in Computer Engineering Islamic Azad University, Tehran Science and Research (Sirjan) Branch, Iran in 2016. He also obtained his Bachelor and Associate degrees in Computer-Software in 2012 and 2010, respectively. From 2011 to 2019, he worked in the industry, IT Department of Jahrom Municipality, Iran as Software Engineer. He is also serving as Editorial Board Member and Reviewer for some international high-ranked journals. His research interests include orchestration of Cloud, Fog and Edge Computing, Serverless Computing, and Autonomous Systems. For more details, please visit his homepage: aslanpour.github.io}\\

\parpic{\includegraphics[width=0.8in,clip,keepaspectratio]{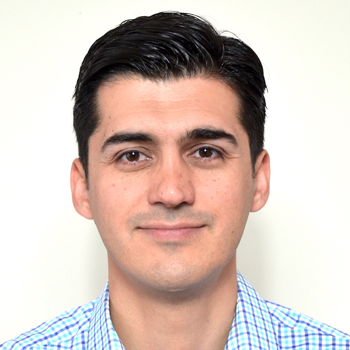}}
\noindent {\footnotesize \sffamily{\bf Adel N. Toosi}
is a senior lecturer (a.k.a. Associate Professor) at Department of Software Systems and Cybersecurity,
Faculty of Information Technology, Monash University, Australia. Before joining Monash, Dr Toosi was a
Postdoctoral Research Fellow at the University of Melbourne from 2015 to 2018. He received his Ph.D. degree in 2015 from the School of Computing and Information Systems at the University of Melbourne. His Ph.D. thesis
was nominated for CORE John Makepeace Bennett Award for the Australasian Distinguished Doctoral
Dissertation and John Melvin Memorial Scholarship for the Best Ph.D. thesis in Engineering. Dr Toosi made
significant contributions to the areas of resource management and software systems for cloud computing. His
research interests include Cloud/Fog/Edge Computing, Software Defined Networking, Green Computing and
Energy Efficiency. Currently, he is working on green energy harvesting for Edge/Fog computing environments.
For further information, please visit his homepage: http://adelndjarantoosi.info
}\\

\parpic{\includegraphics[width=0.8in,clip,keepaspectratio]{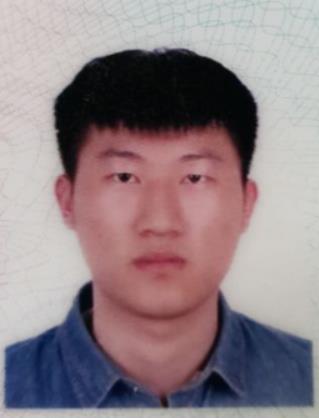}}
\noindent {\footnotesize \sffamily{\bf Junhui Du} received the BSc degree in mathematics from the Nanjing University of Information Science Technology, China, in 2021. He is currently working toward the master’s degree in mathematics with the Center for Applied Mathematics, Tianjin University, China. His research interests include Internet of Things, deep learning, and mobile edge computing.
}\\ \\ \\

\parpic{\includegraphics[width=0.8in,clip,keepaspectratio]{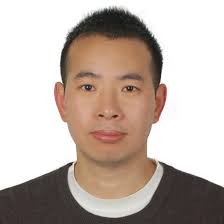}}
\noindent {\footnotesize \sffamily{\bf Huaming Wu} received the BE and MS degrees from the Harbin Institute of Technology, China, in 2009 and 2011, respectively, both in electrical engineering, and the PhD degree with the highest honor in computer science at Freie Universit\"at Berlin, Germany, in 2015. He is currently an associate professor with the Center for Applied Mathematics, Tianjin University, China. His research interests include model-based evaluation, wireless and mobile network systems, mobile cloud computing and deep learning. He is a senior member of IEEE and a member of ACM. For further information, please visit: \url{http://huamingwu.cn}.} \\

\parpic{\includegraphics[width=0.8in,clip,keepaspectratio]{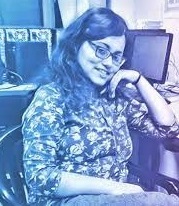}}
\noindent {\footnotesize \sffamily{\bf Shreya Ghosh} is a Postdoctoral Research Fellow at The Pennsylvania State University, Pennsylvania, USA. She received her PhD from the Department of Computer Science and Engineering, IIT Kharagpur, India in 2021. Her current research interests include machine learning, trajectory data mining, cloud computing and Internet of Spatial Things. Shreya is the recipient of the prestigious TCS fellowship. She is the member of AnitaB.org and student member of IEEE and ACM.}

\parpic{\includegraphics[width=0.8in,clip,keepaspectratio]{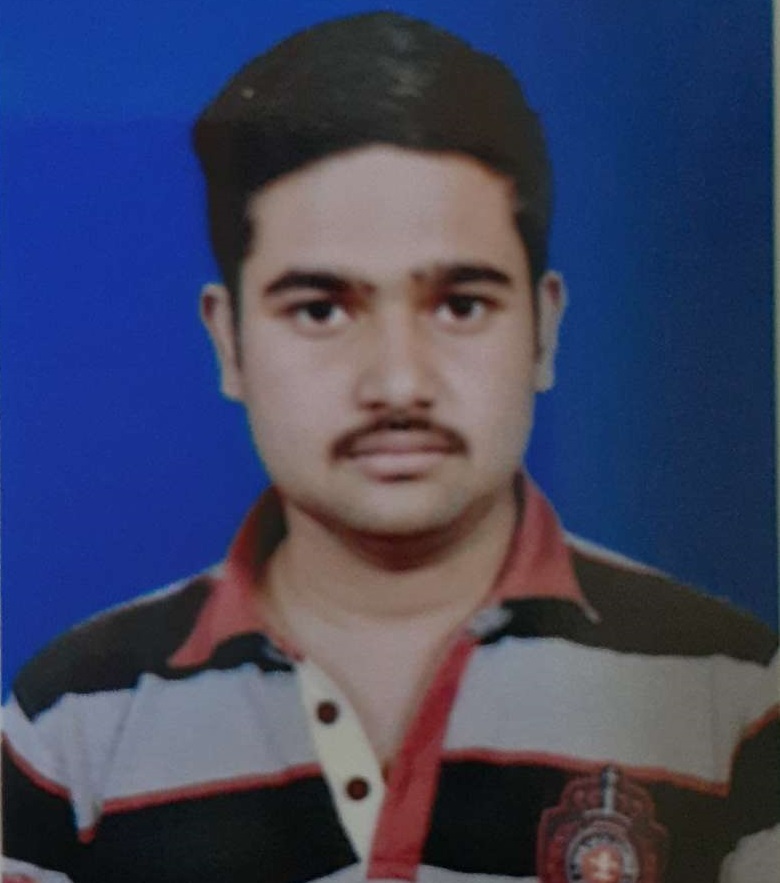}}
\noindent {\footnotesize \sffamily{\bf Deepraj Chowdhury}  is with department of electronics and communication, IIIT- Naya Raipur. He has Co-authored 4 research papers in different conferences like ICACCP 2021, INDICON 2021. He also has 3 Indian copyright registered, and applied for 2 Indian Patent. He is also serving as a reviewer in Wiley Transaction on emerging Telecommunication Technologies.} \\ \\

\parpic{\includegraphics[width=0.8in,clip,keepaspectratio]{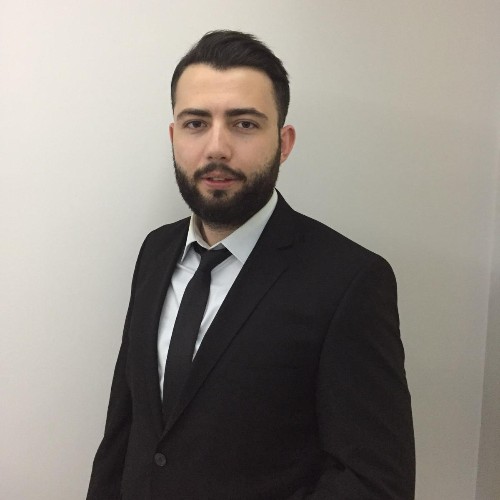}}
\noindent {\footnotesize \sffamily{\bf Muhammed Golec} is a PhD student in Computer Science at Queen Mary University. After his undergraduate graduation, he was awarded the Ministry of Education Scholarship, one of the most prestigious scholarships in his country. Within the scope of this scholarship, he graduated from Queen Mary University of London Computer Science with a high degree (Distinction). His master thesis was found successful and published in IEEE Consumer Electronics Magazine. He worked at Sisecam Company as an Electrical and Electronics Maintenance Engineer for one year to consolidate his academic skills in the private sector. His research interests include AI, Cloud Computing, and Security and Privacy. For further information, please visit \url{https://www.linkedin.com/in/muhammed-golec-b55756119/}.} \\

\parpic{\includegraphics[width=0.8in,clip,keepaspectratio]{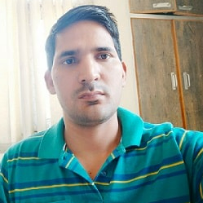}}
\noindent {\footnotesize \sffamily{\bf Mohit Kumar}
is Assistant Professor in the Department of Information Technology at Dr. B R Ambedkar National Institute of Technology, Jalandhar, India. He received his Ph.D. degree from Indian Institute of Technology Roorkee in the field of Cloud Computing, 2018, and M.Tech degree in Computer Science and Engineering from ABV-Indian Institute of Information Technology Gwalior, India in 2013. He has received his B.Tech degree in Computer Science and Engineering from MJP Rohilkhand University Bareilly, 2009. His research topics cover the areas of Cloud computing, Fog computing, Edge Computing, Internet of Things, Soft Computing. He has published more than 20 research articles in reputed journals and international conferences. He has been Session chair and keynotes Speaker of many International conferences, webinars, FDP, STC in India. He has guided two M.Tech Thesis and guided 1 Ph.D. Scholar. He is an active reviewer of several reputed journals and international conferences. He is a member of the IEEE.
}\\

\parpic{\includegraphics[width=0.8in,clip,keepaspectratio]{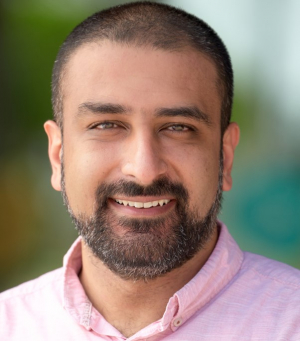}}
\noindent {\footnotesize \sffamily{\bf Ahmed M. Abdelmoniem}
is a Lecturer (Assistant Professor) of Big Data and Distributed Systems at the School of EECS, QMUL and leads the Scalable Adaptive Yet Efficient Distributed (SAYED) Systems Research Group. He has a PhD in Computer Science and Engineering from the Hong Kong University of Science and Technology (HKUST), Hong Kong. His research interests lie in the intersection of distributed systems, computer networks and machine learning. He is an investigator on several UK and international grants totalling nearly USD 1 million in funding. His work appears in top-tier conferences and journals including NeurIPS, AAAI, MLSys, ACM EuroSys, IEEE INFOCOM, IEEE ICDCS, and IEEE/ACM Transactions on Networking. He is interested in supervising students with a background in Computer Networks, Machine Learning, Distributed Systems and Cloud/Edge Computing} \\

\parpic{\includegraphics[width=0.8in,clip,keepaspectratio]{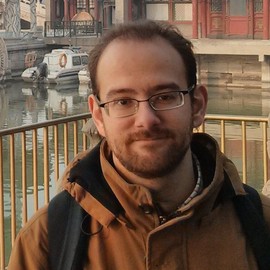}}
\noindent {\footnotesize \sffamily{\bf Felix Cuadrado}
received the Ph.D. degree in telecommunications engineering from the Universidad Politécnica de Madrid (UPM), Spain, in 2009. He is currently a Senior Distinguished Fellow (Beatriz Galindo scheme) with the Universidad Politécnica de Madrid, a Visiting Reader at the Queen Mary University of London, and a fellow of the Alan Turing Institute. He has numerous publications in top tier journals and conferences, including IEEE TRANSACTIONS ON SOFTWARE ENGINEERING, IEEE TRANSACTIONS ON CLOUD COMPUTING, Elsevier JSS, Elsevier FCGS, IEEE ICDCS, and WWW. His research explores the challenges arising from large-scale data-intensive applications through a combination of software engineering, distributed systems, and mathematical approaches
}\\

\parpic{\includegraphics[width=0.8in,clip,keepaspectratio]{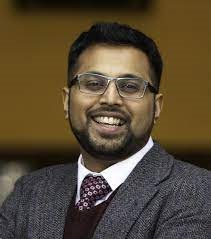}}
\noindent {\footnotesize \sffamily{\bf Blesson Varghese}
received the Ph.D. degree in computer science from the University of Reading, UK on international scholarships. He is a Reader (Associate Professor) in computer science at the University of St Andrews and an honorary faculty member at Queen’s University Belfast. He is the Principal Investigator of the Edge Computing Hub and was a Royal Society Short Industry Fellow to British Telecommunications plc. His interests include developing and analysing novel parallel and distributed systems and applications that span the cloud–edge–device continuum. More information is available from http://www.blessonv.com
}\\

\parpic{\includegraphics[width=0.8in,clip,keepaspectratio]{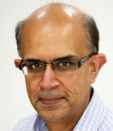}}
\noindent {\footnotesize \sffamily{\bf Omer Rana} is a Professor of Performance Engineering in School of Computer Science \& Informatics at Cardiff University and Deputy Director of the Welsh e-Science Centre. He holds a Ph.D. from Imperial College. His research interests extend to three main areas within computer science: problem solving environments, high performance agent systems and novel algorithms for data analysis and management. Moreover, he leads the Complex Systems research group in the School of Computer Science \& Informatics and is director of the `Internet of Things' Lab, at Cardiff University. He has published over 310 papers in peer-reviewed international conferences and journals. He serves on the Editorial Board of IEEE Transactions on Parallel and Distributed Systems, ACM Transactions on Internet Technology, and ACM Transactions on Autonomous and Adaptive Systems. He has served as a Co-Editor for a number of journals, including Concurrency: Practice and Experience (John Wiley), IEEE Transactions on System, Man, and Cybernetics: Systems, and IEEE Transactions on Cloud Computing.} \\

\parpic{\includegraphics[width=0.8in,clip,keepaspectratio]{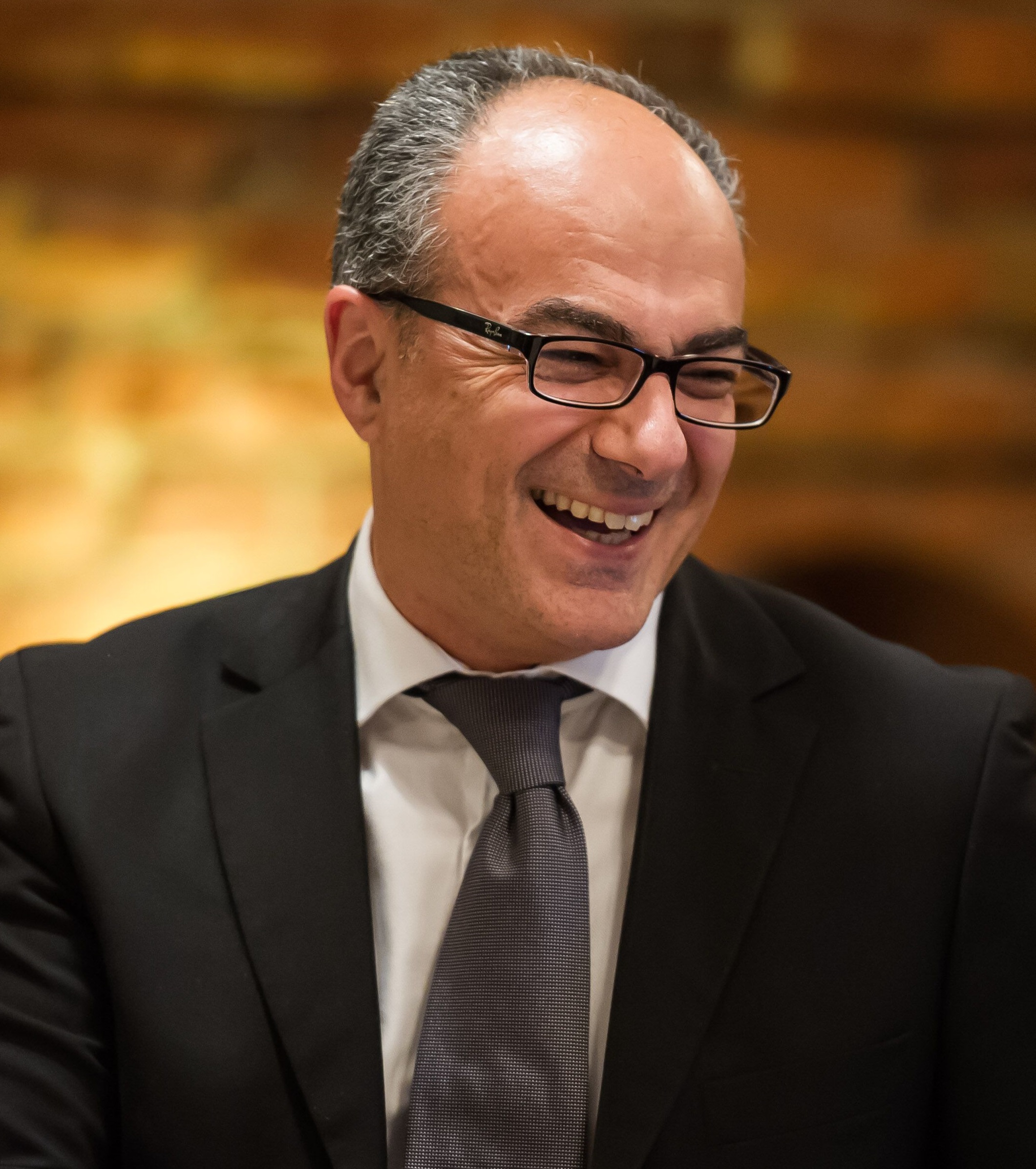}}
\noindent {\footnotesize \sffamily{\bf Schahram Dustdar} is Full Professor of computer science heading the Research Division of Distributed Systems at the TU Wien, Austria. He is founding Co-Editor-in-Chief of the new ACM Transactions on Internet of Things (ACM TIoT) as well as Editor-in-Chief of Computing (Springer). He is an Associate Editor of IEEE Transactions on Services Computing, IEEE Transactions on Cloud Computing, ACM Transactions on the Web, and ACM Transactions on Internet Technology, as well as on the editorial board of IEEE Internet Computing and IEEE Computer. Dustdar is IEEE Fellow (2016), recipient of the ACM Distinguished Scientist Award (2009), the ACM Distinguished Speaker ward (2021), the IBM Faculty Award (2012), an Elected Member of the Academia Europaea: The Academy of Europe, where he is Chairman of the Informatics Section. In 2021 Dustdar was elected EAI Fellow as well as Fellow and President for the Asia-Pacific Artificial Intelligence Association (AAIA).} \\

\parpic{\includegraphics[width=0.8in,clip,keepaspectratio]{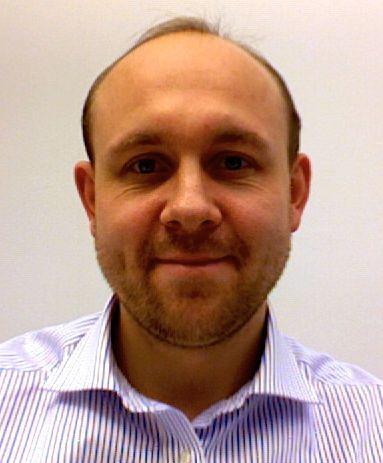}}
\noindent {\footnotesize \sffamily{\bf Steve Uhlig} obtained a Ph.D. degree in Applied Sciences from the University of Louvain, Belgium, in 2004. From 2004 to 2006, he was a Postdoctoral Fellow of the Belgian National Fund for Scientific Research (F.N.R.S.). His thesis won the annual IBM Belgium/F.N.R.S. Computer Science Prize 2005. Between 2004 and 2006, he was a visiting scientist at Intel Research Cambridge, UK, and at the Applied Mathematics Department of University of Adelaide, Australia. Between 2006 and 2008, he was with Delft University of Technology, the Netherlands. Prior to joining Queen Mary University of London, he was a Senior Research Scientist with Technische Universität Berlin/Deutsche Telekom Laboratories, Berlin, Germany. Since January 2012, he has been the Professor of Networks and Head of the Networks Research group at Queen Mary, University of London. Between 2012 and 2016, he was a guest professor at the Institute of Computing Technology, Chinese Academy of Sciences, Beijing, China. With expertise in network monitoring, large-scale network measurements and analysis, and network engineering, during his career he has been published in over 100 peer-reviewed journals, and awarded over £3million in grant funding. Awarded a Turing Fellow, Steve is also the Principal Investigator on a new project funded by the Alan Turing Institute: 'Learning-based reactive Internet Engineering' (LIME). He is currently the Editor in Chief of ACM SIGCOMM Computer Communication Review, the newsletter of the ACM SIGCOMM SIG on data communications. Since December 2020, Steve has also held the position of Head of School of Electronic Engineering and Computer Science. Current Research interests: Internet measurements, software-defined networking, content delivery.} \\
\end{document}